\documentclass[a4paper,fleqn,usenatbib]{mnras}

\usepackage[T1]{fontenc}
\usepackage{ae,aecompl}
\usepackage{tikz}
\usepackage{graphicx}	% Including figure files
\usepackage{amsmath}	% Advanced maths commands
\usepackage{amssymb}	% Extra maths symbols
\usepackage[caption=false]{subfig}
\usepackage{float}

\title[Statistical study of galaxy triplets]{Fundamental parameters of isolated galaxy triplets in the local Universe: Statistical study}

\author[Tawfeek et al.]{
Amira A. Tawfeek,$^{1,2}$\thanks{E-mail: amira$\_$nriag@yahoo.com}
Gamal B. Ali,$^{1}$
Ali Takey,$^{1}$
Zainab Awad,$^{2}$
and Z. M. Hayman$^{2}$
\\
$^{1}$National Research Institute of Astronomy and Geophysics (NRIAG), 11421 Helwan, Cairo, Egypt\\
$^{2}$Astronomy, Space Science and Meteorology Department, Faculty of Science, Cairo University, 12613 Giza, Egypt
}

\date{Accepted XXX. Received YYY; in original form ZZZ}

\pubyear{2017}

\begin{document}
\label{firstpage}
\pagerange{\pageref{firstpage}--\pageref{lastpage}}
\maketitle

% Abstract of the paper
\begin{abstract}

Understanding the dynamics of galaxy triplet systems is one of the significant ways of obtaining insight into the dynamics of large galaxy clusters. Toward that aim, we present a detailed study of all isolated triplet systems (total of 315) taken from the ``SDSS-based catalogue of Isolated Triplets'' (SIT). In addition, we compared our results with those obtained for a sample of triplets from the Local Supercluster (LS), SDSS-triplets, Tully's catalogue, Wide (W) and Compact (K)-triplets. In addition, we performed the correlation between the dynamical parameters and the Large Scale Structure (LSS). Interestingly, we found that there is no correlation between both the mean projected separation for the triplet systems and the LSS and its dynamical parameters. Furthermore, we found that only $3\%$ of these systems can be considered as compact since the mean harmonic separation ($r_h$) is more than 0.4 Mpc for $97\%$ of the population.Thus we may conclude that, mergers might not have played a dominant role in their evolution.	

\end{abstract}

% Select between one and six entries from the list of approved keywords.
% Don't make up new ones.
\begin{keywords}
galaxies: fundamental parameters -- galaxies: groups: general -- galaxies: statistics
\end{keywords}

%%%%%%%%%%%%%%%%%%%%%%%%%%%%%%%%%%%%%%%%%%%%%%%%%%

%%%%%%%%%%%%%%%%% BODY OF PAPER %%%%%%%%%%%%%%%%%%

\section{Introduction}

Galaxies are no longer considered as island universes. They aggregate to form pairs, triplets, groups (not more than 50 galaxies) or clusters (up to 300 galaxies)\citep{Neta96}. Recently, it was found that $54\%$ of the galaxies in the local Universe (with redshifts smaller than 0.1) aggregate into galaxy groups and clusters, while $20\%$ are located in collapsing regions around these groups and clusters. The rest ($26\%$) are field galaxies \citep[][and references therein]{Blanton2003}.

Galaxy triplets were first thought to be formed as  close binaries then a remote galaxy is attracted later as a third body \citep{Valtonen91, Chernin94}. This scenario is no longer held since \citet{Kiseleva2000} found that triplets are  groups of three member galaxies immersed in a dark matter halo without an observed formation of close binaries. Therefore, triplets are a special type of galaxy groups with the smallest number of members \citep{Duplancic15}. Galaxy triplets have nearly the same complexity of larger groups; statistics, dynamics, and evolutionary properties. They are characterised by their proximity and low relative velocities which make them ideal for studying galaxy interactions and merging scenarios \citep{Duplancic13}.

The first triplet catalogue, which included 84 Northern isolated galaxy triplets, was compiled by \citet{Karachentseva1979}. They used visual inspection of the Palomar Sky Survey prints to observe these systems and study some of their observational properties. According to their procedure, triplet systems were constructed with members of apparent magnitudes brighter than 15.7 mag and angular diameter between 0.5 and 2 arcmin. Thus, a given triplet system which satisfied this criterion, is said to be isolated if the distance to its nearest neighbour is 3 times larger than the average distance between its galaxy members. This criteria yielded to select triplet member galaxies regardless of their radial velocity ($v$), and hence a system may be isolated only in projection. For these reasons, \citet{Karachentsev81, Karachentsev1988} conducted spectroscopic observations and numerical studies on this catalogue and found that only $64\%$ of the observed systems can be considered physical triplets with velocity differences $\Delta$$v$ $<$ 500 $km~s^{-1}$.  

Although triplets were thought to represent a collection of unrelated field galaxies rather than physical structures, recent analysis by \citet{Toledo11} showed that these galaxies have signatures of physical interactions which infer that most of them are indeed real physical structures. These interactions have a strong effect on the fundamental properties of their galaxies such as star formation rate, nuclear activity, and morphology \citep{O'Mill12}. 
Moreover, it is well known that interactions are common in triplets and compact systems in which they may end up into a single system \citep{Duplancic13}. Hence, studying these systems is crutial to understand their origin, evolution, intrinsic structure and how the surrounding environment affects their morphological, physical, and dynamical characteristics. 

The key aim of this work is to conduct a statistical study on the 315 triplet systems from ``SDSS-based catalogue of Isolated Triplets''(SIT) \citep{Fernandez15} to estimate their basic parameters; namely, radial velocities ($v$), distances (D), angular separations $(\theta)$, projected separations $(r_p)$, mean harmonic separation $(r_h)$, velocity dispersion $(\sigma)$, virial masses $(M_{vir}/M_ \odot)$, luminosity $(L_r)$, absolute magnitudes $(M_r)$, and mass-to-light ratios $(M_{vir}/L_r)$. Accordingly, we will distinguish between compact (k) and wide (w) triplets in our sample according to their mean harmonic separation ($r_h$). Then, we will compare our results with another documented triplets sample. Finally, we will investigate the correlations between the dynamical parameters of the studied triplet systems and the Large Scale Structure (LSS). 

This paper is organised as follows; in $\S$~\ref{s:Sample} we introduce our sample and the calculated basic parameters for each system and its three members. The results of the statistical study and the correlations between the triplet parameters are presented and discussed in $\S$~\ref{s:Dis} while the main conclusions and remarks are given in $\S$~\ref{s:Conc}.

% This is a simple template for authors to write new MNRAS papers.
% See \texttt{mnras\_sample.tex} for a more complex example, and \texttt{mnras\_guide.tex}
% for a full user guide.

\section{Sample description and their fundamental parameters}
\label{s:Sample}

In this section we briefly describe the chosen sample and discuss in some details the calculations of some fundamental parameters of each system and its galaxy members.

\subsection{The chosen sample}

We conduct a statistical study on 315 galaxy triplet systems taken from the SIT catalogue by \citet{Fernandez15}. This catalogue is based on the Sloan Digital Sky Survey- Tenth Data Release \citep [SDSS-DR10][]{DR10}. Triplet systems (in this selected catalogue) are restricted to have members with spectroscopic information, r-band model magnitudes ($m_r$) in the range 14.5 $\le m_r \le$ 17.7 over a redshift (z) range 0.006 $\le$ z $\le$ 0.080. \citet{Fernandez15} based their isolation criterion on galaxies with no neighbours within radial velocity difference $\Delta v$ $\le$ 500 $km~s^{-1}$ and all galaxies are found within 1 Mpc radius ($d$ $\le$ 1 Mpc). In addition, they performed a star-galaxy separation in order to prevent classifying a star as a galaxy and removed multiple identifications of the same galaxy. Hence, they based their isolation criteria mainly on both the velocity difference ($\Delta v$) and the projected distance space. 
Following this criteria they compiled a catalogue of 315 isolated triplet systems. SDSS coloured images of some triplet systems of these catalogue are illustrated in appendix~\ref{s:appendix}. In these systems, galaxies are considered to be physically bounded to the primary galaxy (chosen to be the brightest one in the member), if $\Delta$ $v$ $\le$ 160 $km~s^{-1}$, and the projected separation ($r_p$) is smaller than 450 kpc. 

We choose this catalogue in our study because (i) it compiles a large sample of triplet galaxies (315 systems), (ii) triplets are isolated with no neighbours for at least 1 Mpc radius and hence there is no interaction with the surrounding environment,(iii) isolated triplets have been homogeneously selected, with a hierarchical structure, using an automated method, and (iv) all triplet galaxies have spectroscopic information and their galaxy parameters (RA, DEC, z) are available.

The observational parameters (apparent magnitude ($m_r$) in the r-band, k-correction, the luminosity distance ($D_L$), the photometric and spectroscopic redshifts) were extracted from CasJobs tool in SDSS DR12 \citep{DR12}. These parameters are helpful in estimating other basic parameters such as; the absolute magnitude ($M_r$) in r-band, and the radial velocity ($v$). The SDSS-DR12 is the final data release of the SDSS-III which contains observations up to July 2014. It covers the complete dataset of the BOSS, APOGEE, and the stellar radial velocity measurements from MARVELS surveys. SDSS-DR12 includes all information in DR11 and previous releases of SDSS \citep{Alam15}. Throughout this study we compute the statistical parameters based on the spectroscopic redshift which is more accurate than the photometric one. Furthermore, spectroscopic redshift does not have the neighbouring effect that influences the photometric measurements \citep{Patton11, Shen16}

\subsection{Galaxy triplets: fundamental parameters }
\label{ss:eqn}

In the following context we describe briefly the calculations of both the dynamical ($\sigma$, $r_h$, $M_{vir}/M_ \odot$, and $M_{vir}/L_r$) and physical parameters ($v$, $\theta$, $r_p$, $M_r$, $L_r$) of the triplet systems and their members, respectively, and show how they are linked together. Each triplet system is characterised by its $\sigma$, $r_h$, $M_{vir}/M_ \odot$, and $M_{vir}/L_r$ while for each member galaxy in the system we calculate the rest of the basic parameters; see Tables~\ref{tbl:TGM} and~\ref{tbl:TGS}. The adopted constants in the calculations throughout this work are: Hubble constant ($H_0$= 70 $km~s^{-1} Mpc^{-1}$), speed of light (c= $2.99792\times10^5$ $km~s^{-1}$), gravitational constant (G= $6.674\times10^{-11}$  $m^3~kg^{-1}~s^{-2}$), the solar absolute magnitude and colour are taken from SDSS  where ($M_\odot,_g$ (in g-band) = $5.12 \pm0.2$), and ($M_\odot$(g-r)=$0.44\pm0.02$). The solar absolute magnitude in r-band $M_\odot,_r$ can be computed by subtracting $M_\odot,_g$ in g-band from $M_\odot$(g-r); thus $M_\odot,_r$= 4.68, and solar mass ($M_\odot$= $1.9891\times10^{30}$ kg). Details on the calculations of $M_\odot,_r$ are given in the website of SDSS \footnote{$www.sdss.org/dr14/algorithms/ugrizVegaSun/$}, and \citep{Blanton2003, Bilir2005, Rodgers2006}. 

Triplet as a unit dynamical system and their members are characterised by the following parameters:\\

\begin{enumerate}
 \item Physical parameters
 \begin{itemize}
  \item\textbf{Angular separation ($\theta_{ik}$) in arcmin}\\
  
   Angular separation can be calculated by two methods: one is applying Equation~\ref{eq:theta} as taken from \citet{Smart65}
\begin{multline}
 \theta_{ik}=\{cos^{-1}[sin(\delta_{ik})sin(\delta_{ik})+cos(\delta_{ik})cos(\delta_{ik})\\ 
            * cos((\alpha_{ik})-(\alpha_{ik}))]\},
   \label{eq:theta}
%  θik=cos-1[sin(δik)sin(δik)+cos(δik)cos(δik)*cos(αik-αik)] ; i,k=1,2,3; i≠k
\end{multline}

where
$i$, $k$=1, 2, 3; $i$ $\not$= $k$,
$\delta$ is the declination and $\alpha$ is the right ascension of the galaxy. The other method is using the python module  \textit{``astCoords.calcAngSepDeg''} in \textit{AstLib} \footnote{http://astlib.sourceforge.net/docs/astLib/astLib.astCoords-module.html}. We obtained the same results using both methods.\\   
 
  \item \textbf{Projected separation $(r_p)$ in Mpc} \\
  Projected separation is defined as the minimum separation between two celestial objects determined by the product of the distance (D=$v/H_0$) and the angular separation ($\theta$) between them \citep{Chernin00, Ali01,Vavilova05, Mac15 }. Hence, we can estimate the projected separation between each two galaxies in the triplet systems by using the generalised equation
\begin{equation}
 r_p=2<v>H_0^{-1}sin(\theta_{ik}/2),  
  \label{eq:rp}
% rp=2<v>H0-1sin(θik/2) ; i,k=1,2,3; i≠k   (Ali G., 2001)
\end{equation}
where $v$ is the radial velocity defined as the product of speed of light ($c$) and the redshift ($z$) in $km~s^{-1}$ ($ v=cz$) \citep{Dekel99, Schneider06}, $\theta_{ik}$ is the angular separation between the members of each triplet system, $i$, $k$=1, 2, 3; $i$ $\not$= $k$. 

 \end{itemize}

 \item \textbf { Dynamical parameters}
 \begin{itemize}
  \item \textbf {Velocity dispersion $(\sigma)$ in $km~s^{-1}$}\\
 
After estimating the radial velocity ($v$) of the three galaxies in a triplet system, we can compute the system velocity dispersion $(\sigma)$ following the expression
\begin{equation}
 \sigma=[<(v)^2>-(<v>)^2]^{\frac{1}{2}}, 
   \label{eq:sigma}
\end{equation}
where
$<>$ denotes the average overall the members of the system \citep [][]{Chernin00, Schneider06, Duplancic15}.\\

  \item \textbf{Mean harmonic separation $(r_h)$ in Mpc}\\

The mean harmonic separation ($r_h$) is a remarkable parameter suitable for cosmological investigations since it is not specified for a certain type of groups or clusters. It represents the angular size of the core of a galaxy group or cluster. It depends on the position of the brightest galaxy in the group or cluster, and thus it is important in studying very distant clusters where only the brightest galaxies could be visible. Since, this parameter represents the core size of the group or cluster, it is strongly related to their gravitational radius and therefore has a statistical power \citep{Hickson77_1, Hickson77_2}. Harmonic separation is based mainly on the projected separation ($r_p$, see above) between members of a triplet system. It indicates whether the members are close to each other (i.e. compact) or far from each other (i.e. loose). Therefore, determination of $r_h$ is important to categorise the systems into wide (w) and compact (k) following \citet{Chernin00} as

$$r_h (Mpc)=
\begin{cases}
 \mbox{< 0.04 \quad compact system (k)}\\
 \\
 \mbox{> 0.04 \quad  wide system (w)}
\end{cases}
$$

The general formula of the mean harmonic separation ($r_h$) is expressed in terms of N,the number of galaxies in a given group or cluster, as

\begin{equation}
 r_h=[ \,(2/N(N-1))\Sigma r_{p_{ik}}^{-1}] \,^{-1}, 
  \label{eq:rh}
%  r_h=[(1/3)\Sigma r_p_{ik}^{-1}]^{-1} ; i,k=1,2,3 ; i \not= k 
%  rh= [(1/3) ∑ rp(ik)-1]-1 ; i,k=1,2,3; i≠k
\end{equation}
where  i, k=1, 2,....,N ; i $\not$= k \citep{Hickson77_1}

In the case of triplet galaxies, N=3, Equation~\ref{eq:rh} will be re-written as

\begin{equation}
 r_h=[ \,(1/3)\Sigma r_{p_{ik}}^{-1}] \,^{-1}, 
  \label{eq:rh_2}
%  r_h=[(1/3)\Sigma r_p_{ik}^{-1}]^{-1} ; i,k=1,2,3 ; i \not= k 
%  rh= [(1/3) ∑ rp(ik)-1]-1 ; i,k=1,2,3; i≠k
\end{equation}
where i, k=1, 2, 3; i$\not$=k \citep{Chernin00, Vavilova05, Feng16}.\\

 \item \textbf{Virial Mass $(M_{vir}/M_\odot)$}\\
 
For a system of \textit{N} members occupied in a circle of minimum radius \textit{R}, that contains the centres of these members, the virial mass is given by 
\begin{equation}
 M_{vir}=3\pi N(N-1)^{-1}G^{-1} \sigma^2 R, 
 \end{equation}
 For triplets, N=3, we have
\begin{equation} 
 M_{vir}= (9\pi/2G)~R~ \sigma^2, 
  \label{eq:Mvir}
  \end{equation}
%  Mvir= 3Π N (N-1)-1 G-1 σ2 R = (9Π/2G) Rσ2

where \textit{G} is the gravitational constant, \textit{R} can be considered to be the mean harmonic separation $r_h$ in meters, and $\sigma$ is the velocity dispersion in $m~s^{-1}$ \citep{Chernin00,Vavilova05,Liljeblad12,Duplancic15}.\\
 
 \item \textbf{Mass-to-Light ratio $(M_{vir}/L_r)$}\\
 
 To compute the mass-to-light ratio of a system of galaxies, we first have to calculate the luminosity of each galaxy ($L_r$, in r-band) using the following equation
 \begin{equation}
 L_r=dex[0.4(M_\odot,_r-M_r)],
  \label{eq:Lr}
\end{equation}
where $M_\odot,_r$ and $M_r$ are the solar and galactic absolute magnitudes in the r-band, respectively \citep{Binney1987,Ali01,Schneider06}.\\

The absolute magnitude of a galaxy in the r-band is given by
\begin{equation}
 M_r=m_r -5(log(D_L -1)) -k_{corr}, 
  \label{eq:Mr}
\end{equation}
where
$m_r$ is the apparent magnitude in r-band corrected from reddening, $D_L$ is the luminosity distance in pc, and $k_{corr}$  is the k-correction in the r-band. These parameters are obtained from the SDSS database.

Therefore, the mass-to-light ratio ($M_{vir}/L_r$) is
\begin{equation}
M_{vir}/L_r= M_{vir}/dex[0.4(M_\odot,_r-M_r)],
  \label{eq:Mvirl}
\end{equation}

where in Equation~\ref{eq:Mvirl}, $L_r$ is the total luminosity of the three members in each system ($L_r=\sum_{n=1}^3 L_{r,n}$).
 \end{itemize}
\end{enumerate}

Tables ~\ref{tbl:TGM} and ~\ref{tbl:TGS} summarise our calculations of both the physical and dynamical parameters of the triplet systems and their members for the studied catalogue SIT \citep{Fernandez15}. In Table ~\ref{tbl:TGS}, the maximum value of the mass-to-light ratio is computed for 313 triplet systems only as we excluded two systems (SIT 99 and SIT 168) due to their very large ratios (567 and 810, respectively). Each of these two systems compose of very faint galaxies (G1, G2, and G3), especially the third galaxy (G3),  which make their total luminosity very small, and hence, the mass-to-light ratio becomes greater than usual. On the other hand, the other dynamical parameters ($\sigma$, $M_{vir}$) for these systems fall in the normal range. The computed dynamical parameters of the 315 studied triplet systems from SIT are listed in Table ~\ref{tbl:TG} (see Appendix~\ref{s:appendix}).

\begin{table*}
  \begin{center}
    \tabcolsep 5.8pt
    \normalsize
    \caption{Ranges of the calculated basic parameters of the member galaxies in the studied sample of triplets \label{tbl:TGM}}
\begin{tabular}{|l|l|l|l|} 
\hline
  \multicolumn{1}{|c|}{Parameters} &
  \multicolumn{1}{c|}{Units} &
  \multicolumn{2}{c|}{Range}\\
%   \multicolumn{1}{c|}{}\\
  \multicolumn{1}{|c|}{(members)} &
  \multicolumn{1}{c|}{} &
  \multicolumn{1}{c|}{Min.} &
  \multicolumn{1}{c|}{Max.}\\ 
%  Parameters (members)                 &  Units   &  Min.Range      & Max. Range \\  \hline
\hline
 Radial velocity ($v$)               & $km~s^{-1}$   & 1819.7           & 24010.3 \\  
%  Angular separation ($\theta$)         & arcmin        & 0.11             & 52.39  \\  
 Projected separation ($r_p$)          & Mpc           & 0.009            & 0.847   \\  
 Absolute magnitude in r-band ($M_r$)  & --            & -22.8            & -14.5 \\  
 Luminosity ($L_r$)                    & $L_{\odot}$   & $4.7\times10^7$  & $9.7\times10^{10}$ \\  
%  Luminosity distance ($D_L$)           & Mpc           & 26.14            & 363.95  \\  
 \hline
\end{tabular}
\end{center}
\end{table*}

\begin{table*}
  \begin{center}
    \tabcolsep 5.8pt
    \small 
    \caption{\textbf{Statistical parameters calculated for} the 315 triplet systems selected from the SIT catalogue. STD refers to the standard deviation}  \label{tbl:TGS}
    \begin{tabular}{|l|l|l|l|l|l|l|}
\hline
 \multicolumn{1}{|c|}{Parameters} &
  \multicolumn{1}{c|}{Units} &
  \multicolumn{2}{c|}{Range} &
  \multicolumn{3}{c|}{Statistical Parameters}\\
  \multicolumn{1}{|c|}{(systems)} &
  \multicolumn{1}{c|}{} &
  \multicolumn{1}{c|}{Min.} &
  \multicolumn{1}{c|}{Max.}&
  \multicolumn{1}{c|}{Median} &
  \multicolumn{1}{c|}{Mean} &
  \multicolumn{1}{c|}{STD}\\ 
%  Parameters (members)                 &  Units   &  Min.Range      & Max. Range \\  \hline
\hline
 
 Velocity dispersion ($\sigma$) & $km~s^{-1}$   & 6.2       & 123.6       & 43.8   & 45.4   & 21.9 \\  
 Mean harmonic separation($r_h$) & Mpc           & 0.017     & 0.508       & 0.17   & 0.19   & 0.10 \\  
 Virial Mass $(M_{vir})$         & $M_\odot$ & $10.0\times10^9$& $1.2\times10^{13}$& $1.5\times10^{12}$& $1.5\times10^{12}$& $1.8\times10^{12}$ \\  
 Mass-to-Light ratio $(M_{vir}/L_r)$& --        & 0.1         & 308      & 41.4  & 41.4  & 50.6   \\  
 \hline
  \end{tabular}
\end{center}
\end{table*}

\begin{table*}
  \begin{center}
    \tabcolsep 5.8pt
    \small 
    \caption{Comparison between the median values of the dynamical parameters ($\sigma$, $r_h$, $M_{vir}$, $M_{vir}/L_r$) of our sample and other triplet samples \label{tbl:compare}. NTS refers to the Number of Triplet Systems}
    \begin{tabular}{|l|l|l|l|l|l|l|}
\hline
 \multicolumn{1}{|c|}{Sample}&
 \multicolumn{1}{|c|}{NTS} &
  \multicolumn{1}{c|}{$\sigma$} &
  \multicolumn{1}{c|}{$r_h$} &
%   \multicolumn{1}{c|}{}&
  \multicolumn{1}{c|}{$M_{vir}$}&
  \multicolumn{1}{c|}{$(M_{vir}/L_r)$}&
  \multicolumn{1}{c|}{References}\\
  \multicolumn{1}{|c|}{} &
  \multicolumn{1}{|c|}{} &
  \multicolumn{1}{c|}{$(km~s^{-1})$} &
  \multicolumn{1}{c|}{$(kpc)$} &
  \multicolumn{1}{c|}{($M_\odot \times10^{12}$)}&
  \multicolumn{1}{c|}{($M_\odot/L_\odot$)}&
  \multicolumn{1}{c|}{}\\ 
%  Parameters (members)                 &  Units   &  Min.Range      & Max. Range \\  \hline
\hline
 
 SIT                     & 315      & 44         & 166       & 1.50     & 23   & This work\\  
 LS                      & 176      & 30         & 160       & 0.36     & 35   &\citet{Vavilova05}\\  
 MK2000                  & 156      & 33         & 178       & 0.52     & 35   & \citet{MK2000}\\
 Tully                   & 56       & 56         & 269       & 2.63     & 111  & \citet{Tully87}\\
 Wide                    & 37       & 66         & 396       & 4.47     & 173  & \citet{Trofimov95}\\
 SDSS                    & 92       & 119        & 67        & 1.00     & --   & \citet{Duplancic15} \\
 K-triplets              & 37       & 114        & 54        & 1.00     & --   & \citet{Duplancic15} \\
%  Mock triplets & 159   & 126.9$\pm$13.2 & 42.7$\pm$1.3 & 11.8$\pm$0.1 &--   \\  
 \hline
    \end{tabular}
  \end{center}
\end{table*}

\begin{table*}
  \begin{center}
    \tabcolsep 5.8pt
    \small 
    \caption{Comparison between the ranges of the dynamical parameters ($\sigma$, $M_{vir}$, $M_{vir}/L_r$) of galaxy triplets in our study and more denser groups and clusters of galaxies \label{tbl:compare2}. NG is the number of galaxies in a given system, group, or cluster.}
    \begin{tabular}{|l|l|l|l|l|l|l|}
\hline
 \multicolumn{1}{|c|}{Sample}&
 \multicolumn{1}{|c|}{NG} &
  \multicolumn{1}{c|}{size} &
  \multicolumn{1}{c|}{$\sigma$} &
  
%   \multicolumn{1}{c|}{}&
  \multicolumn{1}{c|}{$M_{vir}$}&
  \multicolumn{1}{c|}{$(M_{vir}/L_r)$}&
  \multicolumn{1}{c|}{References}\\
  \multicolumn{1}{|c|}{} &
  \multicolumn{1}{|c|}{} &
   \multicolumn{1}{|c|}{Mpc} &
  \multicolumn{1}{c|}{$(km~s^{-1})$} &
%   \multicolumn{1}{c|}{$(kpc)$} &
  \multicolumn{1}{c|}{($M_\odot \times 10^{12}$)}&
  \multicolumn{1}{c|}{($M_\odot/L_\odot$)}&
  \multicolumn{1}{c|}{}\\ 
%  Parameters (members)                 &  Units   &  Min.Range      & Max. Range \\  \hline
\hline
 
 SIT                     & 3           & 1       & 6-124        & 0.01-10       & 0.1-308   & This work\\  
 groups                  & $<$ 50      & 2       & 150-500      & 20-32         & 300-400   &\citet{Carroll06, Bohringer06}\\  
 clusters                & 50-1000     & 6-8     & 800-2000     & 1000-3200     & $>$500   & \citet{Carroll06, Bohringer06}\\
 \hline
    \end{tabular}
  \end{center}
\end{table*}

%%%%%%%%%%%%%%%%%%%%%%%%%%%%%%%%%%%%%%%%%%%%%%%%%%%%%%%%%%%%%%%%%%%%

%%%%%%%%%%%%%%%%%%%%%%%%%%%%%%%%%%%%%%%%%%%%%%%%%%%%%%%%%%%%%%%%%%%%

\section{Results and Discussion}
\label{s:Dis}

Statistical studies of galaxy triplet systems are crucial in understanding the main properties and the evolution of galaxies in groups. Analysis of such poor populated groups (3 members) is an important step to study more populated groups. In this section, we present the distribution of the basic parameters of triplet systems that control their properties. Then, we demonstrate the satellite to central galaxy ratios of both absolute magnitude ($M_r$) and luminosity ($L_r$), in the r-band. In addition, the correlation between the dynamical parameters of the studied triplets and their mean projected separation ($\overline{r_p}$) is determined by calculating their correlation coefficients $(f_{xy})$. Finally, we illustrate the correlation between triplets' dynamical parameters and the surrounding Large-Scale Structure (LSS).

\subsection{Distributions of galaxy triplets parameters}

% Table~\ref{tbl:TGM} lists the ranges of the computed parameters for triplet galaxies while Table~\ref{tbl:TGS} gives the range, median, mean, and the standard deviation of the dynamical parameters for the triplet systems.
% From Tables ~\ref{tbl:TGM} and~\ref{tbl:TGS} we notice that $M_{vir}/L_r$ of wide triplet systems are larger than those of compact systems. In addition, values of both velocity dispersion and virial masses are close to low-mass loose groups computed by \citet{Merch05}.

\begin{figure*}
%  \label{fig:histo}
 %\centering
\subfloat{\includegraphics[width=3.00in]{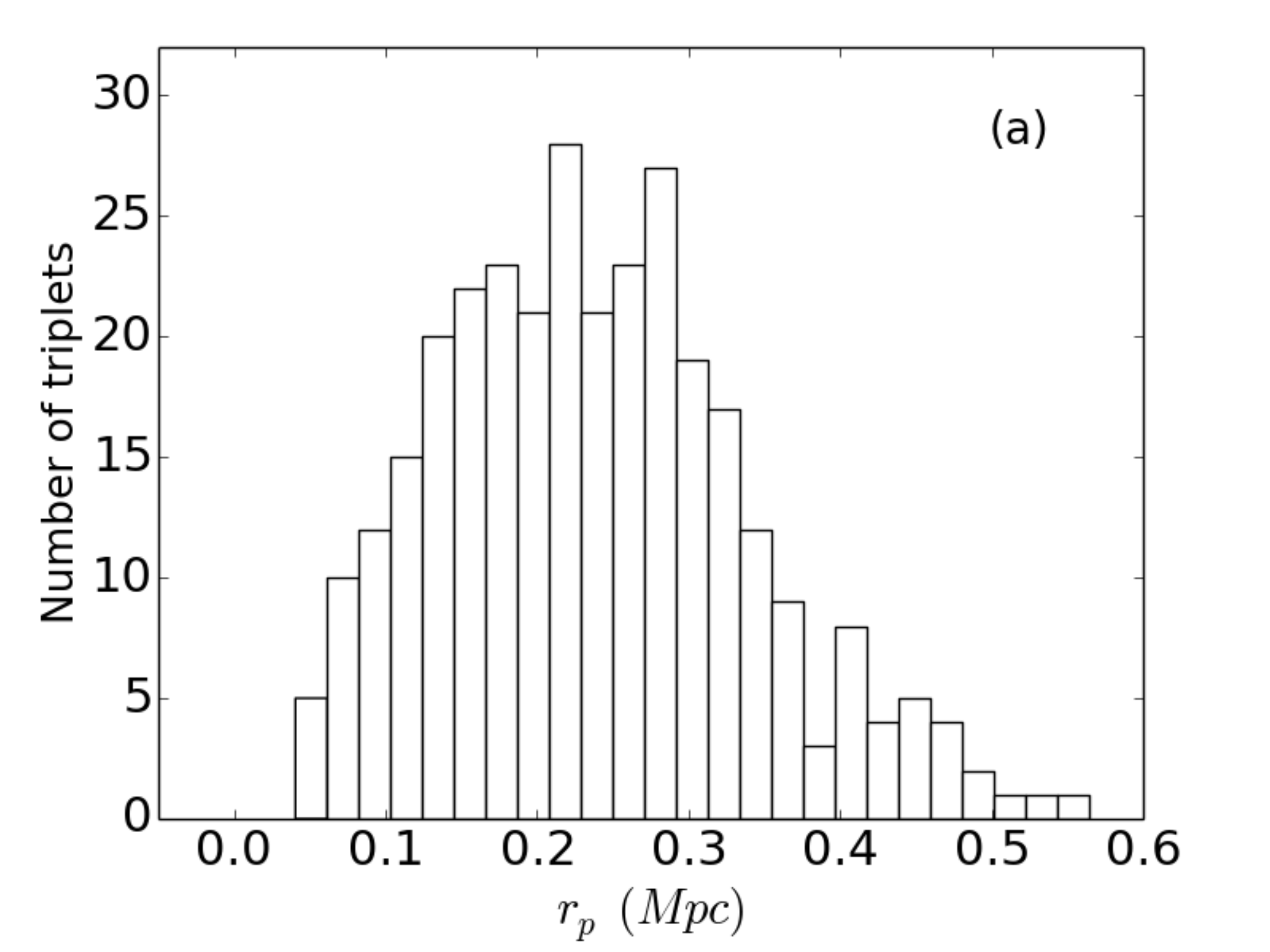}\label{subfig:rp}}
\subfloat{\includegraphics[width=3.00in]{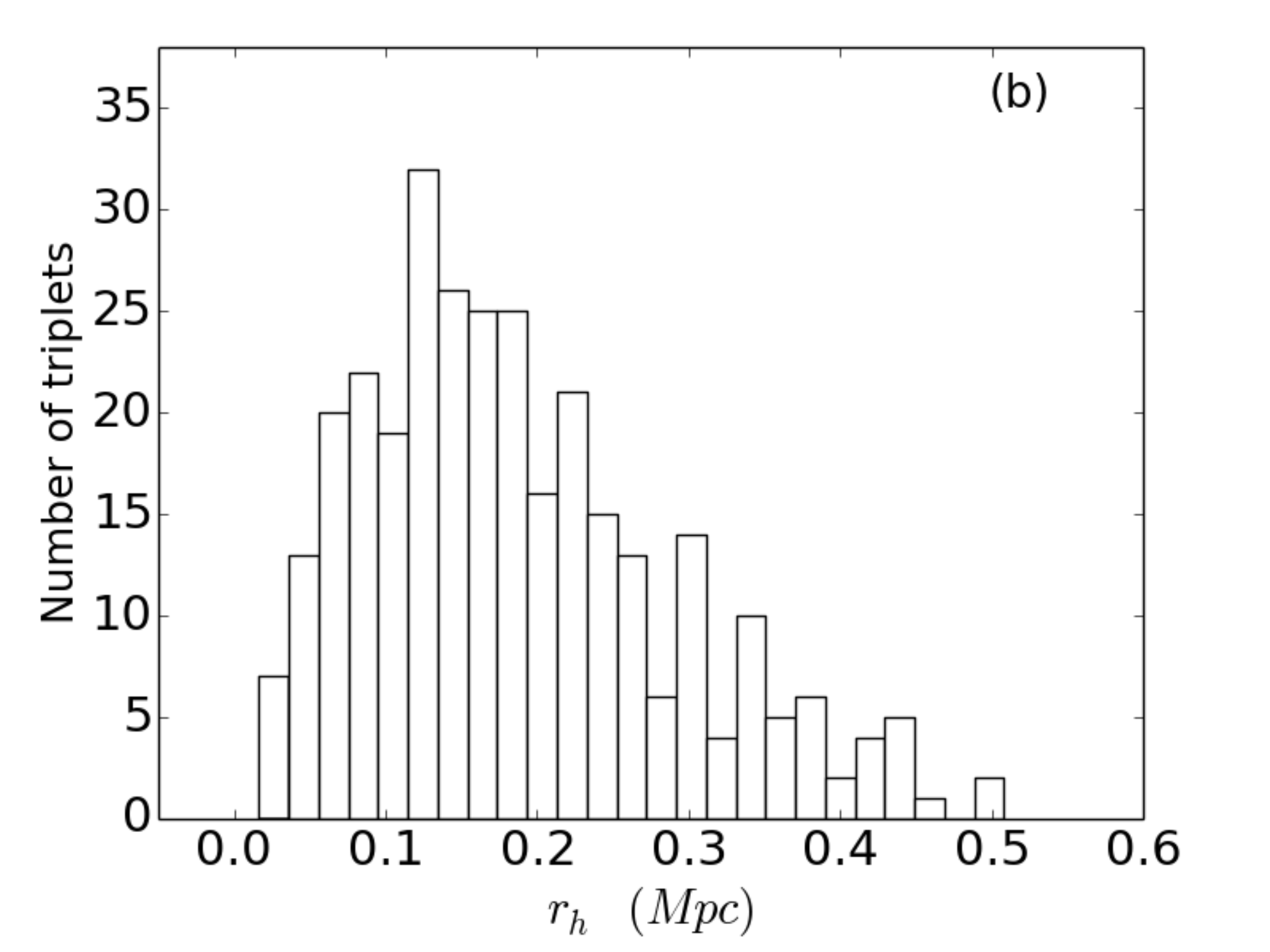}\label{subfig:rh}}\\
\subfloat{\includegraphics[width=3.00in]{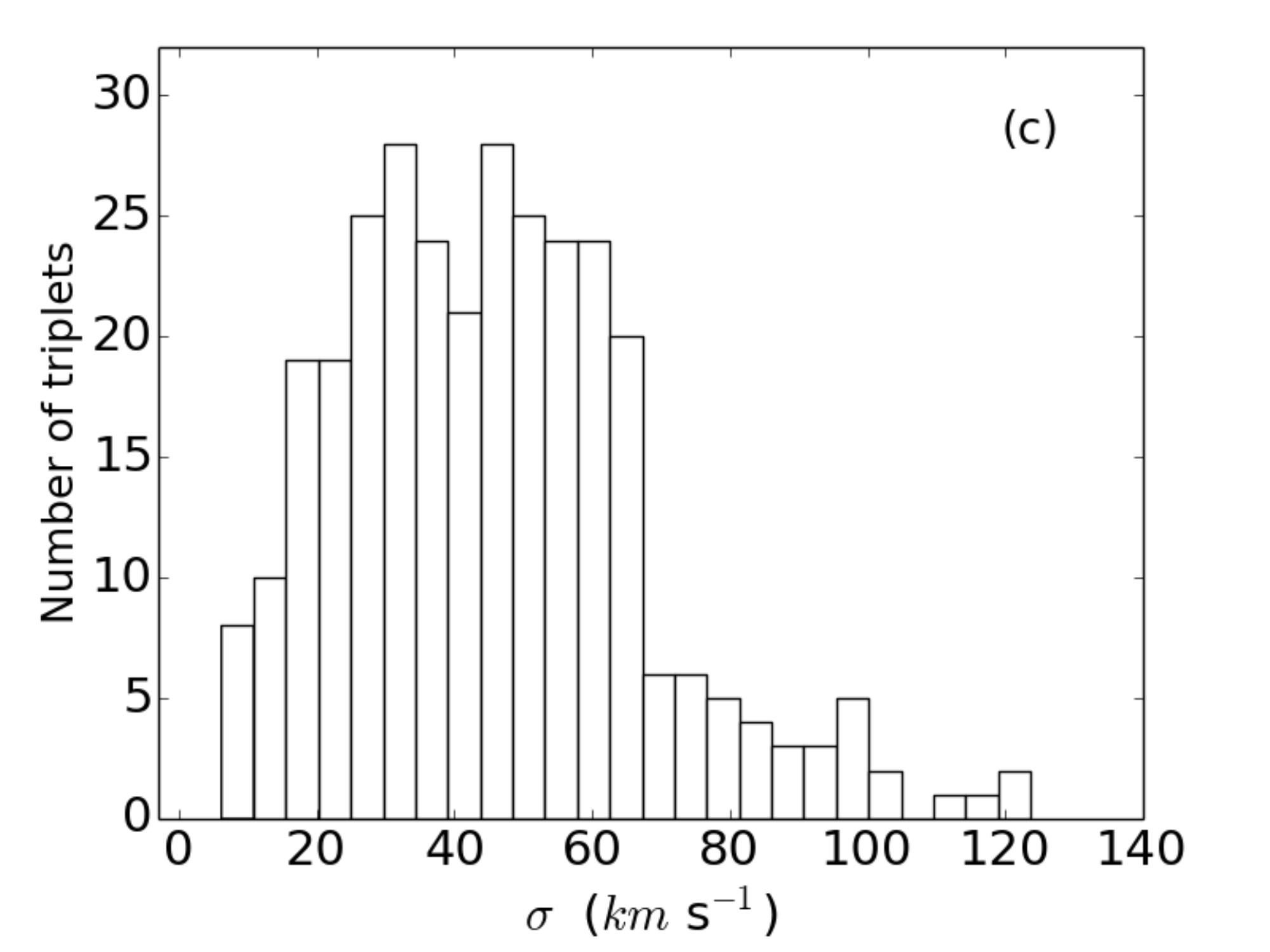}\label{subfig:sigma}}
\subfloat{\includegraphics[width=3.00in]{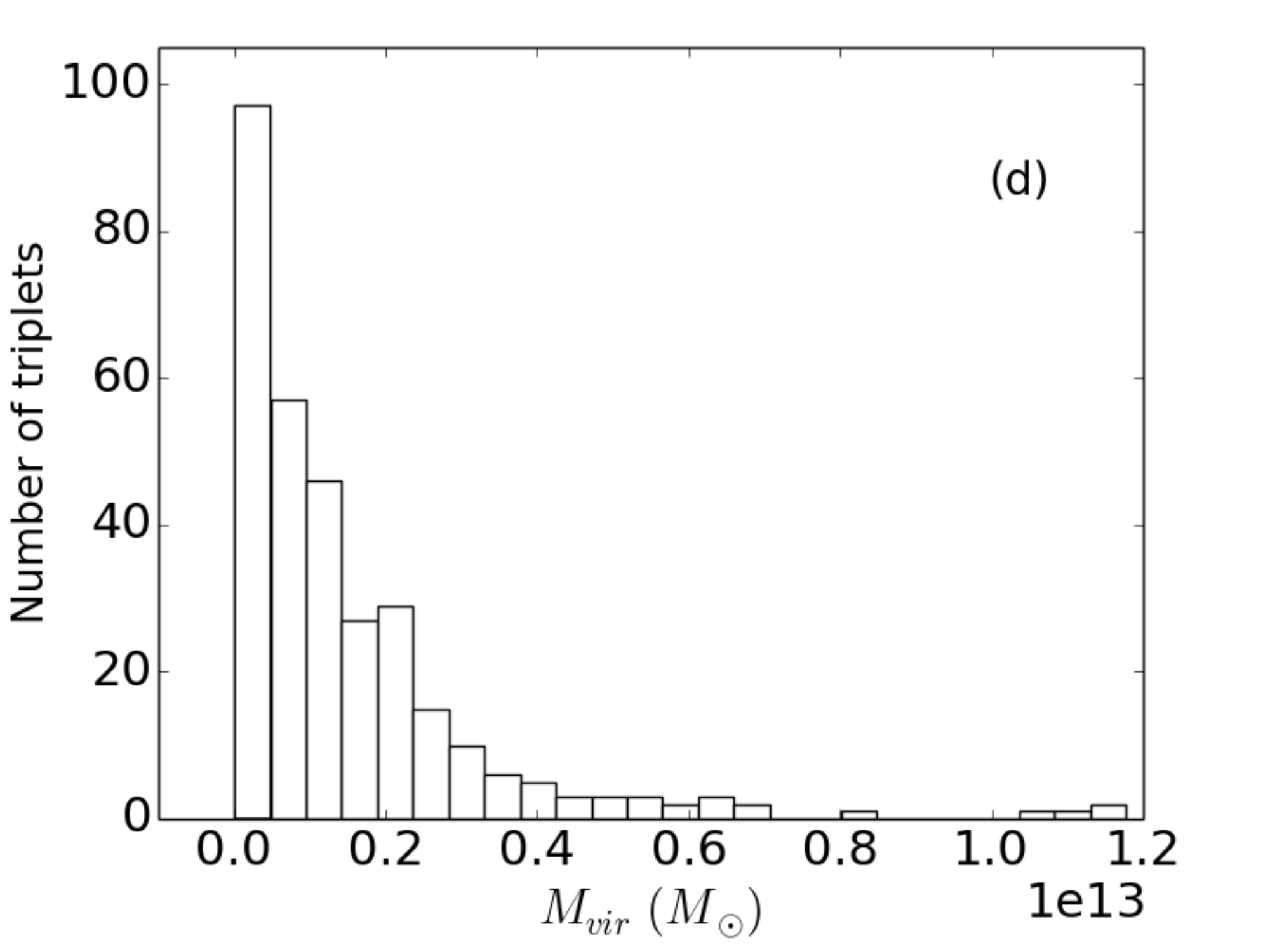} \label{subfig:Mvir}}\\
\subfloat{\includegraphics[width=3.00in]{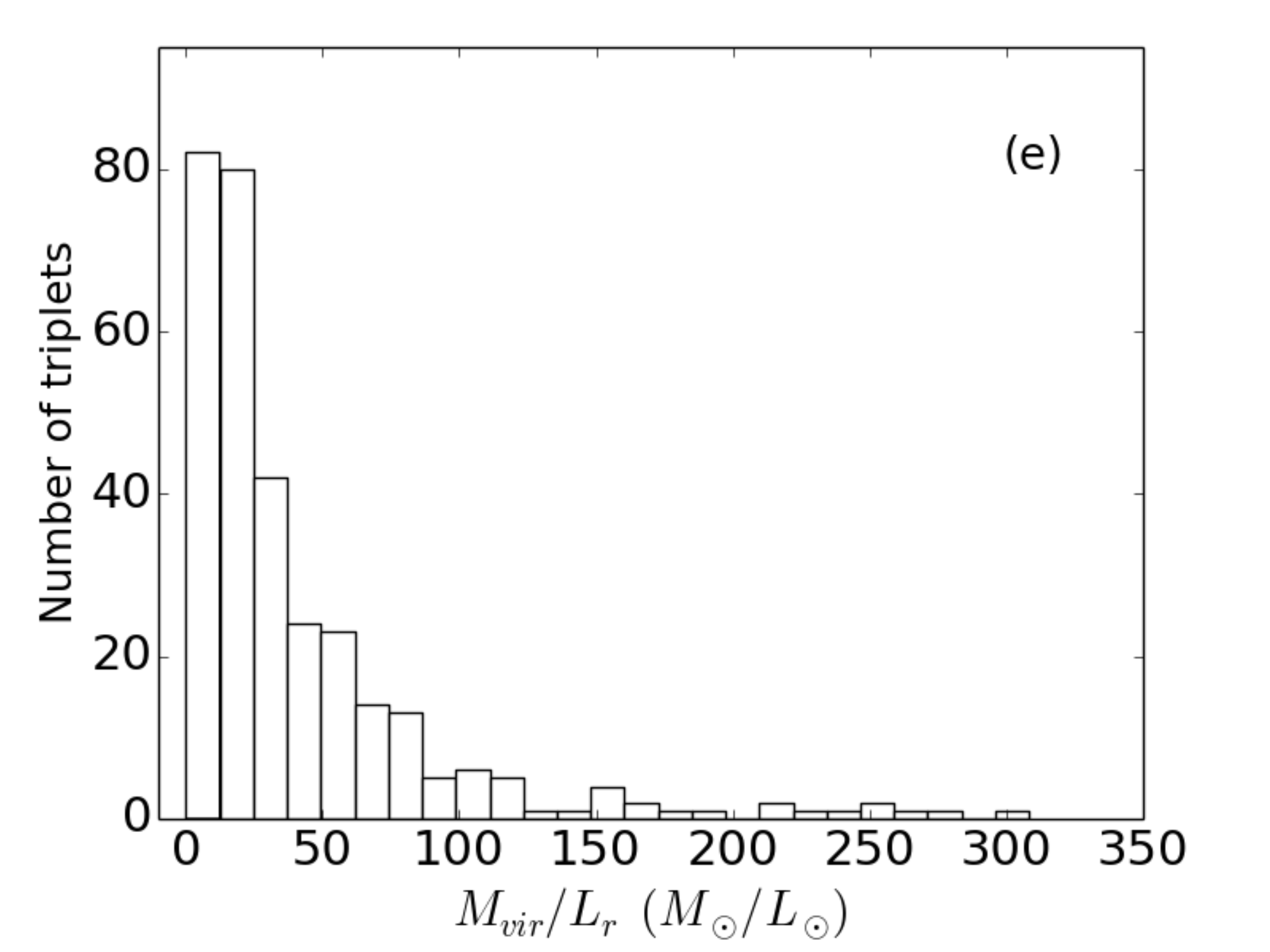}\label{subfig:ML}}\\
%  \subfloat[\scriptsize Redshift distribution]{\includegraphics[scale=0.3, viewport=0 0 540 420, clip]{eps_figures/z_histo.eps}\label{subfig:z}}
% \newline
\caption{Histograms represent the distribution of the computed parameters for triplet systems and their members. (a) the projected separation ($r_p$), (b) the mean harmonic separation ($r_h$), (c) the velocity dispersion ($\sigma$), (d) the virial mass ($M_{vir}$), and (e) the mass-to-light ratio ($M_{vir}/L_r$ )}
\label{fig:histograms}
\end{figure*}

Fig.~\ref{fig:histograms} represents the distribution of the calculated basic parameters of the studied 315 triplet systems.
Fig.~\ref{subfig:rp} demonstrates the distribution of the projected separation ($r_p$) showing a multi-modal distribution with a mean value 0.238 Mpc, and a standard deviation of 0.102. In addition, the majority of systems  have projected separation between 0.12 and 0.35 Mpc with two peaks around 0.21 Mpc, and 0.29 Mpc.

Fig.~\ref{subfig:rh} shows the distribution of the mean harmonic separation with a right-skewed or asymmetric distribution with highest population (73 systems) between 0.1 to 0.2 Mpc and a lowest population (5 systems) with nearly 0.5 Mpc. We also noticed that only 10 systems $(3.2\%)$ are compact triplets, with $r_h < 0.04$ Mpc, while the other 305 $(96.8\%)$ are wide galaxy triplets ($r_h>0.04$ Mpc). 

Fig.~\ref{subfig:sigma} illustrates the distribution of velocity dispersion showing a right-skewed histogram with two peaks with the same number of triplets (28 systems) at the ranges between 30 $km~s^{-1}$ to 35 $km~s^{-1}$ and 45 $km~s^{-1}$ to 50 $km~s^{-1}$, respectively. In addition, most systems fall into velocity dispersion of 20 to 60 $km~s^{-1}$. Only 5 systems have velocity dispersion between 100 and 120 $km~s^{-1}$. There is no occurrence from 105 to 110 $km~s^{-1}$.

Fig.~\ref{subfig:Mvir} and Fig.~\ref{subfig:ML} represent the distribution of the virial mass and the mass-to-light ratio, respectively. The distribution of virial mass shows a maximum number of 170 systems with virial masses smaller than $10^{12}M_\odot$, and about 75 systems with virial masses between $0.1\times10^{13}$ and $0.2\times10^{13} M_\odot$. In contrast, there is no occurrence of systems between $0.8\times10^{13}$ and $1.1\times10^{13}M_\odot$, and only 7 systems have virial masses between $1.1\times10^{13}$ to $1.2\times10^{13}M_\odot$. This range is wider than that obtained in previous studies \citep{Vavilova05}.

In order to benchmark our result with previous studies, we compared our computed dynamical parameters with other triplet samples, see Table~\ref{tbl:compare}. For comparison we used a list of 176 galaxy triplets from the Local Supercluster (LS) compiled by \citet{Vavilova05} on the basis of LEDA2003 under the restriction of radial velocity ($v<$ 3100  $km~s^{-1}$) in the zone $\vert b \vert >$ 10$^{\circ}$, the 156 triplet sample on the basis of LEDA1996 compiled by \citet{MK2000} (MK2000) with a criteria based on the Karachentsev algorithm \citep{karachen94}, a list of 56 triplets in Tully's catalogue with radial velocity less than 3000  $km~s^{-1}$ \citep{Tully87}, wide triplets compiled by \citet{Trofimov95} with velocity different $<$ 600 $km~s^{-1}$, SDSS-triplets which composed of 92 triplets with absolute magnitude brighter than $M_r$= -20.5 in the redshift range 0.01$\le z\le $ 0.14 \citep{Duplancic15}, K-triplets which composed of 37 compact triplets selected from the original catalogue of \citet{Karachentsev1988} with redshift range 0$\le z\le $0.05 and radial velocity difference $<$ 500 $km~s^{-1}$ \citep{Duplancic15}.

From this comparison we found that the median values of our sample are close to those of LS and MK2000 although they are selected under different conditions and criteria. The median of the mass-to-light ratio ($M_{vir}/L_r$) of our sample is 1.5 times smaller than that of LS and MK2000, and 4-7 times smaller than wide and Tully samples. This is probably due to the small range of apparent magnitude of SIT sample (11 $\le m_r \le$ 15.7). On the other hand, wide sample shows the largest mean harmonic separation ($r_h$), and therefore, the largest virial mass ($M_{vir}$) and mass-to-light ratio ($M_{vir}/L$). Moreover, the SDSS and the K-triplets have the largest $\sigma$, and the smallest $r_h$ because they are identified as compact triplets with small projected separations ($r_p$). These different ranges are probably due to different selection criteria used to compile these different samples.

Since most of the systems in SIT catalogue are wide (loose), we expect that their properties will match those of loose groups. This expectation has been confirmed after confronting our systems with loose groups in the sample studied by \citet{Merch05}.
 
In addition, we compared the ranges of the dynamical parameters of our triplets sample ($\sigma$, $M_{vir}$, and $M_{vir}/L$) with more denser groups and clusters, see Table~\ref{tbl:compare2}. This comparison showed that the ranges of the dynamical parameters increase steeply from low to denser groups and clusters of galaxies. The velocity dispersion starts from 6 $km\, s^{-1}$ for galaxy triplets up to 2000 $km\, s^{-1}$ for rich clusters. Whereas, the virial mass increases from 0.01 $M_\odot \times 10^{12}$  in case of galaxy triplets up to 3200 $M_\odot \times 10^{12}$ for rich clusters. Finally, the mass-to-light ratio shows an ascending manner starting from 0.1 for triplets to reach its maximum at 500 for rich clusters.

% Finally, the distribution of redshift is shown in Fig.~\ref{subfig:z} with multi-modal histogram. Most systems fall approximately between 0.025 and 0.057.  

\subsection{Satellite-to-central galaxy parameters ratios}
\begin{figure*}
%  \label{fig:histo}
 \centering
 \subfloat{\includegraphics[scale=0.4, viewport=0 0 540 440, 
clip]{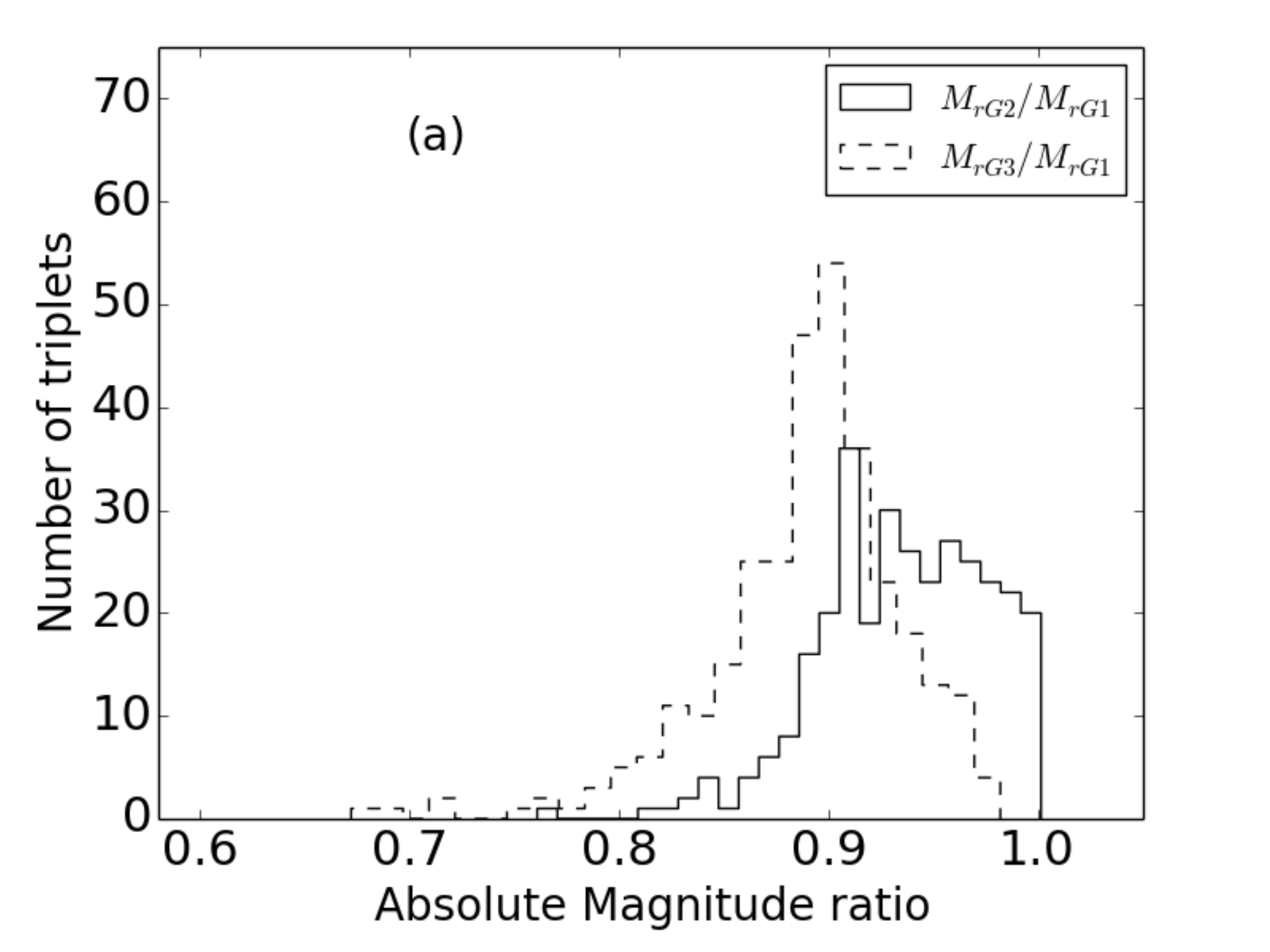}\label{subfig:M123}}
 \subfloat{\includegraphics[scale=0.4, viewport=0 0 540 440, 
clip]{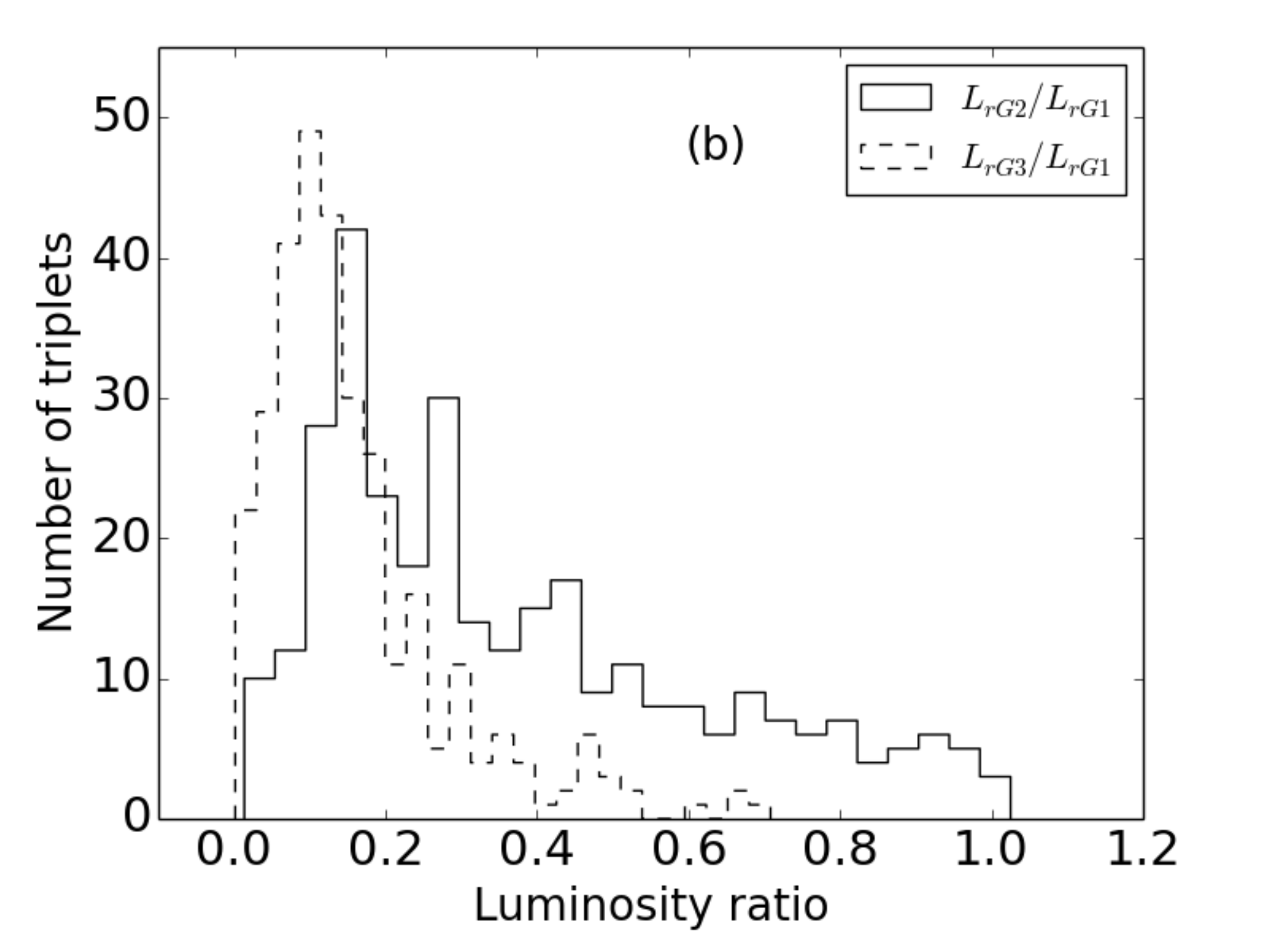}\label{subfig:L123}}
 
%  \subfloat[\scriptsize Redshift distribution]{\includegraphics[scale=0.3, viewport=0 0 540 420, clip]{eps_figures/z_histo.eps}\label{subfig:z}}
% \newline
\caption{The ratio between the absolute magnitude ($M_r$), panel a, and the luminosity ($L_r$), panel b, between the galaxies G2 and G1 (solid lines) and G3 and G1 (dashed lines) for the studied SIT sample.}
\label{fig:historatio}
\end{figure*}

In Fig.~\ref{fig:historatio}, we illustrate the ratio of both absolute magnitude ($M_r$), panel a, and luminosity ($L_r$) , panel b, in the r-band for the primary galaxy (G1), the brightest in the system, and the two satellite galaxies G2 and G3. It is notable that the absolute magnitude ratio of G2$/$G1 (solid line) has a mean value of 1$/$1.06, while that for G3$/$G1 (dashed line) is 1$/$1.12. Similarly, the luminosity of G2$/$G1 (solid line) has a mean value of 1$/$3.45 and that of G3$/$G1 (dached line) is 1$/$7.7. Thus, both the absolute magnitude and the luminosity ratio of G3$/$G1 are much smaller than those of G2$/$G1. This indicates that isolated triplets usually show a hierarchical structure, and it is less probably to find a triplet system with two similar absolute magnitude and luminosity for its members. We conclude that in triplets the central (primary) galaxy is always brighter and more luminous than the two satellite galaxies. These results are in line with \citet{Fernandez15} findings in which the typical satellite-to-primary mass ratio is 1$/$100. They also found that the central galaxy is the most massive galaxy of the three members in the triplet system. The statistical parameters of the absolute magnitude and luminosity ratios are summarised in Table~\ref{tbl:ratios}.
\begin{table*}
  \begin{center}
    \tabcolsep 5.8pt
    \small 
    \caption{Statistical values for the ratios of absolute magnitude ($M_r$) and luminosity ($L_r$) in the r-band of the satellite galaxies (G2 and G3) to those of the primary galaxy (G1). STD refers to the standard deviation.} \label{tbl:ratios}
    \begin{tabular}{|l|l|l|l|l|l|l|}
\hline
 \multicolumn{1}{|c|}{Parameters} &
%   \multicolumn{1}{c|}{Units} &
%   \multicolumn{2}{c|}{Range} &
  \multicolumn{2}{c|}{Range}&
  \multicolumn{3}{c|}{Statistical Values}\\
  \multicolumn{1}{c|}{} &
  \multicolumn{1}{c|}{Min.} &
  \multicolumn{1}{c|}{Max.}&
  \multicolumn{1}{c|}{Median} &
  \multicolumn{1}{c|}{Mean} &
  \multicolumn{1}{c|}{STD} \\
  
%   \multicolumn{1}{c|}{}
   
%  Parameters (members)                 &  Units   &  Min.Range      & Max. Range \\  \hline
\hline
 
 $M_{rG2}/M_{rG1}$   & 0.76   & 1.00  & 0.94   & 0.94   & 0.04 \\  
 $M_{rG3}/M_{rG1}$   & 0.67   & 0.98  & 0.89   & 0.89   & 0.05 \\  
 $L_{rG2}/L_{rG1}$   & 0.01   & 1.03  & 0.37   & 0.29   & 0.25 \\  
 $L_{rG3}/L_{rG1}$   & 0.001  & 0.71  & 0.16   & 0.13   & 0.12 \\ 
 \hline
    \end{tabular}
  \end{center}
\end{table*}

\subsection{Correlations of the mean separation between triplet members ($\overline{r_p}$) with the systems dynamical parameters ($\sigma$, $M_{vir}$, $M_{vir}/L_r$)}
\begin{figure}
% \label{fig:srprs}
\centering
 \subfloat{\includegraphics[scale=0.4, viewport=0 0 554 420, 
clip]{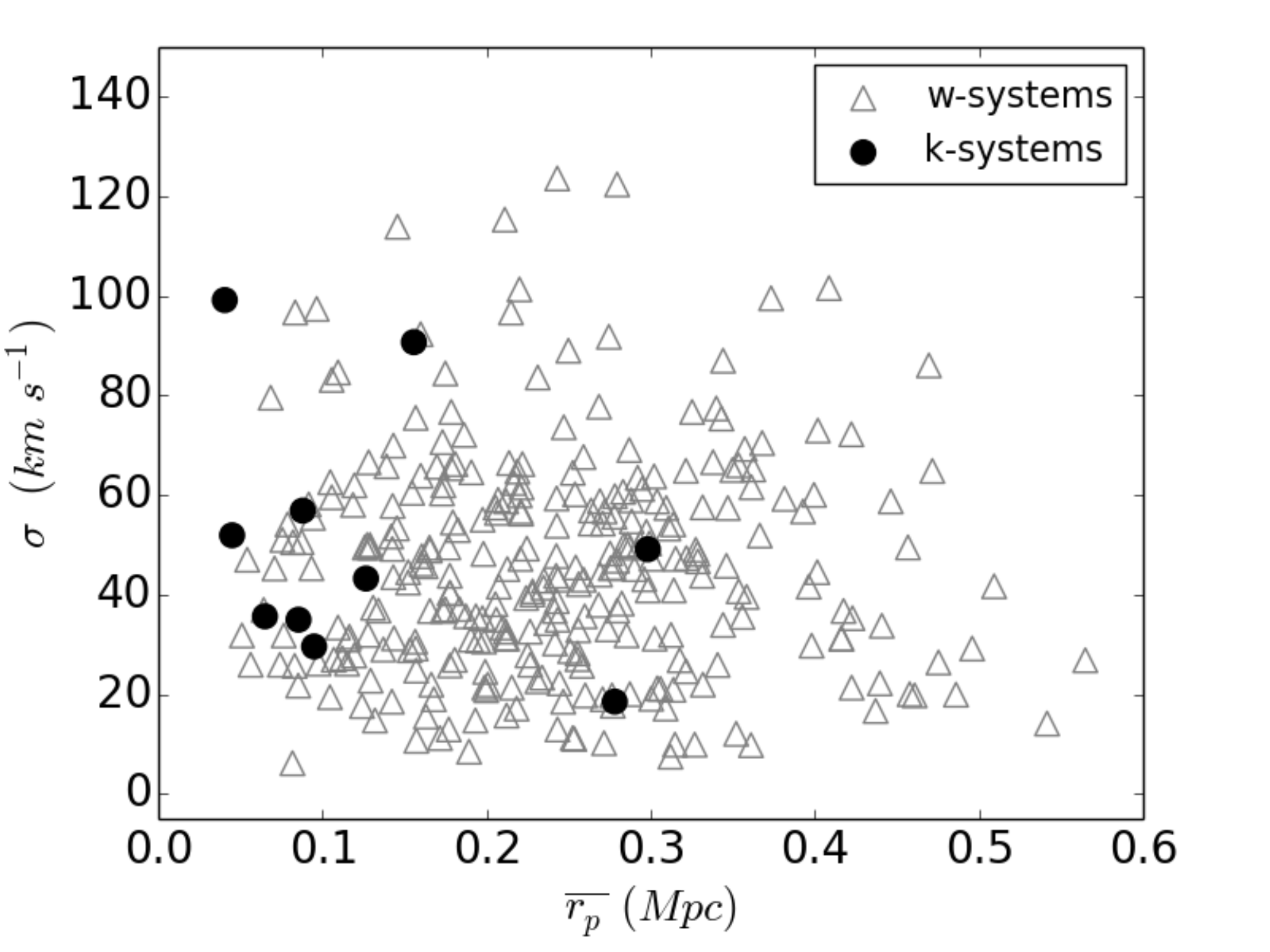}\label{subfig:sigmarp}}
 \newline
\caption{Correlation between the velocity dispersion ($\sigma$) and the mean projected separation ($\overline{r_p}$) for the 315 triplet systems of SIT catalogue. Circles represent the ten compact systems (k-systems), while triangles represent the wide systems (w-systems) of this sample, as shown in the legend.}
\label{fig:vrprs}
\end{figure}

\begin{figure}
\centering
 \subfloat{\includegraphics[scale=0.4, viewport=0 0 554 430, 
clip]{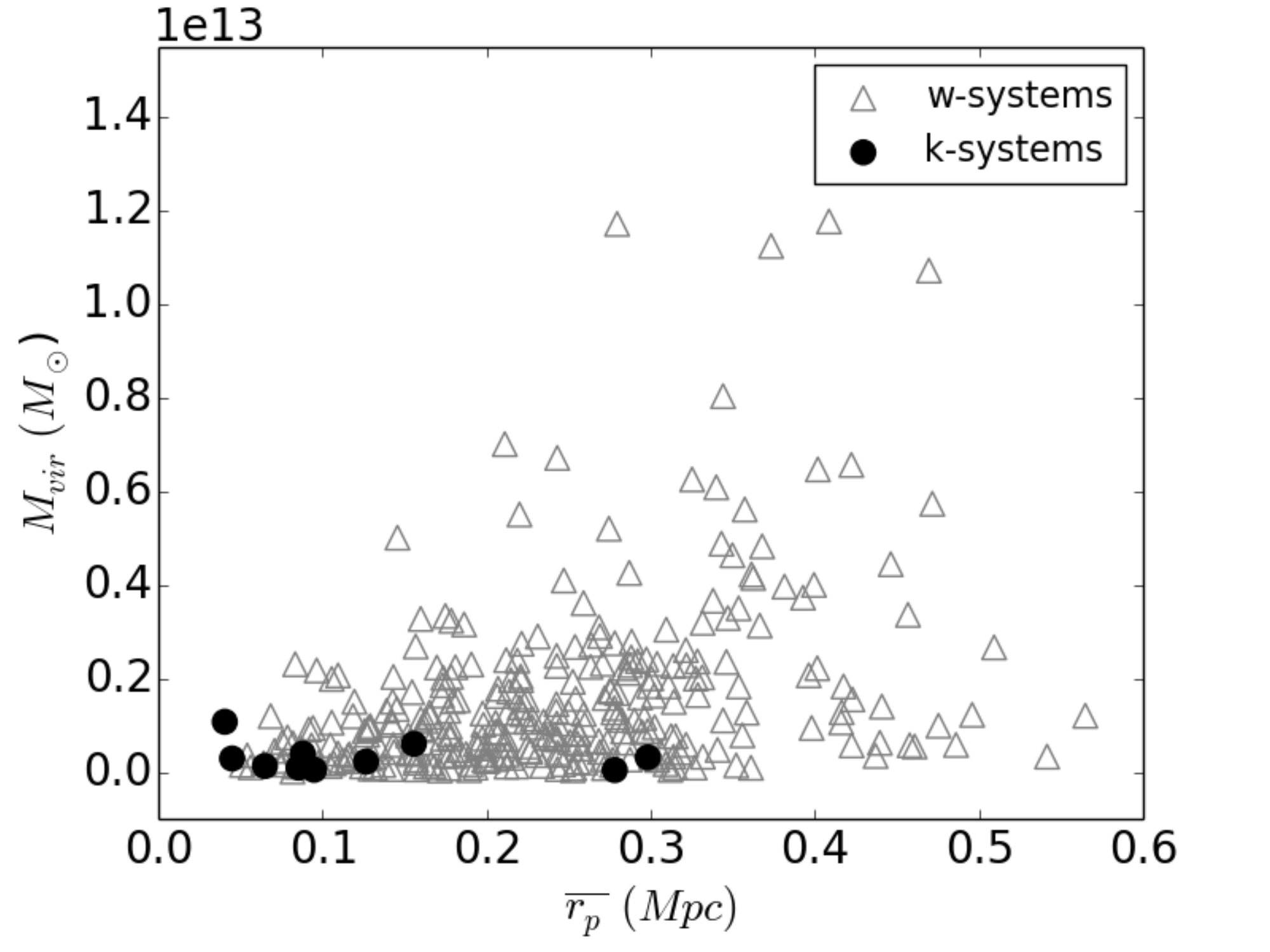}\label{subfig:Mrp}}
 \newline
\caption{Correlation of virial mass ($M_{vir}$) with mean projected separation ($\overline{r_p}$) for the 315 triplet systems of SIT. Symbols are the same as in Fig.~\ref{subfig:sigmarp}.}
\label{fig:Mrprs}
\end{figure}

\begin{figure}
\centering
 \subfloat{\includegraphics[scale=0.4, viewport=0 0 554 430, 
clip]{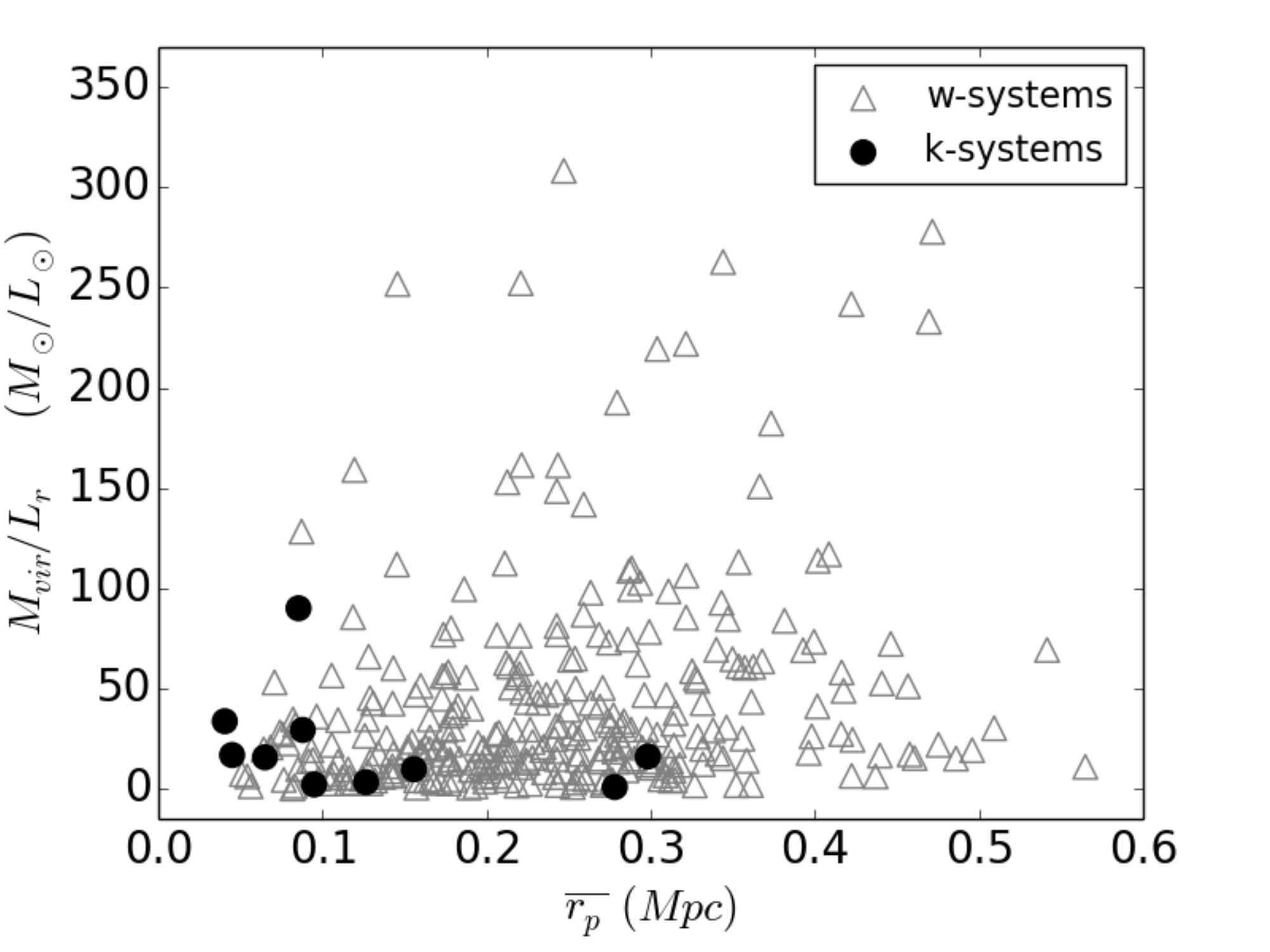}\label{subfig:MLrp}}
  \newline
\caption{Correlation of mass-to-light ratio ($M_{vir}/L_r$) with mean projected ($\overline{r_p}$) separation for the wide (w, triangles) and the compact (k, circles) systems in our sample.}
\label{fig:MLrprs}
\end{figure}

% \label{Sec:}
Correlation coefficients quantify the correlation between two or more random variables measured for a system of galaxies. In this study, we examined the correlation between the dynamical parameters ($\sigma$, $M_{vir}$, $M_{vir}/L_r$) and the mean values of the projected separations,$r_p$, (see Figs.~\ref{fig:vrprs},~\ref{fig:Mrprs}, and~\ref{fig:MLrprs}) by estimating their correlation coefficients.  

% \subsubsection{Velocity dispersions $(\sigma)$ versus projected $(r_p)$ and real separation $(r_s)$}
% \label{SubSec:MethodChoATT}

Fig.~\ref{fig:vrprs} demonstrate the relationship between the velocity dispersion ($\sigma$) and $\overline{r_p}$. It is clear that there is no correlation between $\sigma$ and $\overline{r_p}$ ($f_{rp\sigma}$=-0.03). We also found that the compact systems, in this catalogue, are distributed at lower $\overline{r_p}$ indicating that $\overline{r_p}$ decreases directly with $r_h$. In addition, the majority of these systems are concentrated at lower $\sigma$, smaller than 60 $km$ $s^{-1}$.

% \subsubsection{Virial Mass $(M_{vir})$ versus projected $(r_p)$ and real separation $(r_s)$}

The correlation of $M_{vir}$ with $\overline{r_p}$, in Fig.~\ref{fig:Mrprs}, shows a random distribution with a weak correlation coefficient ($f_{r_pM_{vir}}$= 0.37). Compact systems tend to be concentrated in a small range of $M_{vir}$ ($0.004\le M_{vir} ( \times 10^{13} M_\odot) \le 0.11 $) and show a nearly linear relation with $\overline{r_p}$.

% \subsubsection{Mass-to-Light ratio $(M_{vir}/L_r)$ versus projected $(r_p)$ and real separation $(r_s)$}
Fig.~\ref{fig:MLrprs} illustrates the marginal correlation between $\overline{r_p}$ and $M_{vir}/L_r$ with $f_{r_pM_{vir}/L_r}$= 0.06. It is hard to identify a specific relation from this random distribution. However, we found that compact systems (solid circles) are concentrated at lower $M_{vir}/L_r$ ($0.84\le M_{vir}/L_r \le 90.03)$ compared to wide systems (open triangles).

The weak correlations obtained indicate that triplets in this catalogue follow the characteristics and properties of Tully \citep{Tully87} and Wide triplets \citep{Trofimov95} confirming that these systems are far from the state of virial equilibrium since they have the largest mean harmonic separations and mass-to-light ratios that is in-line with \cite{Vavilova05}. Hence, we may conclude that merging is hard to take place between members of such wide triplet systems. 

In addition, we found that compact systems have nearly constant $M_{vir}$ along the variation of $\overline{r_p}$ and are characterised by small values of $M_{vir}$ and $M_{vir}/L_r$. They are always fall at the small range ($0.04\le \overline{r_p}\le 0.3$). This defines the small halo where members of compact systems are embedded.

\subsection{Correlations between triplets dynamical parameters and the Large-Scale Structure (LSS)}

\begin{figure*}
\centering
 \subfloat{\includegraphics[scale=0.4, viewport=0 0 554 430, 
clip]{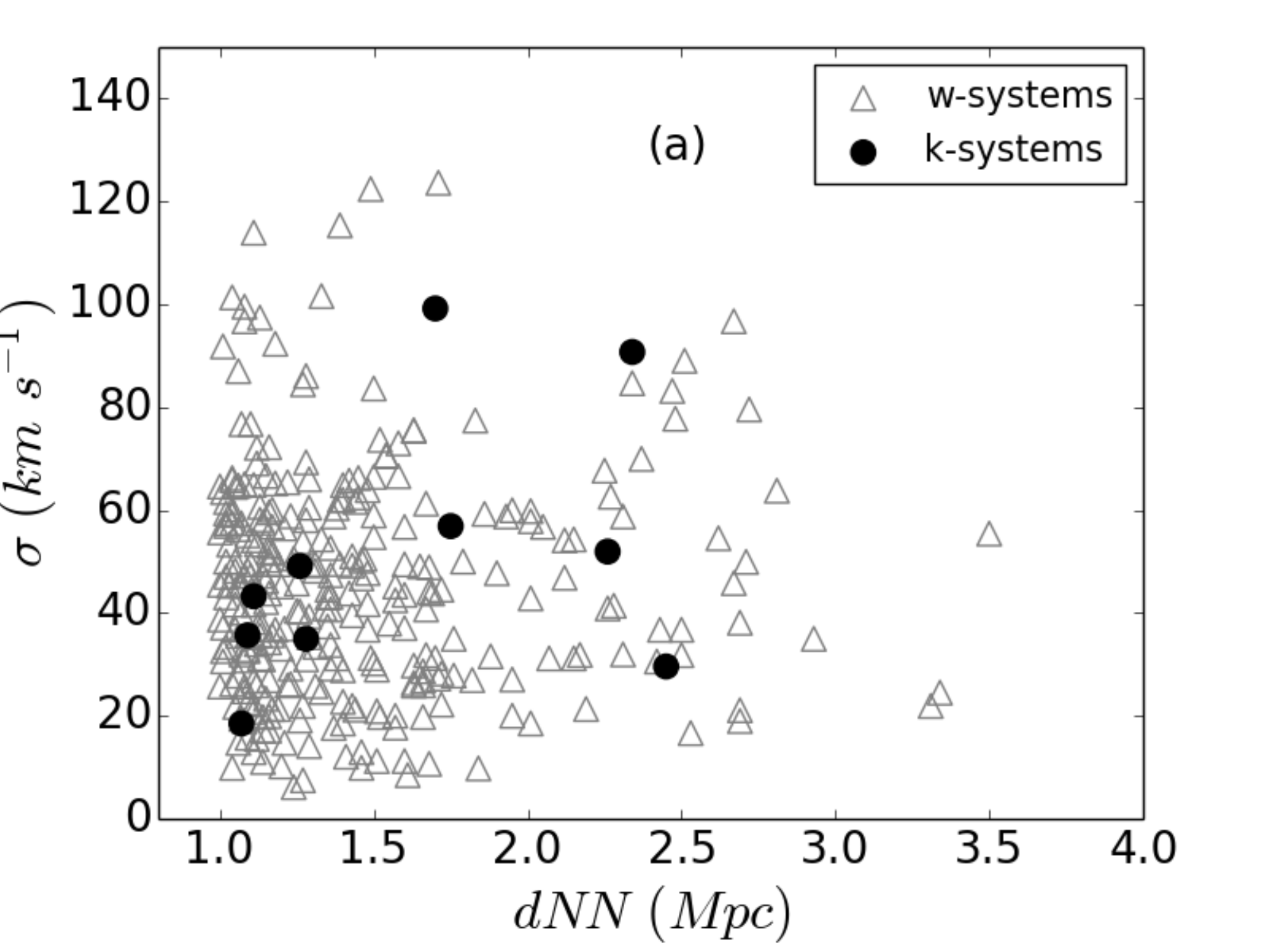}\label{subfig:dNNs}}
 \subfloat{\includegraphics[scale=0.4, viewport=0 0 554 430, 
clip]{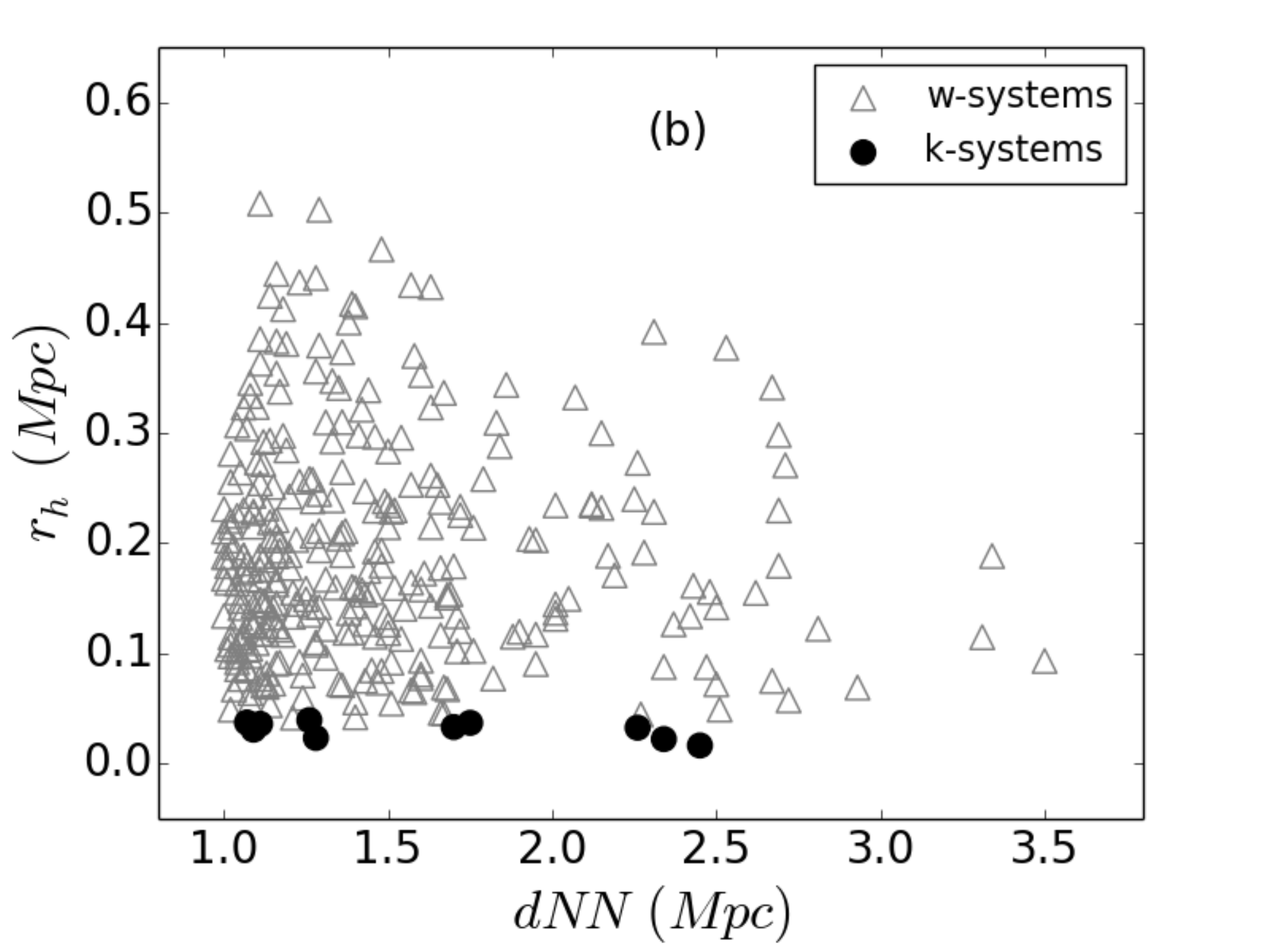}\label{subfig:dNNrh}}\\
 \subfloat{\includegraphics[scale=0.4, viewport=0 0 554 430, 
clip]{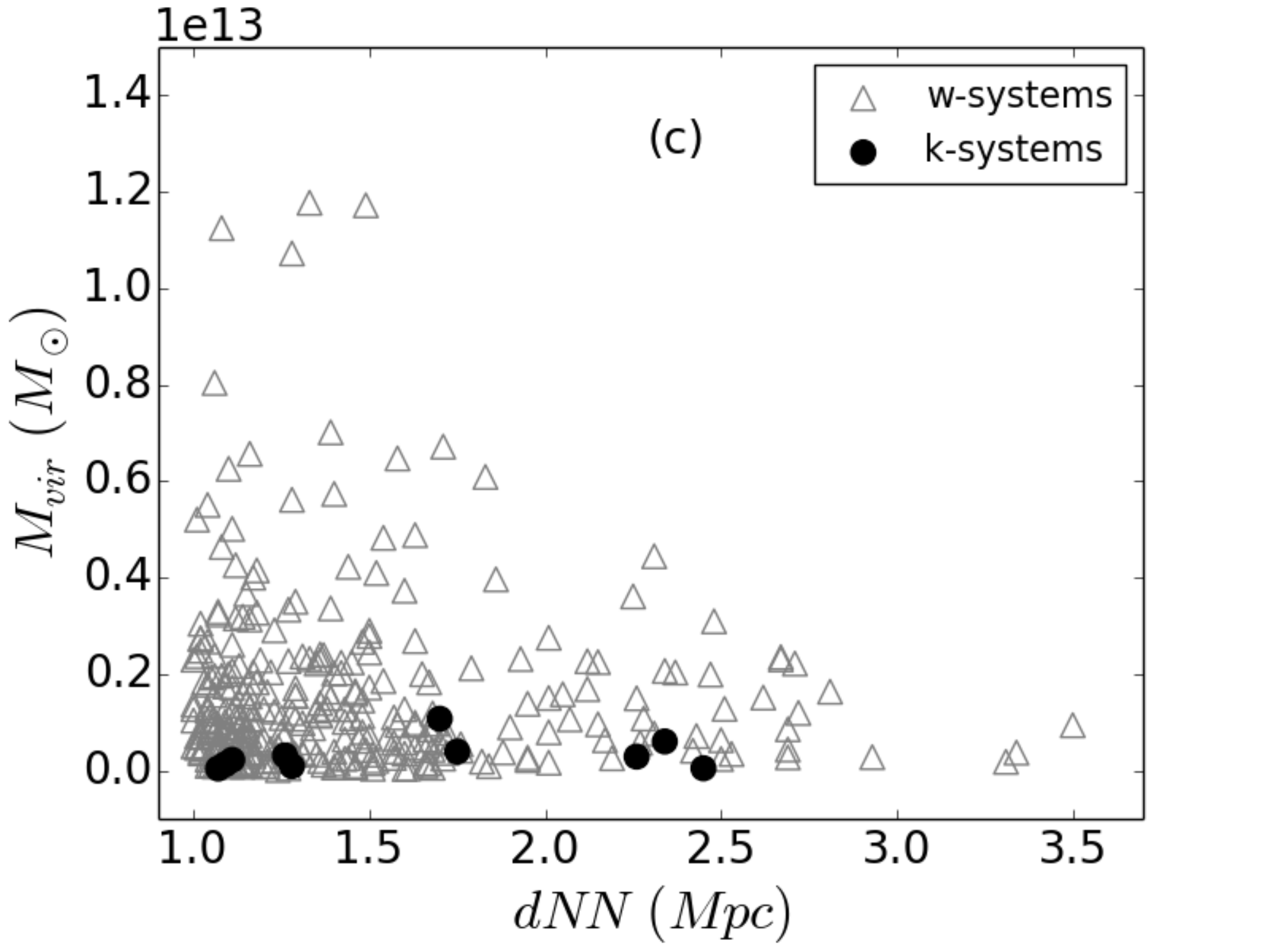}\label{subfig:dNNM}}
 \subfloat{\includegraphics[scale=0.4, viewport=0 0 554 430, 
clip]{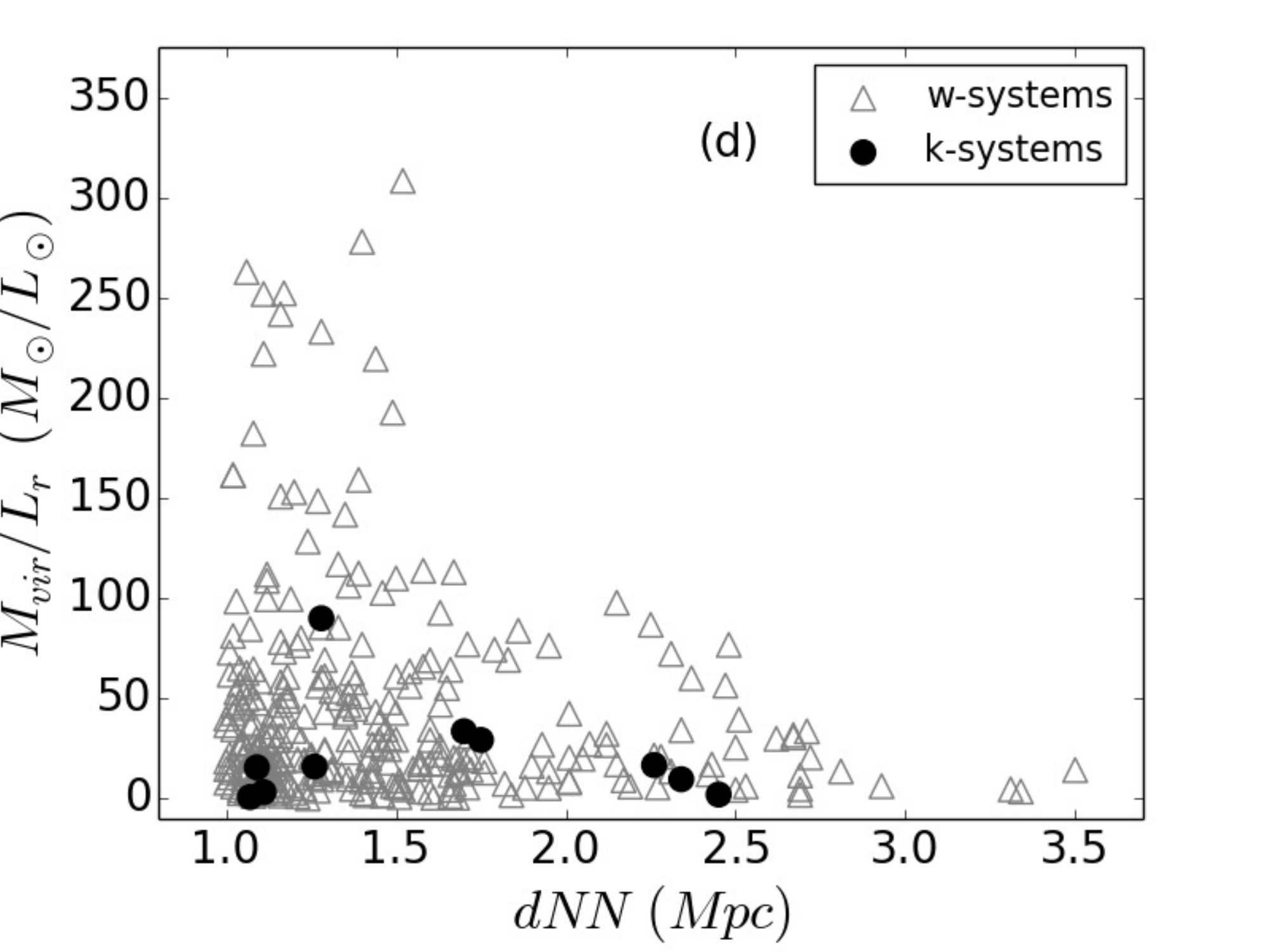}\label{subfig:dNNML}}
 \newline
\caption{Distance from the brightest galaxy to the first neighbour ($d_{NN}$) plotted versus (a) velocity dispersion ($\sigma$), (b) mean harmonic separation ($r_h$), (c) virial mass ($M_{vir}$), and (d) mass-to-light ratio ($M_{vir}$/$L_r$) for the 313 galaxy triplet systems. Wide and compact systems are categorised as shown in the legend.}
\label{fig:dNNsrM}
\end{figure*}

\begin{figure*}
\centering
 \subfloat{\includegraphics[scale=0.4, viewport=0 0 554 430, 
clip]{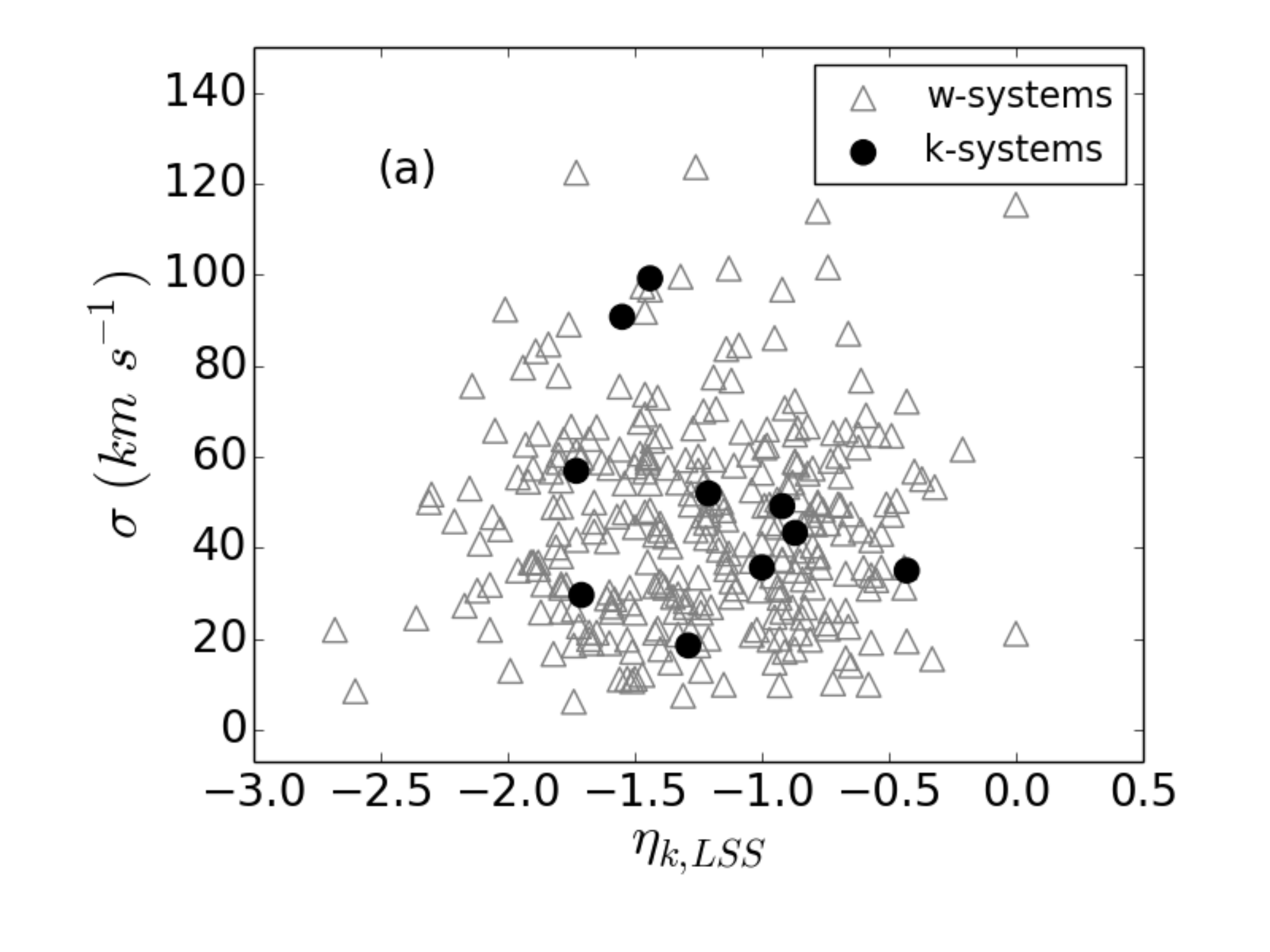}\label{subfig:sklss}}
 \subfloat{\includegraphics[scale=0.4, viewport=0 0 554 430, 
clip]{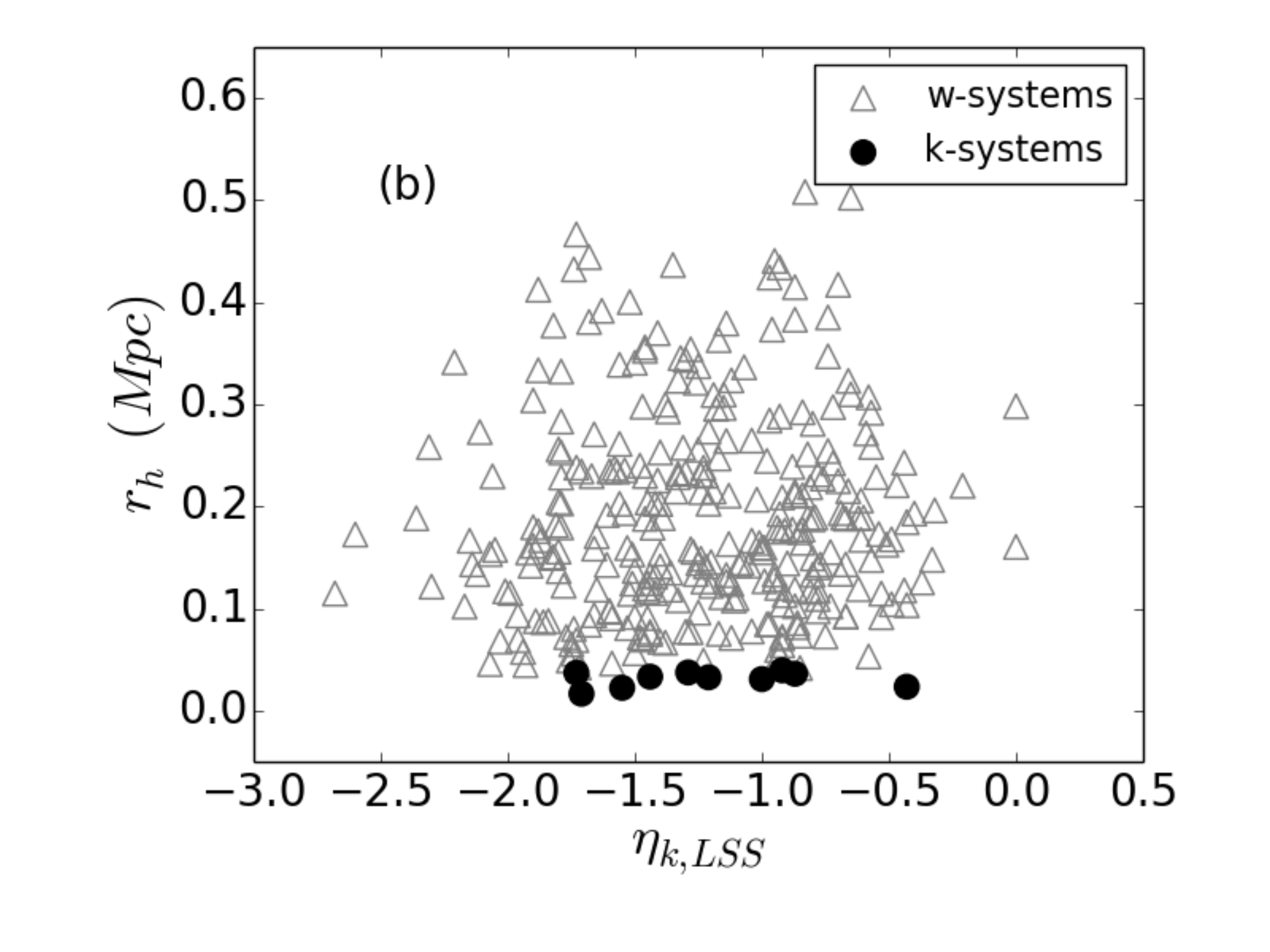}\label{subfig:rhklss}}\\
 \subfloat{\includegraphics[scale=0.4, viewport=0 0 554 430, 
clip]{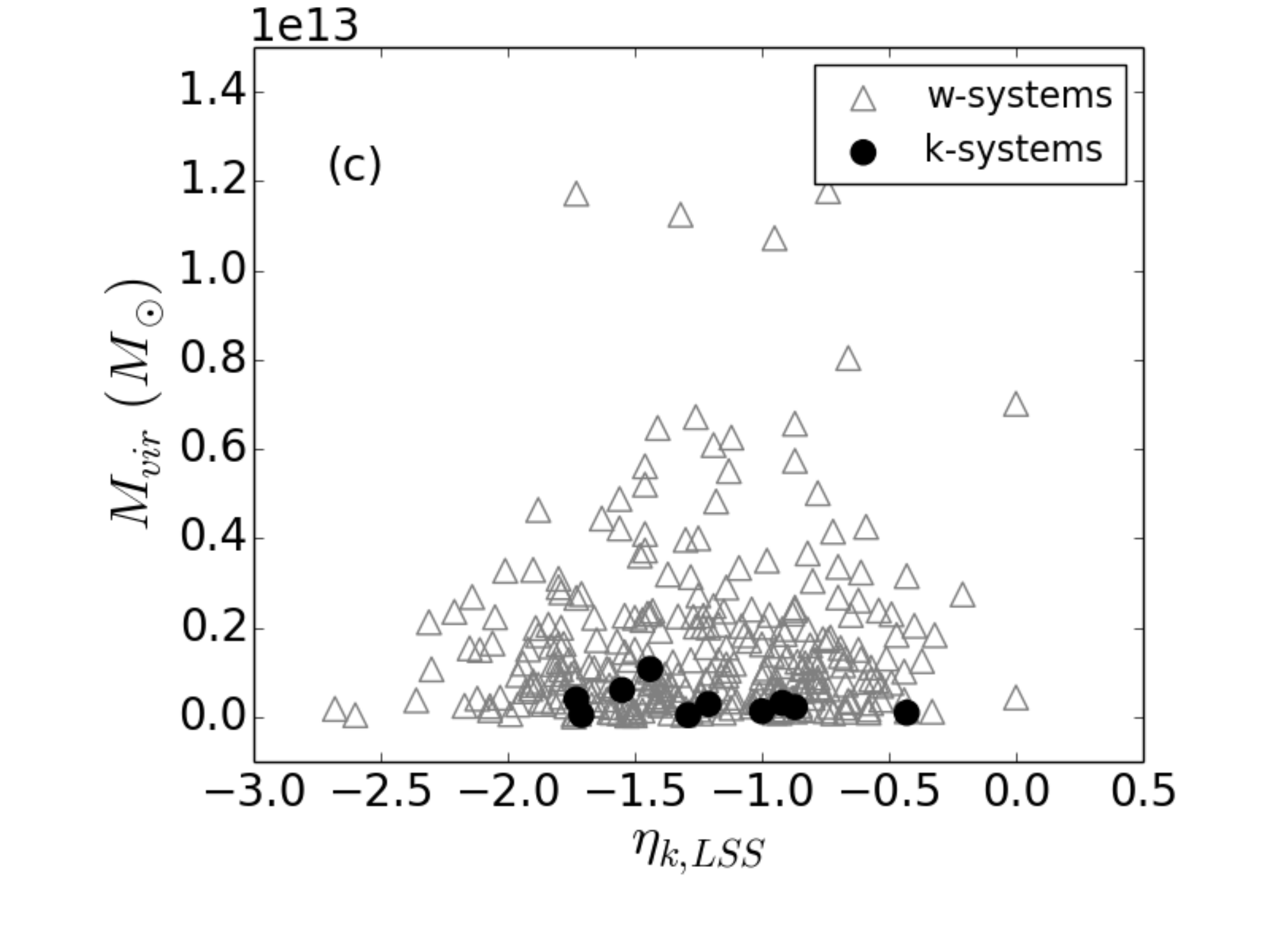}\label{subfig:Mklss}}
 \subfloat{\includegraphics[scale=0.4, viewport=0 0 554 430, 
clip]{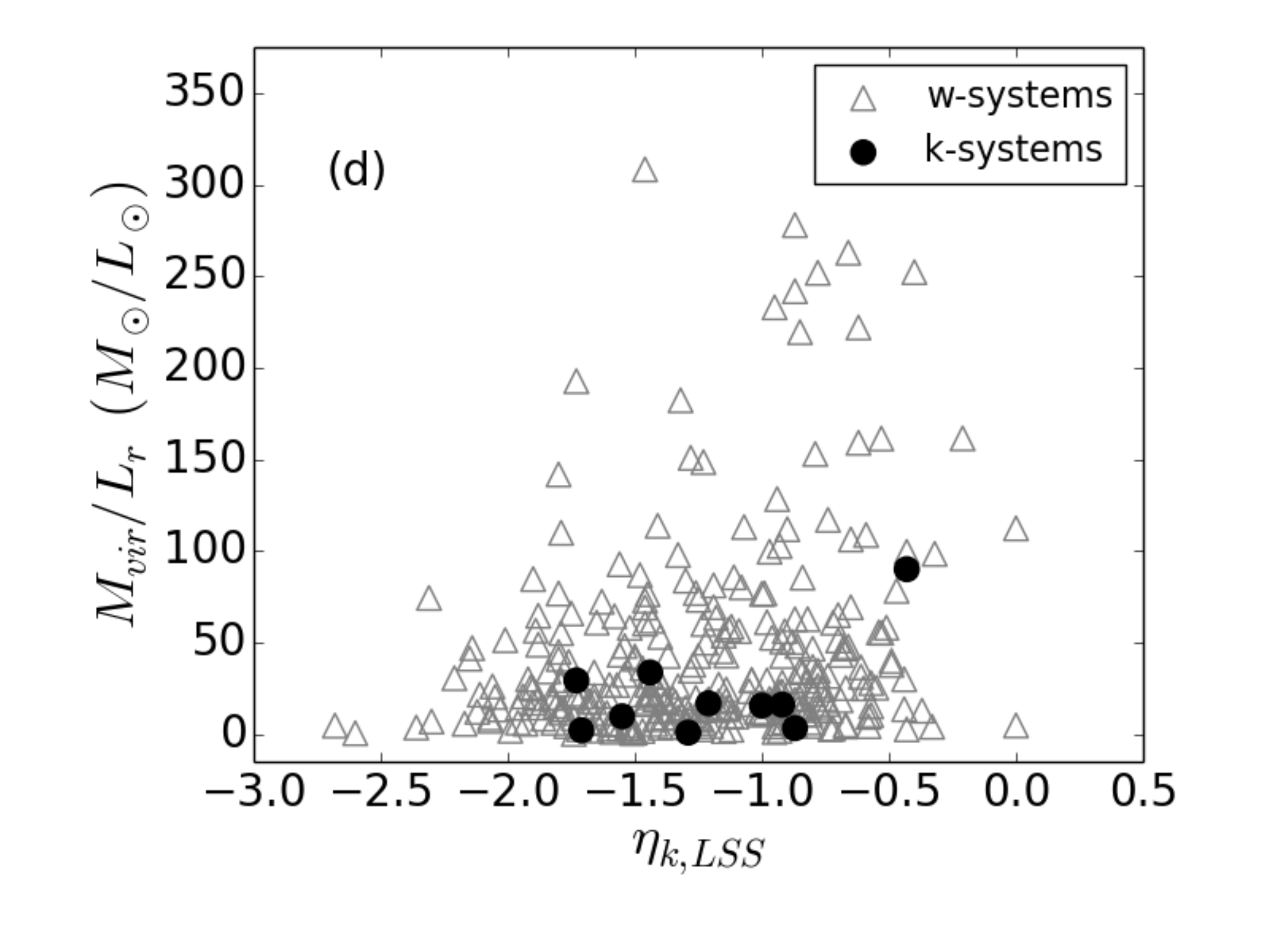}\label{subfig:MLklss}}
 \newline
\caption{Scatter plots between projected density of the LSS ($\eta_{k,LSS}$) and (a) velocity dispersion ($\sigma$), (b) mean harmonic separation ($r_h$), (c) virial mass ($M_{vir}$), and (d) mass-to-light ratio ($M_{vir}$/$L_r$) for the 313 galaxy triplets systems. Circles represent the ten compact systems (k-systems), while triangles represent the wide systems (w-systems) of this sample.}
\label{fig:klss}
\end{figure*}

\begin{figure*}
\centering
 \subfloat{\includegraphics[scale=0.4, viewport=0 0 554 430, 
clip]{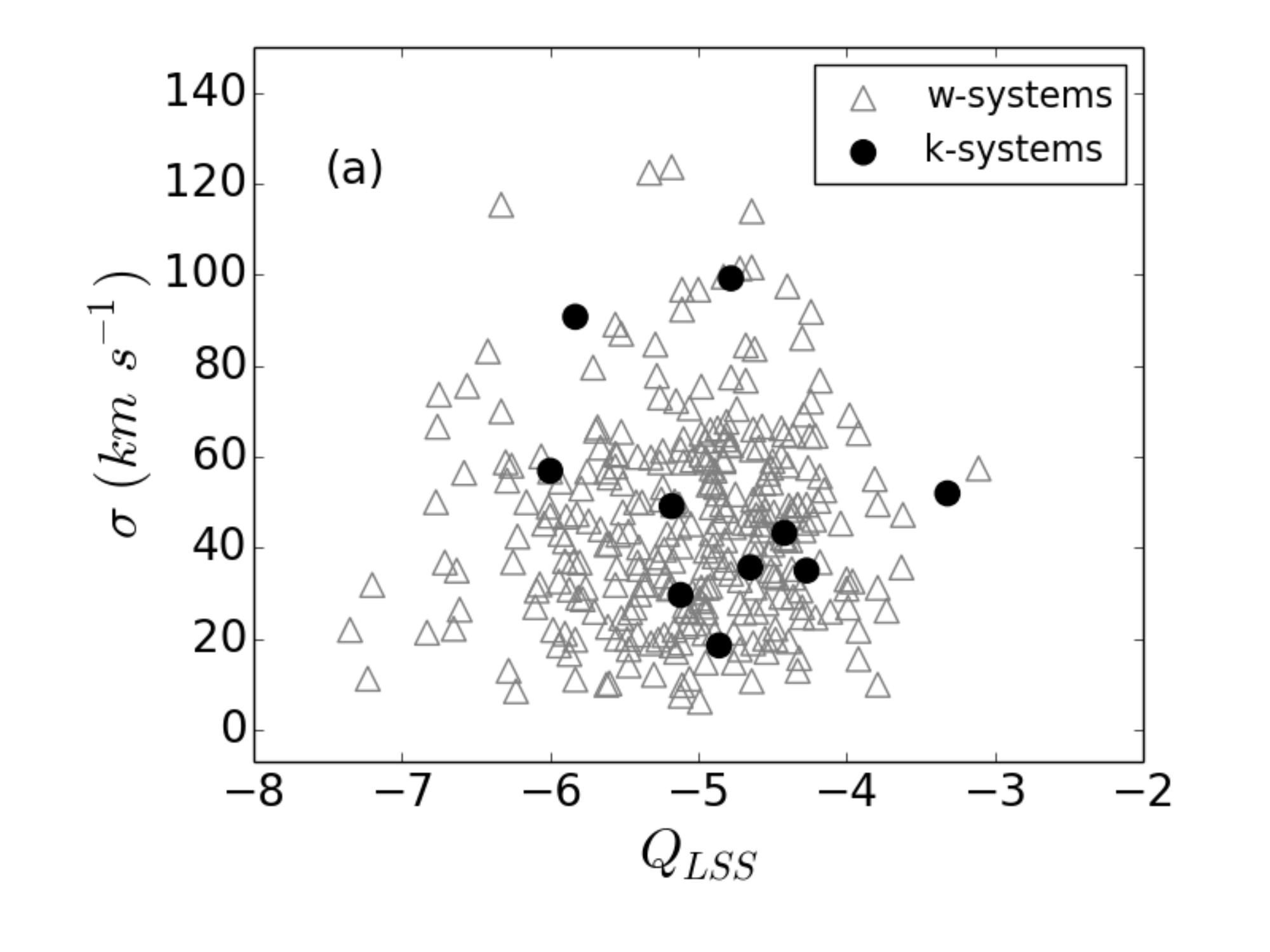}\label{subfig:QQs}}
 \subfloat{\includegraphics[scale=0.4, viewport=0 0 554 430, 
clip]{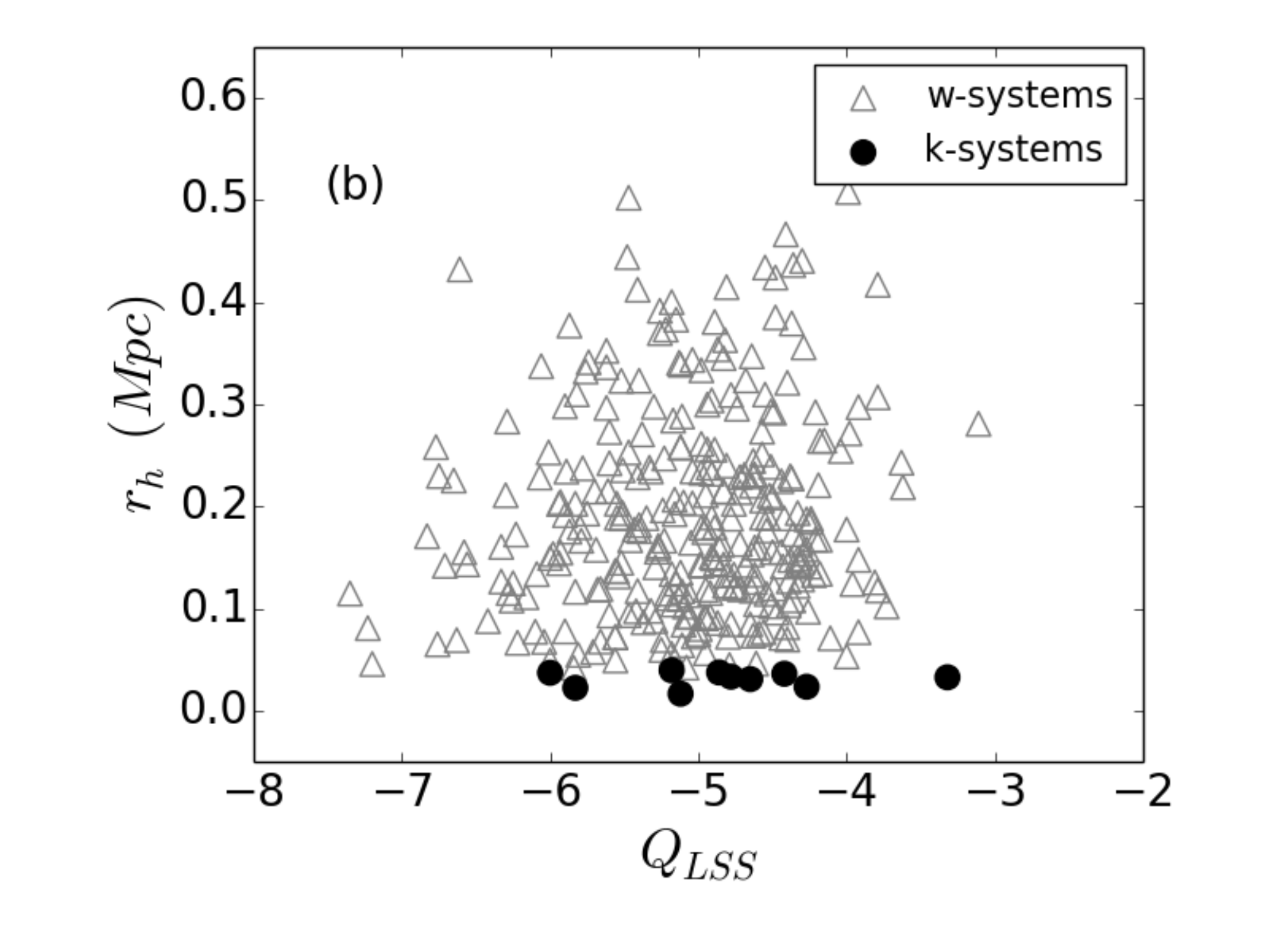}\label{subfig:QQrh}}\\
 \subfloat{\includegraphics[scale=0.4, viewport=0 0 554 430, 
clip]{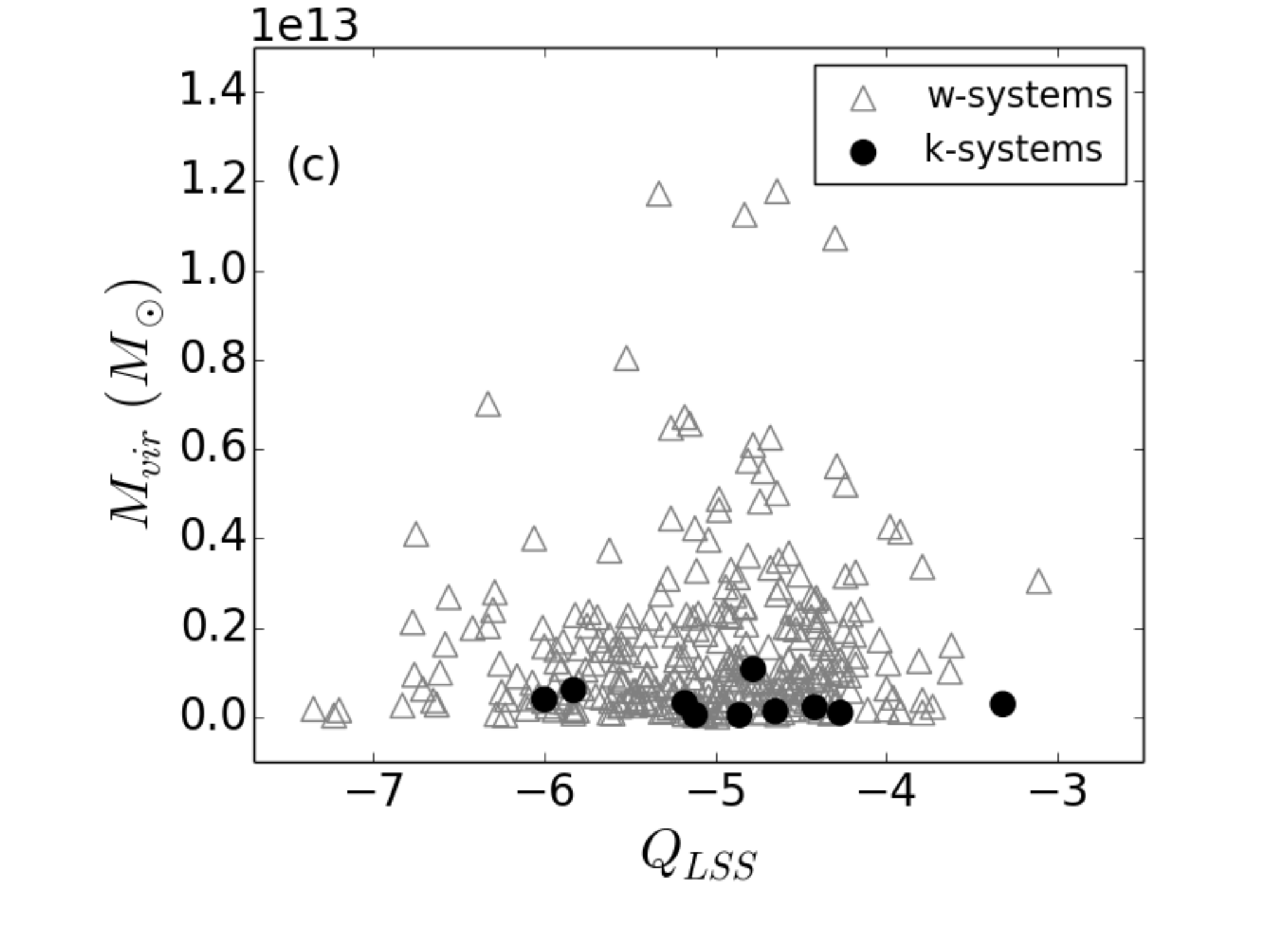}\label{subfig:QQM}}
 \subfloat{\includegraphics[scale=0.4, viewport=0 0 554 430, 
clip]{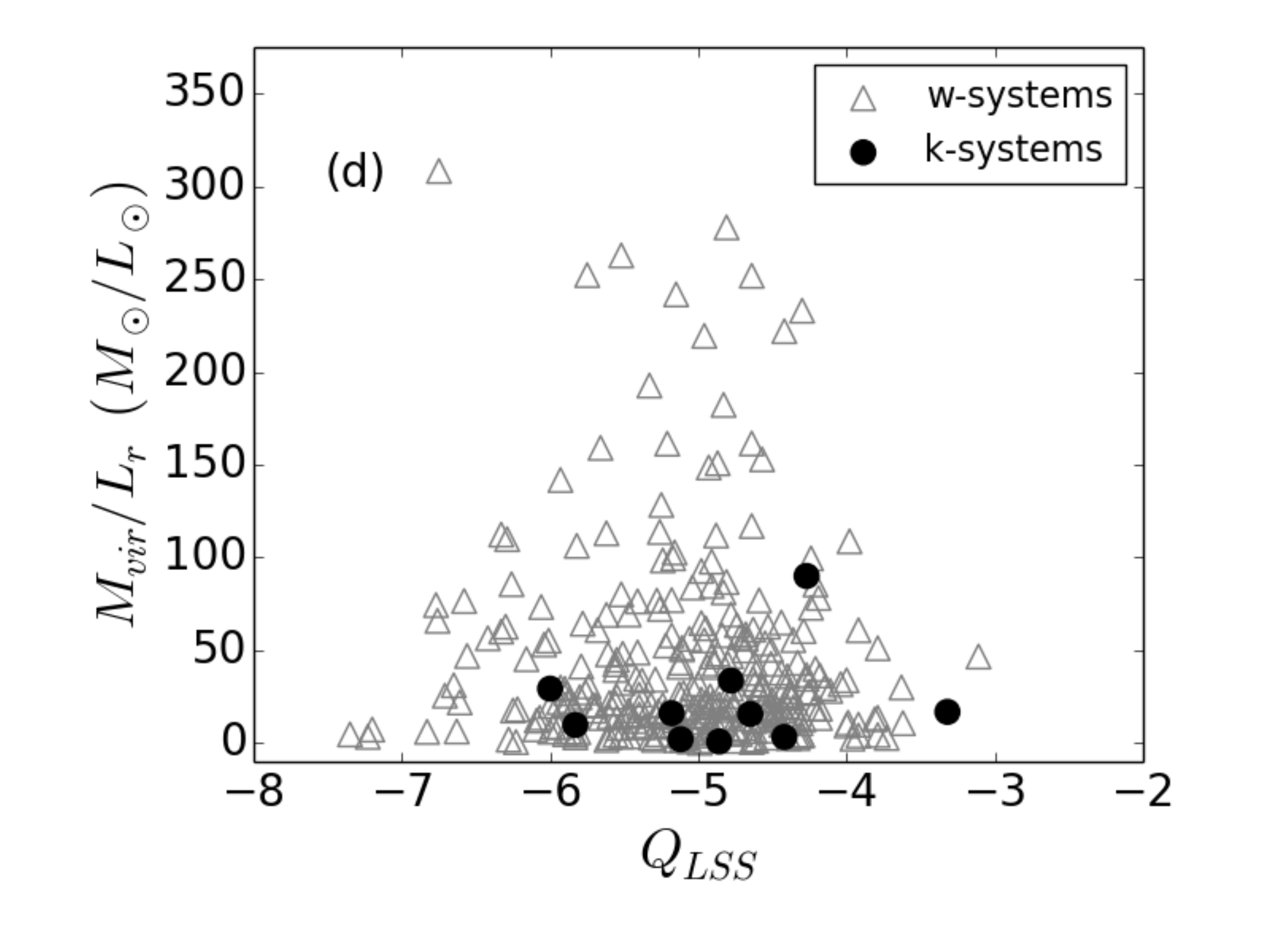}\label{subfig:QQML}}
 \newline
\caption{Scatter plots between tidal strength estimation of the LSS ($Q_{LSS}$) and (a) velocity dispersion ($\sigma$), (b) mean harmonic separation ($r_h$), (c) virial mass ($M_{vir}$), and (d) mass-to-light ratio ($M_{vir}$/$L_r$) for the 313 galaxy triplets systems. Symbols are the same like in Fig.~\ref{fig:klss}.}
\label{fig:QQsrM}
\end{figure*}

\begin{figure*}
\centering
 \subfloat{\includegraphics[scale=0.4, viewport=0 0 554 430, 
clip]{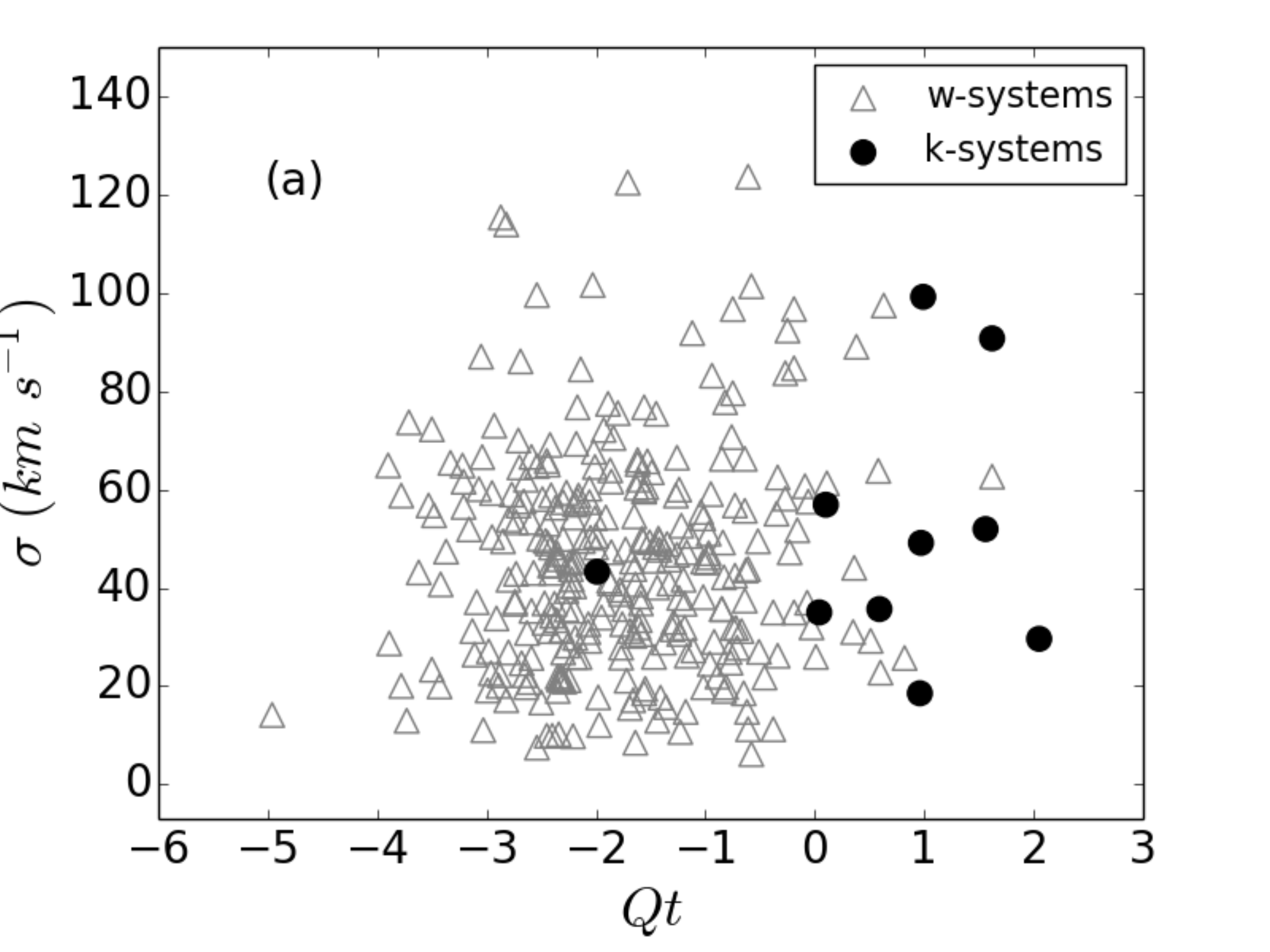}\label{subfig:Qts}}
 \subfloat{\includegraphics[scale=0.4, viewport=0 0 554 430, 
clip]{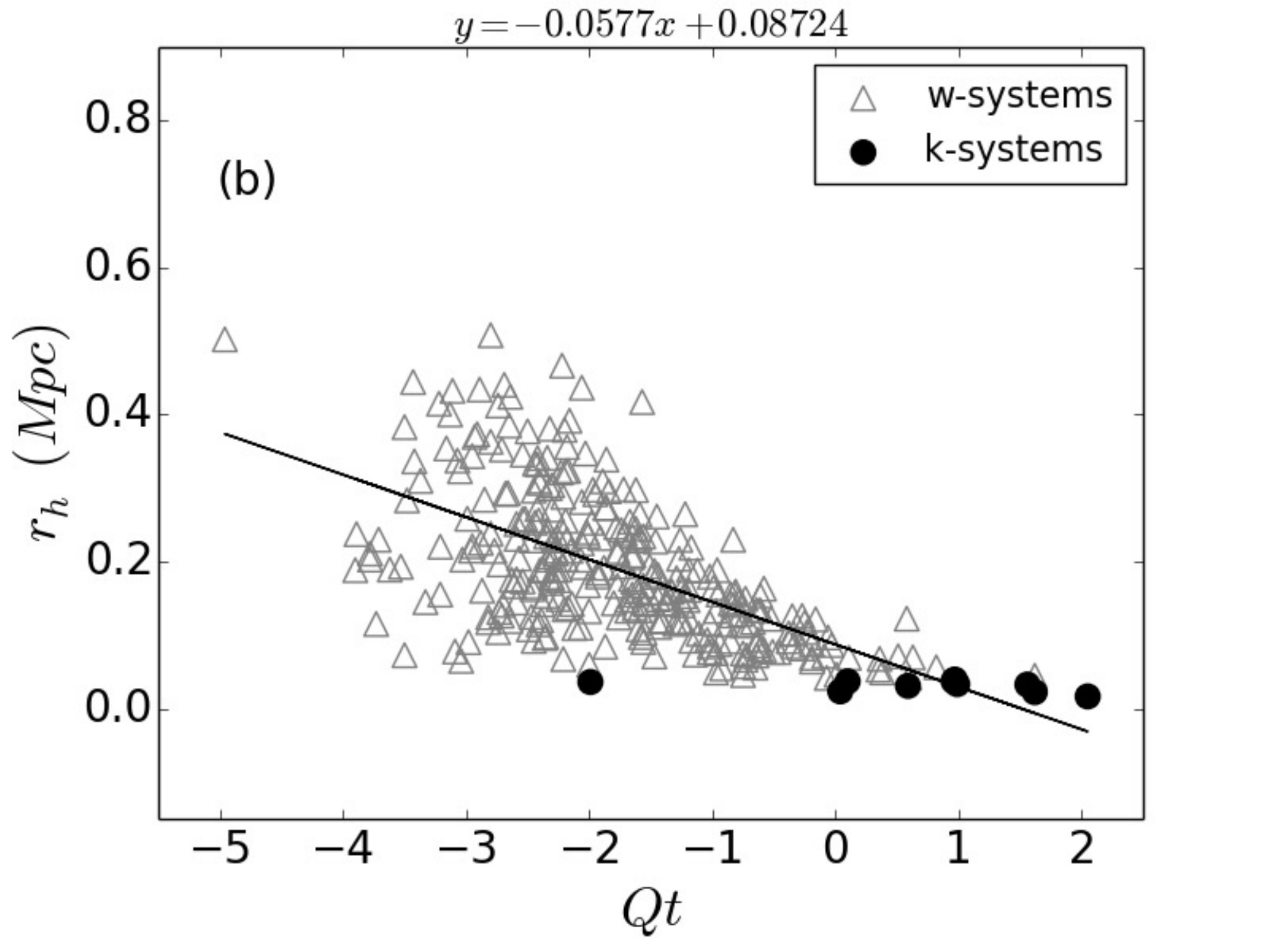}\label{subfig:Qtrh}}\\
 \subfloat{\includegraphics[scale=0.4, viewport=0 0 554 430, 
clip]{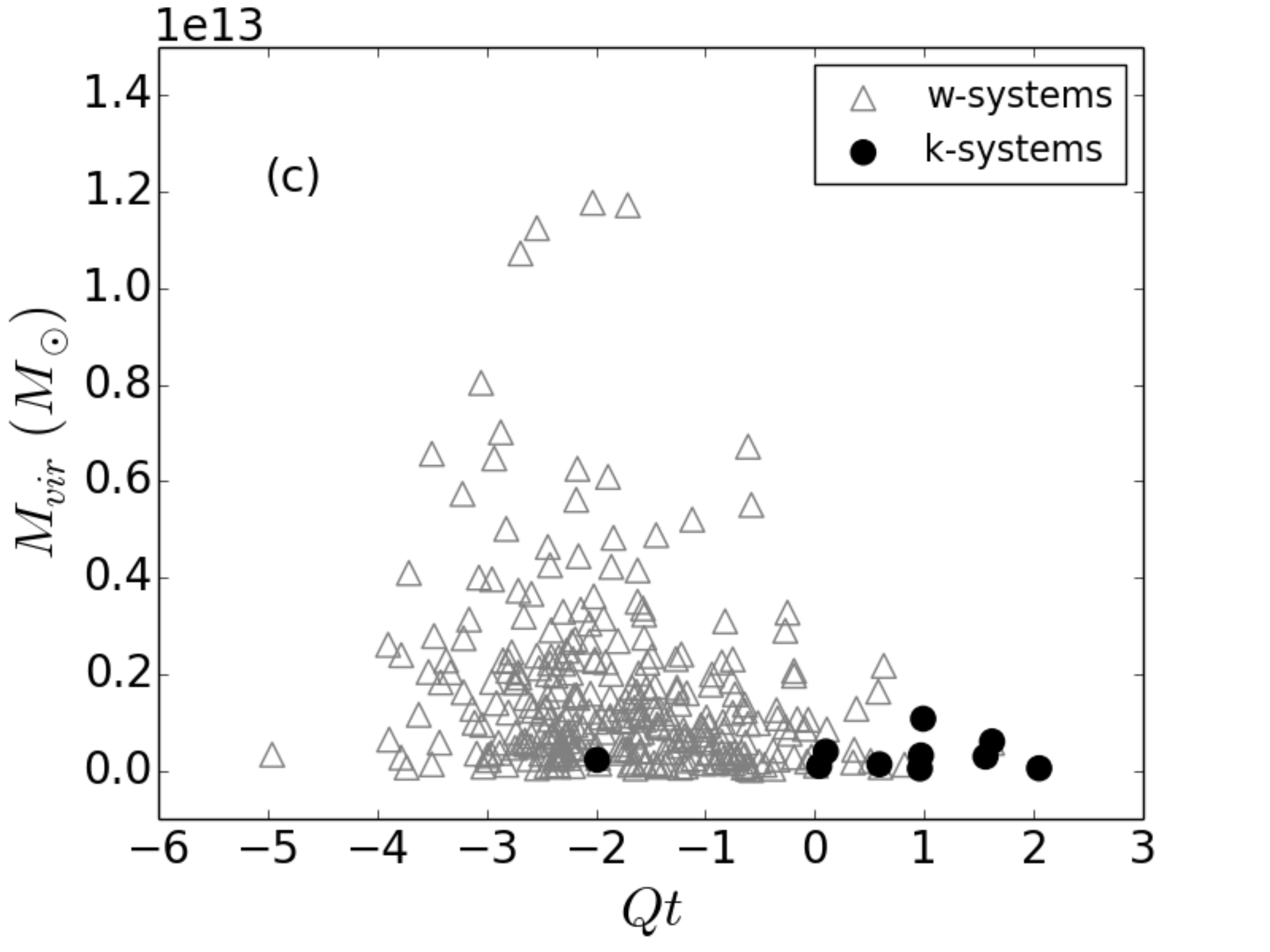}\label{subfig:QtM}}
 \subfloat{\includegraphics[scale=0.4, viewport=0 0 554 430, 
clip]{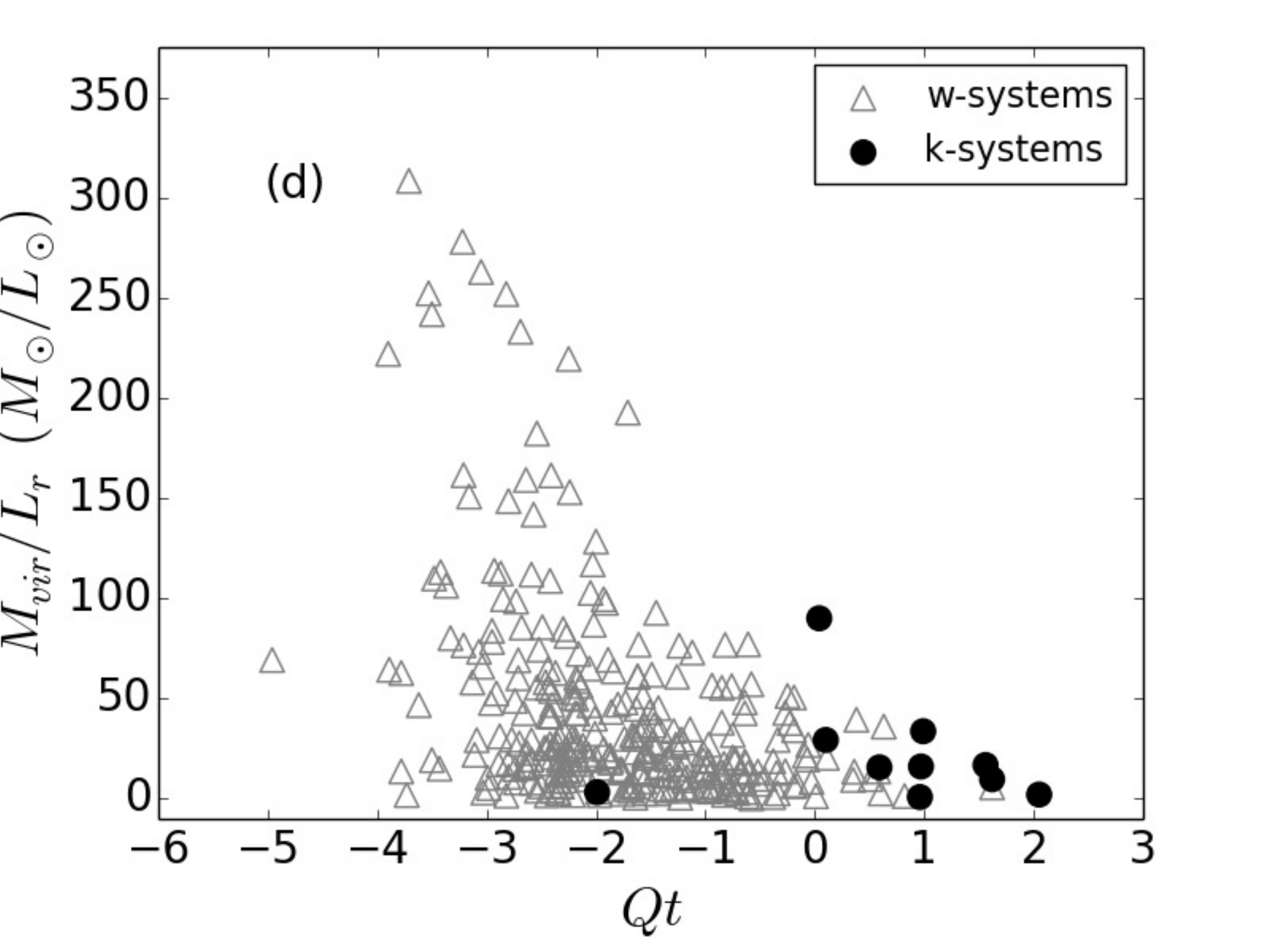}\label{subfig:QtML}}
 \newline
\caption{Correlation between tidal strength estimated on the galaxy (G1) exerted by the G2 and G3 galaxies ($Qt$) and (a) velocity dispersion ($\sigma$), (b) mean harmonic separation ($r_h$), (c) virial mass ($M_{vir}$), and (d) mass-to-light ratio ($M_{vir}$/$L_r$) for the galaxy triplet systems. Circles represent the ten compact systems (k-systems), while triangles represent the wide systems (w-systems) of this sample.}
\label{fig:QtsrM}
\end{figure*}

 \citet{Fernandez15} specified in their final primary isolated galaxy triplets sample to have radial velocity differences range -500$ \le \Delta v \le$ 500 $km~s^{-1}$ (and within 1 Mpc). They have also characterised the Large-Scale Structure (LSS) around each primary galaxy (G1) to be $d_{NN} \le$ 5 Mpc, where $d_{NN}$ denotes the distance to the first nearest neighbour from G1 and is specified to be from 1 to 5 Mpc in their criteria. They, also, defined the average density along line of sight through the LSS as the projected number density parameter ($\eta_{k,LSS}$) and the tidal strength generated by the galaxies in the LSS ($Q_{LSS}$). These parameters are quantified to investigate the effect of the LSS environment on the properties of triplet galaxies. Hence, it is important to explore the corelation between $d_{NN}$ (in Mpc), $\eta_{k,LSS}$ and $Q_{LSS}$ and the triplets' dynamical parameters ($\sigma$, $r_h$, $M_{vir}$, and $M_{vir}$/$L_r$); see Figures ~\ref{fig:dNNsrM}, ~\ref{fig:klss} and ~\ref{fig:QQsrM}. In addition, the relation between the tidal strength ($Qt$), defined as the total gravitational interaction strength produced by the neighbour galaxies (G2 and G3) on the central galaxy (G1), and the triplets dynamical parameters has been investigated (Fig.~\ref{fig:QtsrM}).
 
The correlation between the distance from G1 to the first nearest neighbour of the LSS  ($d_{NN}$) in Mpc and the velocity dispersion ($\sigma$), the mean harmonic separation ($r_h$), the virial mass ($M_{vir}$), and the mass-to-light ratio ($M_{vir}$/$L_r$) of the 313 triplet systems (after excluding the two systems (SIT 99 and SIT 168) mentioned before in Sec.~\ref{ss:eqn}) is illustrated in Fig.~\ref{fig:dNNsrM}. It is clear that there is no correlation between $d_{NN}$ and the dynamical parameters of the studied sample, where the correlation coefficients were found to be 0.018, -0.06, -0.075, and -0.0013, respectively. Interestingly, we found that most of compact systems (8 from 10) are located at lower $\sigma$ ($18.5\le \sigma \le 56.9$) and are all concentrated toward smaller $M_{vir}$, and $M_{vir}$/$L_r$. $M_{vir}$ of compact systems have nearly the same value representing an almost linear distribution with the LSS parameters. They are always occupied the $d_{NN}$ range $1.1 \le d_{NN} \le 2.4$ indicating that the probability of an isolated triplet system to have neighbours increases in compact systems. This means that wide triplets seems to be more isolated than compact triplets.

Similarly, Figs.~\ref{fig:klss} , and ~\ref{fig:QQsrM}, show a poor correlation between $Q_{LSS}$ and $\eta_{k,LSS}$ with $\sigma$, $r_h$, $M_{vir}$, and $M_{vir}$/$L_r$ where we obtained correlation coefficients with $Q_{LSS}$ of 0.055, 0.089, 0.094, and 0.049, respectively, and 0.036, 0.058, 0.076, and 0.015, respectively with $\eta_{k,LSS}$ . Compact systems are always fall in the range $-6 \le Q_{LSS} \le -3$, and $-1.7 \le \eta_{k,LSS} \le -0.5$, and follow the same dynamical characteristics as mentioned in their correlation with $d_{NN}$.

Generally, triplet systems are more popular at lower values of $M_{vir}$ and $M_{vir}$/$L_r$ with all LSS parameters $d_{NN}$, $\eta_{k,LSS}$, and $Q_{LSS}$. 

In Fig.~\ref{fig:QtsrM}, $Qt$ does not show any relation neither with $\sigma$ ($f_{Qt\sigma}$ = 0.074), $M_{vir}$ ($f_{QtM_{vir}}$ = -0.234), nor with $M_{vir}$/$L_r$ ($f_{QtM_{vir}/L_r}$ = -0.0247). We observed that the scattering is more pronounced at lower $Qt$ ($\le$-2) and compact systems are concentrated at higher $Qt$ ($>$0) except one system at $Qt=-2$. This means that tidal strength in compact systems is larger than that in wide systems; i.e. compact systems are more bounded than wide systems. On the other hand, a fair negative correlation between $Qt$ and $r_h$ is found with $f_{Q_{LSS}r_h}$ = -0.614 and can be expressed by the following equation

\begin{equation}
 r_h = -0.057(\pm0.004)\times Qt+ 0.088 .
  \label{eq:Qtrh}
\end{equation}

% \textbf{From Fig.~\ref{fig:dNNsrM} it is clearly noticed that as the distance to the first nearest neighbour decreases, concentration at low $M_{vir}$ and $M_{vir}$/$L_r$ increases. Similarly, more concentration are noticed at low $M_{vir}$ and $M_{vir}$/$L_r$ while correlated with the tidal strength of the LSS $Q_{LSS}$ as shown in Fig.~\ref{fig:QQsrM}}.

Discussions of Figs.~\ref{fig:dNNsrM} and ~\ref{fig:QQsrM} indicate that all galaxy triplets in the studied catalogue are perfectly isolated, and hence has no relation with the LSS. On the other hand, $Qt$ shows a relation with only $r_h$ (see Fig.~\ref{fig:QtsrM}) which means that as the gravitational radius between the members of a triplet system increases, the tidal strength between the triplet galaxies decreases, and vise-versa. Thus suggesting that the dynamical evolution in isolated triplets is mainly due to the galaxies within the same halo and it does not depend on the location of the system in the LSS.
This result is supported by the findings of \citet{Fernandez15} in which they found that isolated triplets have a common origin in their formation and evolution and most of them belong to the outer parts of filaments, walls, and clusters, and they are not located in voids.

% \begin{figure*}
% \centering
%  \subfloat{\includegraphics[scale=0.4, viewport=0 0 554 430, clip]{eps_figures/sigma_KLSS.eps}\label{subfig:Ks}}
%  \subfloat{\includegraphics[scale=0.4, viewport=0 0 554 430, clip]{eps_figures/rh_KLSS.eps}\label{subfig:Krh}}\\
%  \subfloat{\includegraphics[scale=0.4, viewport=0 0 554 430, clip]{eps_figures/Mvir_KLSS.eps}\label{subfig:KM}}
%  \subfloat{\includegraphics[scale=0.4, viewport=0 0 554 430, clip]{eps_figures/Mvirl_KLSS.eps}\label{subfig:KML}}
%  \newline
% \caption{Correlation between the projected density estimation of the LSS ($K_{LSS}$) and (a) velocity dispersion ($\sigma$), (b) mean harmonic separation ($r_h$), (c) virial mass ($M_{vir}$), and (d) mass-to-light ratio ($M_{vir}$/$L_r$) for the galaxy triplet systems}
% \label{fig:KsrM}
% \end{figure*}

%%%%%%%%%%%%%%%%%%%%%%%%%%%%%%%%%%%%%%%%%%%%%%%%%%%%%%%%%%%%%%%%%%%%

%%%%%%%%%%%%%%%%%%%%%%%%%%%%%%%%%%%%%%%%%%%%%%%%%%%%%%%%%%%%%%%%%%%%
\section{Conclusions}
\label{s:Conc}

We carried out a statistical study on 315 triplet systems chosen from the recently published catalogue of isolated galaxy triplets ``SDSS-based catalogue of Isolated Triplets'' (SIT), \citep{Fernandez15}. For these systems and their members,   we computed the physical (radial velocities ($v$), angular separations $(\theta)$, projected separations $(r_p)$, luminosity $(L_r)$, and absolute magnitudes $(M_r)$) as well as the dynamical (velocity dispersion $(\sigma)$, the mean harmonic separation ($r_h$), virial masses $(M_{vir}/M_ \odot)$, and mass-to-light ratios $(M_{vir}/L_r)$) parameters. We then distinguished between wide and compact triplet systems by estimating their mean harmonic separation $(r_h)$. In addition, we investigated the correlation between the estimated parameters and the triplets parameters given in the published catalogue. 
From the results of this study we may summaries our main conclusions as follows:

1. It is difficult to predict merging between members of triplet systems, in this study, as a step through their evolution.

2. The selected catalogue includes only $3.2\%$ compact systems and the rest are wide which might be the main reason that we found larger median values of physical and dynamical parameters than other previous studies.

3. Denser groups ($>$ 3 members) and clusters posses higher values of dynamical parameters.

4. SIT systems are more bounded (high tidal strength) at lower values of gravitational radii ($r_h$). 

5. Compact systems in SIT catalogue record the smallest mean values of dynamical parameters, comparing to K-triplet sample. This may imply that SIT compact systems are special and further investigation of such systems in other catalogues is required for future work.

6. SIT systems are perfectly isolated from their surrounding. This indicates that their dynamical evolution is independent of their location in the Universe, but depends on the evolution of the systems' own members. Therefore, we may confirm that isolated triplets have a common origin in their formation and evolution and most of them belong to the outer parts of filaments, walls, and clusters, and they are not located in voids similar to the findings by \citet{Fernandez15}.
  
7. Parameters of primary (G1) and satellite galaxies (G2 and G3) show that G1 is always brighter and more luminous than G2 and G3. This confirm the hierarchical structure of isolated triplets \citep{Toledo11, Fernandez15, Duplancic15}. 

The most important thesis beyond this work is to highlight the importance of statistical studies of galaxy triplets as a first essential step toward understanding the formation and evolution of groups and large scale structures.
More analysis of these poor populated systems such as decomposition and photometric studies are requested for better understanding of their properties. These further analyses are kept for our future work.

%%%%%%%%%%%%%%%%%%%%%%%%%%%%%%%%%%%%%%%%%%%%%%%%%%%%%%%%%%%%%%%%%%%%

%%%%%%%%%%%%%%%%%%%%%%%%%%%%%%%%%%%%%%%%%%%%%%%%%%%%%%%%%%%%%%%%%%%%
\section*{Acknowledgements}
This work was impossible to present without the effort made by our colleague Gamal Bakr Ali who passed away just before submitting this work. we dedicate this work for his soul.

Amira A. Tawfeek thanks Samuel Boissier \footnote{$(samuel.boissier@lam.fr)$ [Laboratoire d'Astrophysique de Marseille, France]} and Henri Plana \footnote{$(plana@uesc.br)$ [Laboratorio de Astrofisica Teorica e Observacional, Universidade Estadual de Santa Cruz - 45650-000, Ilheus-BA, Brasil]}, for their helpful comments and suggestions , during a scientific visit to the Laboratoire d'Astrophysique de Marseille (LAM) that improved this work. Amira A. Tawfeek also thanks Sonali Sachdeva \footnote{$(sonali@iucaa.in)$ [Inter-University Centre for Astronomy and Astrophysics (IUCAA), Pune, India]} for her great effort in improving the abstract in our manuscript, during her scientific visit to the Inter-University Centre for Astronomy and Astrophysics.

Funding for SDSS-III has been provided by the Alfred P. Sloan Foundation, the Participating Institutions, the National Science Foundation, and the U.S. Department of Energy Office of Science. The SDSS-III web site is http://www.sdss3.org/.

SDSS-III is managed by the Astrophysical Research Consortium for the Participating Institutions of the SDSS-III Collaboration including the University of Arizona, the Brazilian Participation Group, Brookhaven National Laboratory, Carnegie Mellon University, University of Florida, the French Participation Group, the German Participation Group, Harvard University, the Instituto de Astrofisica de Canarias, the Michigan State/Notre Dame/JINA Participation Group, Johns Hopkins University, Lawrence Berkeley National Laboratory, Max Planck Institute for Astrophysics, Max Planck Institute for Extraterrestrial Physics, New Mexico State University, New York University, Ohio State University, Pennsylvania State University, University of Portsmouth, Princeton University, the Spanish Participation Group, University of Tokyo, University of Utah, Vanderbilt University, University of Virginia, University of Washington, and Yale University.

%%%%%%%%%%%%%%%%%%%%%%%%%%%%%%%%%%%%%%%%%%%%%%%%%%

%%%%%%%%%%%%%%%%%%%% REFERENCES %%%%%%%%%%%%%%%%%%

% The best way to enter references is to use BibTeX:

\bibliographystyle{mnras}
\bibliography{refbib_amira} % if your bibtex file is called example.bib

\begin{thebibliography}{}
\makeatletter
\relax
\def\mn@urlcharsother{\let\do\@makeother \do\$\do\&\do\#\do\^\do\_\do\%\do\~}
\def\mn@doi{\begingroup\mn@urlcharsother \@ifnextchar [ {\mn@doi@}
  {\mn@doi@[]}}
\def\mn@doi@[#1]#2{\def\@tempa{#1}\ifx\@tempa\@empty \href
  {http://dx.doi.org/#2} {doi:#2}\else \href {http://dx.doi.org/#2} {#1}\fi
  \endgroup}
\def\mn@eprint#1#2{\mn@eprint@#1:#2::\@nil}
\def\mn@eprint@arXiv#1{\href {http://arxiv.org/abs/#1} {{\tt arXiv:#1}}}
\def\mn@eprint@dblp#1{\href {http://dblp.uni-trier.de/rec/bibtex/#1.xml}
  {dblp:#1}}
\def\mn@eprint@#1:#2:#3:#4\@nil{\def\@tempa {#1}\def\@tempb {#2}\def\@tempc
  {#3}\ifx \@tempc \@empty \let \@tempc \@tempb \let \@tempb \@tempa \fi \ifx
  \@tempb \@empty \def\@tempb {arXiv}\fi \@ifundefined
  {mn@eprint@\@tempb}{\@tempb:\@tempc}{\expandafter \expandafter \csname
  mn@eprint@\@tempb\endcsname \expandafter{\@tempc}}}

\bibitem[\protect\citeauthoryear{{Ahn} et~al.,}{{Ahn} et~al.}{2014}]{DR10}
{Ahn} C.~P.,  et~al., 2014, \mn@doi [\apjs] {10.1088/0067-0049/211/2/17}, \href
  {http://adsabs.harvard.edu/abs/2014ApJS..211...17A} {211, 17}

\bibitem[\protect\citeauthoryear{{Alam} et~al.,}{{Alam} et~al.}{2015a}]{DR12}
{Alam} S.,  et~al., 2015a, \mn@doi [\apjs] {10.1088/0067-0049/219/1/12}, \href
  {http://adsabs.harvard.edu/abs/2015ApJS..219...12A} {219, 12}

\bibitem[\protect\citeauthoryear{{Alam} et~al.,}{{Alam} et~al.}{2015b}]{Alam15}
{Alam} S.,  et~al., 2015b, \mn@doi [\apjs] {10.1088/0067-0049/219/1/12}, \href
  {http://adsabs.harvard.edu/abs/2015ApJS..219...12A} {219, 12}

\bibitem[\protect\citeauthoryear{{Ali}}{{Ali}}{2001}]{Ali01}
{Ali} G.,  2001, in On the Physical and Geometrical Properties of Paired and
  Interacting Galaxies, PhD thesis, Cairo University, Faculty of Science.

\bibitem[\protect\citeauthoryear{{Argudo-Fern{\'a}ndez}
  et~al.,}{{Argudo-Fern{\'a}ndez} et~al.}{2015}]{Fernandez15}
{Argudo-Fern{\'a}ndez} M.,  et~al., 2015, \mn@doi [\aap]
  {10.1051/0004-6361/201526016}, \href
  {http://adsabs.harvard.edu/abs/2015A%26A...578A.110A} {578, A110}

\bibitem[\protect\citeauthoryear{{Bahcall}}{{Bahcall}}{1996}]{Neta96}
{Bahcall} N.~A.,  1996, ArXiv astro-ph/9611148, \href
  {http://adsabs.harvard.edu/abs/1996astro.ph.11148B} {}

\bibitem[\protect\citeauthoryear{{Bilir}, {Karaali}  \& {Tun{\c c}el}}{{Bilir}
  et~al.}{2005}]{Bilir2005}
{Bilir} S.,  {Karaali} S.,   {Tun{\c c}el} S.,  2005, \mn@doi [Astronomische
  Nachrichten] {10.1002/asna.200510358}, \href
  {http://adsabs.harvard.edu/abs/2005AN....326..321B} {326, 321}

\bibitem[\protect\citeauthoryear{{Binney} \& {Tremaine}}{{Binney} \&
  {Tremaine}}{1987}]{Binney1987}
{Binney} J.,  {Tremaine} S.,  1987, \nat, \href
  {http://adsabs.harvard.edu/abs/1987Natur.326..219B} {326, 219}

\bibitem[\protect\citeauthoryear{{Blanton} et~al.,}{{Blanton}
  et~al.}{2003}]{Blanton2003}
{Blanton} M.~R.,  et~al., 2003, \mn@doi [\apj] {10.1086/375776}, \href
  {http://adsabs.harvard.edu/abs/2003ApJ...592..819B} {592, 819}

\bibitem[\protect\citeauthoryear{{Bohringer}}{{Bohringer}}{2007}]{Bohringer06}
{Bohringer} H.,  2007, {The Universe in X-Rays}.
 Vol. 23

\bibitem[\protect\citeauthoryear{{Carroll} \& {Ostlie}}{{Carroll} \&
  {Ostlie}}{2006}]{Carroll06}
{Carroll} B.~W.,  {Ostlie} D.~A.,  2006, {An introduction to modern
  astrophysics and cosmology}

\bibitem[\protect\citeauthoryear{{Chernin}, {Ivanov}, {Trofimof}  \&
  {Mikkola}}{{Chernin} et~al.}{1994}]{Chernin94}
{Chernin} A.~D.,  {Ivanov} A.~V.,  {Trofimof} A.~V.,   {Mikkola} S.,  1994,
  \aap, \href {http://adsabs.harvard.edu/abs/1994A%26A...281..685C} {281, 685}

\bibitem[\protect\citeauthoryear{{Chernin}, {Dolgachev}  \&
  {Domozhilova}}{{Chernin} et~al.}{2000}]{Chernin00}
{Chernin} A.~D.,  {Dolgachev} V.~P.,   {Domozhilova} L.~M.,  2000, \mn@doi
  [\mnras] {10.1046/j.1365-8711.2000.03909.x}, \href
  {http://adsabs.harvard.edu/abs/2000MNRAS.319..851C} {319, 851}

\bibitem[\protect\citeauthoryear{{Dekel} \& {Ostriker}}{{Dekel} \&
  {Ostriker}}{1999}]{Dekel99}
{Dekel} A.,  {Ostriker} J.~P.,  1999, in {Cambridge University Press} ed.,
  Formation of Structure in the Universe. p.~164

\bibitem[\protect\citeauthoryear{{Duplancic}, {O'Mill}, {Lambas}, {Sodr{\'e}}
  \& {Alonso}}{{Duplancic} et~al.}{2013}]{Duplancic13}
{Duplancic} F.,  {O'Mill} A.~L.,  {Lambas} D.~G.,  {Sodr{\'e}} L.,   {Alonso}
  S.,  2013, \mn@doi [\mnras] {10.1093/mnras/stt985}, \href
  {http://adsabs.harvard.edu/abs/2013MNRAS.433.3547D} {433, 3547}

\bibitem[\protect\citeauthoryear{{Duplancic}, {Alonso}, {Lambas}  \&
  {O'Mill}}{{Duplancic} et~al.}{2015}]{Duplancic15}
{Duplancic} F.,  {Alonso} S.,  {Lambas} D.~G.,   {O'Mill} A.~L.,  2015, \mn@doi
  [\mnras] {10.1093/mnras/stu2518}, \href
  {http://adsabs.harvard.edu/abs/2015MNRAS.447.1399D} {447, 1399}

\bibitem[\protect\citeauthoryear{{Feng}, {Shao}, {Shen},
  {Argudo-Fern{\'a}ndez}, {Wu}, {Lam}, {Yang}  \& {Yuan}}{{Feng}
  et~al.}{2016}]{Feng16}
{Feng} S.,  {Shao} Z.-Y.,  {Shen} S.-Y.,  {Argudo-Fern{\'a}ndez} M.,  {Wu} H.,
  {Lam} M.-I.,  {Yang} M.,   {Yuan} F.-T.,  2016, \mn@doi [Research in
  Astronomy and Astrophysics] {10.1088/1674-4527/16/5/072}, \href
  {http://adsabs.harvard.edu/abs/2016RAA....16...72F} {16, 72}

\bibitem[\protect\citeauthoryear{{Hern{\'a}ndez-Toledo},
  {M{\'e}ndez-Hern{\'a}ndez}, {Aceves}  \&
  {Olgu{\'{\i}}n}}{{Hern{\'a}ndez-Toledo} et~al.}{2011}]{Toledo11}
{Hern{\'a}ndez-Toledo} H.~M.,  {M{\'e}ndez-Hern{\'a}ndez} H.,  {Aceves} H.,
  {Olgu{\'{\i}}n} L.,  2011, \mn@doi [\aj] {10.1088/0004-6256/141/3/74}, \href
  {http://adsabs.harvard.edu/abs/2011AJ....141...74H} {141, 74}

\bibitem[\protect\citeauthoryear{{Hickson}}{{Hickson}}{1977a}]{Hickson77_1}
{Hickson} P.,  1977a, \mn@doi [\apj] {10.1086/155546}, \href
  {http://adsabs.harvard.edu/abs/1977ApJ...217...16H} {217, 16}

\bibitem[\protect\citeauthoryear{{Hickson}}{{Hickson}}{1977b}]{Hickson77_2}
{Hickson} P.,  1977b, \mn@doi [\apj] {10.1086/155643}, \href
  {http://adsabs.harvard.edu/abs/1977ApJ...217..964H} {217, 964}

\bibitem[\protect\citeauthoryear{{Karachentsev}}{{Karachentsev}}{1994}]{karachen94}
{Karachentsev} I.,  1994, Astronomical and Astrophysical Transactions, \href
  {http://adsabs.harvard.edu/abs/1994A%26AT....6....1K} {6, 1}

\bibitem[\protect\citeauthoryear{{Karachentsev} \&
  {Karachentseva}}{{Karachentsev} \& {Karachentseva}}{1981}]{Karachentsev81}
{Karachentsev} I.~D.,  {Karachentseva} V.~E.,  1981, Astrofizika, \href
  {http://adsabs.harvard.edu/abs/1981Afz....17....5K} {17, 5}

\bibitem[\protect\citeauthoryear{{Karachentsev}, {Karachentsev}  \&
  {Lebedev}}{{Karachentsev} et~al.}{1988}]{Karachentsev1988}
{Karachentsev} V.~E.,  {Karachentsev} I.~D.,   {Lebedev} V.~S.,  1988,
  Astrofizicheskie Issledovaniia Izvestiya Spetsial'noj Astrofizicheskoj
  Observatorii, \href {http://adsabs.harvard.edu/abs/1988AISAO..26...42K} {26,
  37}

\bibitem[\protect\citeauthoryear{{Karachentseva}, {Karachentsev}  \&
  {Shcherbanovsky}}{{Karachentseva} et~al.}{1979}]{Karachentseva1979}
{Karachentseva} V.~E.,  {Karachentsev} I.~D.,   {Shcherbanovsky} A.~L.,  1979,
  Astrofizicheskie Issledovaniia Izvestiya Spetsial'noj Astrofizicheskoj
  Observatorii, \href {http://adsabs.harvard.edu/abs/1979AISAO..11....3K} {11,
  3}

\bibitem[\protect\citeauthoryear{{Kiseleva}}{{Kiseleva}}{2000}]{Kiseleva2000}
{Kiseleva} L.,  2000, in {Valtonen} M.~J.,  {Flynn} C.,  eds,  Astronomical
  Society of the Pacific Conference Series Vol. 209, IAU Colloq. 174: Small
  Galaxy Groups. p.~388

\bibitem[\protect\citeauthoryear{{Liljeblad}}{{Liljeblad}}{2012}]{Liljeblad12}
{Liljeblad} E.,  2012, in {E.Liljeblad} ed., Estimating masses of galaxy
  clusters. pp~3--5

\bibitem[\protect\citeauthoryear{{Makarov} \& {Karachentsev}}{{Makarov} \&
  {Karachentsev}}{2000}]{MK2000}
{Makarov} D.~I.,  {Karachentsev} I.~D.,  2000, in {Valtonen} M.~J.,  {Flynn}
  C.,  eds,  Astronomical Society of the Pacific Conference Series Vol. 209,
  IAU Colloq. 174: Small Galaxy Groups. p.~40 (\mn@eprint {}
  {astro-ph/9909343})

\bibitem[\protect\citeauthoryear{{Merch{\'a}n} \& {Zandivarez}}{{Merch{\'a}n}
  \& {Zandivarez}}{2005}]{Merch05}
{Merch{\'a}n} M.~E.,  {Zandivarez} A.,  2005, \mn@doi [\apj] {10.1086/427989},
  \href {http://adsabs.harvard.edu/abs/2005ApJ...630..759M} {630, 759}

\bibitem[\protect\citeauthoryear{{O'Mill}, {Duplancic}, {Garc{\'{\i}}a Lambas},
  {Valotto}  \& {Sodr{\'e}}}{{O'Mill} et~al.}{2012}]{O'Mill12}
{O'Mill} A.~L.,  {Duplancic} F.,  {Garc{\'{\i}}a Lambas} D.,  {Valotto} C.,
  {Sodr{\'e}} L.,  2012, \mn@doi [\mnras] {10.1111/j.1365-2966.2012.20301.x},
  \href {http://adsabs.harvard.edu/abs/2012MNRAS.421.1897O} {421, 1897}

\bibitem[\protect\citeauthoryear{{Patton}, {Ellison}, {Simard}, {McConnachie}
  \& {Mendel}}{{Patton} et~al.}{2011}]{Patton11}
{Patton} D.~R.,  {Ellison} S.~L.,  {Simard} L.,  {McConnachie} A.~W.,
  {Mendel} J.~T.,  2011, \mn@doi [\mnras] {10.1111/j.1365-2966.2010.17932.x},
  \href {http://adsabs.harvard.edu/abs/2011MNRAS.412..591P} {412, 591}

\bibitem[\protect\citeauthoryear{{Rodgers}, {Canterna}, {Smith}, {Pierce}  \&
  {Tucker}}{{Rodgers} et~al.}{2006}]{Rodgers2006}
{Rodgers} C.~T.,  {Canterna} R.,  {Smith} J.~A.,  {Pierce} M.~J.,   {Tucker}
  D.~L.,  2006, \mn@doi [\aj] {10.1086/505864}, \href
  {http://adsabs.harvard.edu/abs/2006AJ....132..989R} {132, 989}

\bibitem[\protect\citeauthoryear{{Schneider}}{{Schneider}}{2006}]{Schneider06}
{Schneider} P.,  2006, in Extragalactic Astronomy and Cosmology.

\bibitem[\protect\citeauthoryear{{Shen} et~al.,}{{Shen} et~al.}{2016}]{Shen16}
{Shen} S.-Y.,  et~al., 2016, \mn@doi [Research in Astronomy and Astrophysics]
  {10.1088/1674-4527/16/3/043}, \href
  {http://adsabs.harvard.edu/abs/2016RAA....16...43S} {16, 43}

\bibitem[\protect\citeauthoryear{{Smart}}{{Smart}}{1965}]{Smart65}
{Smart} W.~M.,  1965, {Text-book on spherical astronomy}

\bibitem[\protect\citeauthoryear{{Trofimov} \& {Chernin}}{{Trofimov} \&
  {Chernin}}{1995}]{Trofimov95}
{Trofimov} A.~V.,  {Chernin} A.~D.,  1995, \azh, \href
  {http://adsabs.harvard.edu/abs/1995AZh....72..308T} {72, 308}

\bibitem[\protect\citeauthoryear{{Tully}}{{Tully}}{1987}]{Tully87}
{Tully} R.~B.,  1987, \mn@doi [\apj] {10.1086/165629}, \href
  {http://adsabs.harvard.edu/abs/1987ApJ...321..280T} {321, 280}

\bibitem[\protect\citeauthoryear{{Valtonen} \& {Mikkola}}{{Valtonen} \&
  {Mikkola}}{1991}]{Valtonen91}
{Valtonen} M.~J.,  {Mikkola} S.,  1991, in Bulletin of the American
  Astronomical Society. p.~1393

\bibitem[\protect\citeauthoryear{{Vavilova}, {Karachentseva}, {Makarov}  \&
  {Melnyk}}{{Vavilova} et~al.}{2005}]{Vavilova05}
{Vavilova} I.~B.,  {Karachentseva} V.~E.,  {Makarov} D.~I.,   {Melnyk} O.~V.,
  2005, Kinematika i Fizika Nebesnykh Tel, \href
  {http://adsabs.harvard.edu/abs/2005KFNT...21a...3V} {21, 3}

\makeatother
\end{thebibliography}

% Alternatively you could enter them by hand, like this:
% This method is tedious and prone to error if you have lots of references
% \begin{thebibliography}{99}
% \bibitem[\protect\citeauthoryear{Author}{2012}]{Author2012}
% Author A.~N., 2013, Journal of Improbable Astronomy, 1, 1
% \bibitem[\protect\citeauthoryear{Others}{2013}]{Others2013}
% Others S., 2012, Journal of Interesting Stuff, 17, 198
% \end{thebibliography}

%%%%%%%%%%%%%%%%%%%%%%%%%%%%%%%%%%%%%%%%%%%%%%%%%%

%%%%%%%%%%%%%%%%% APPENDICES %%%%%%%%%%%%%%%%%%%%%

\appendix

\section{}
\label{s:appendix}

\begin{figure*}
\centering
 \subfloat[\scriptsize SDSS J232500.25-000008.1]{\includegraphics[scale=0.35, 
viewport=0 0 417 417, 
clip]{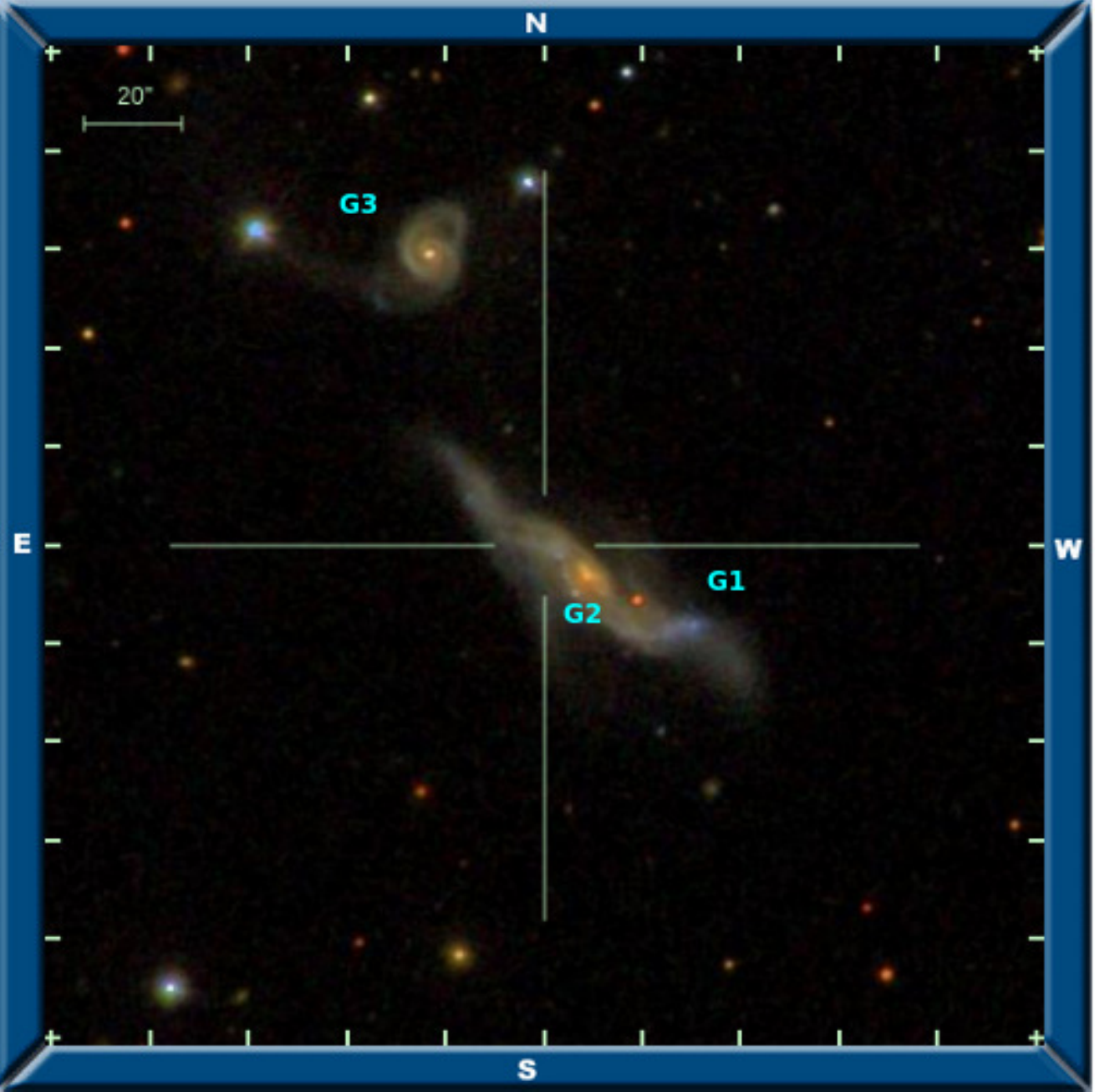}\label{subfig:TK1}}
 \subfloat[\scriptsize SDSS J132008.30+302700.3]{\includegraphics[scale=0.35, 
viewport=0 0 417 417, 
clip]{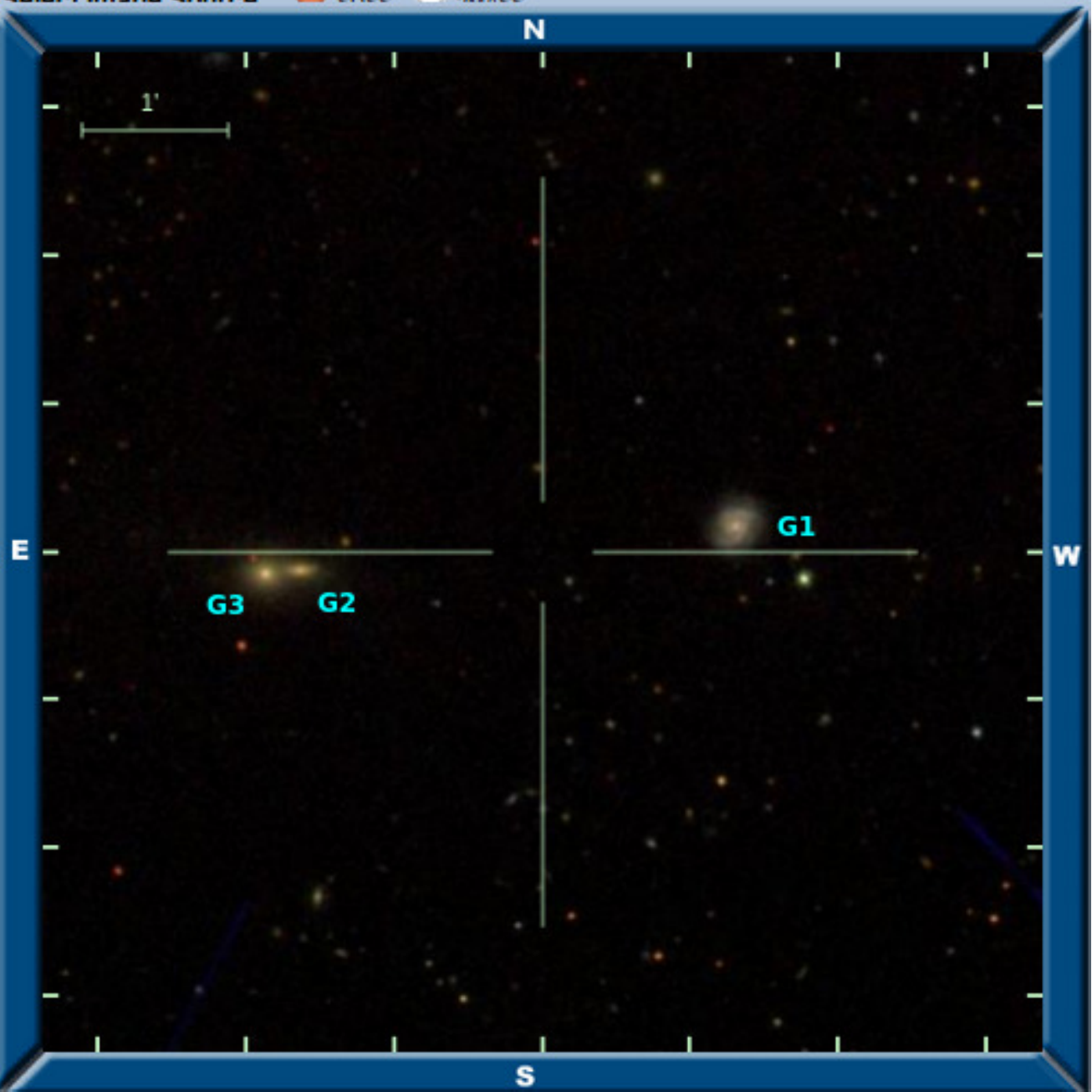}\label{subfig:TK2}}
 \subfloat[\scriptsize SDSS J073432.09+411254.9]{\includegraphics[scale=0.35, 
viewport=0 0 417 417, 
clip]{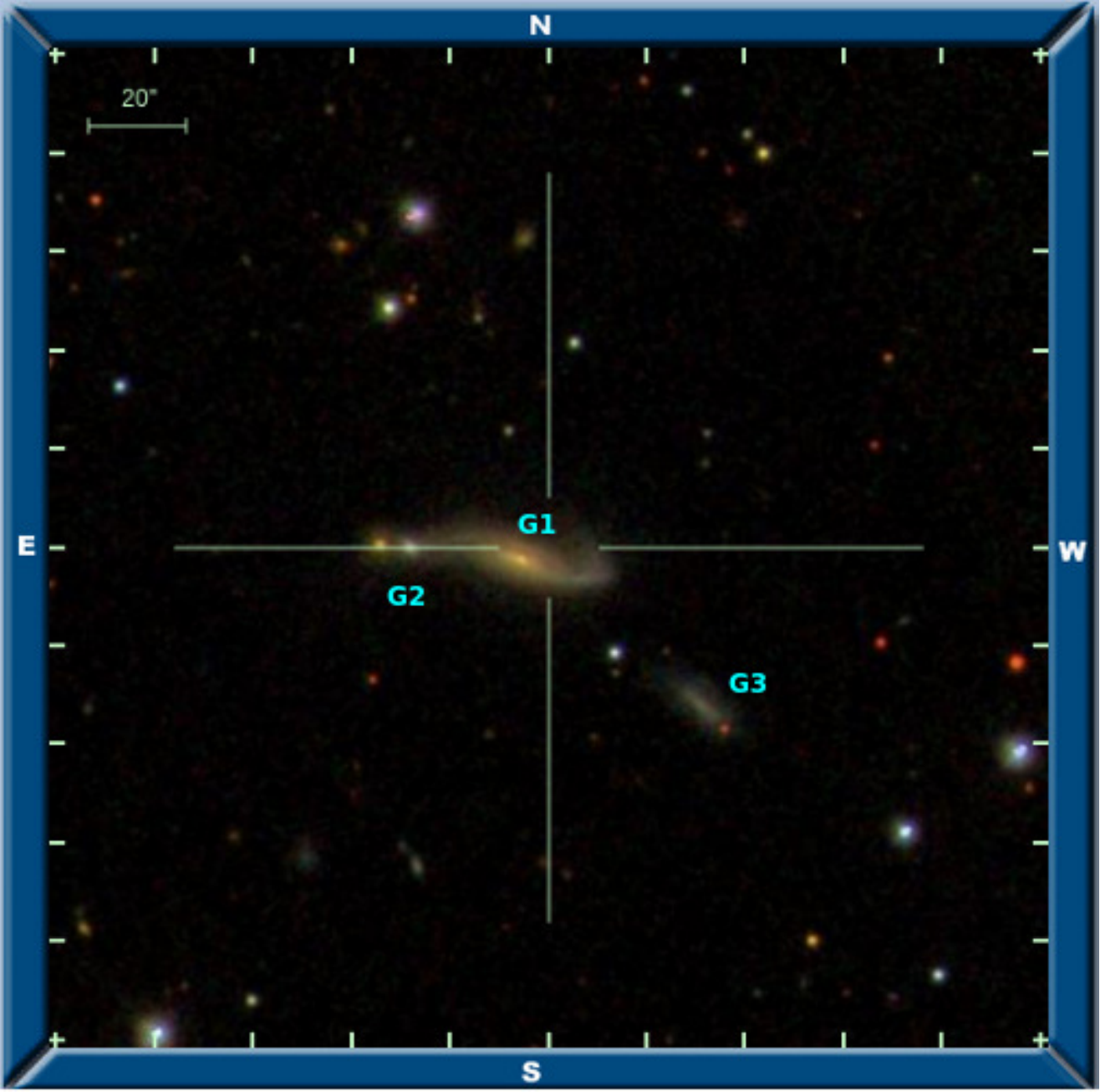}\label{subfig:TK3}}
 \newline
\caption{SDSS color images of three compact triplet systems from the selected catalogue of isolated galaxy triplet}
\label{fig:TK}
\end{figure*}

\begin{figure*}
\centering
 \subfloat[\scriptsize SDSS J112515.70-020434.0]{\includegraphics[scale=0.35, 
viewport=0 0 417 417, 
clip]{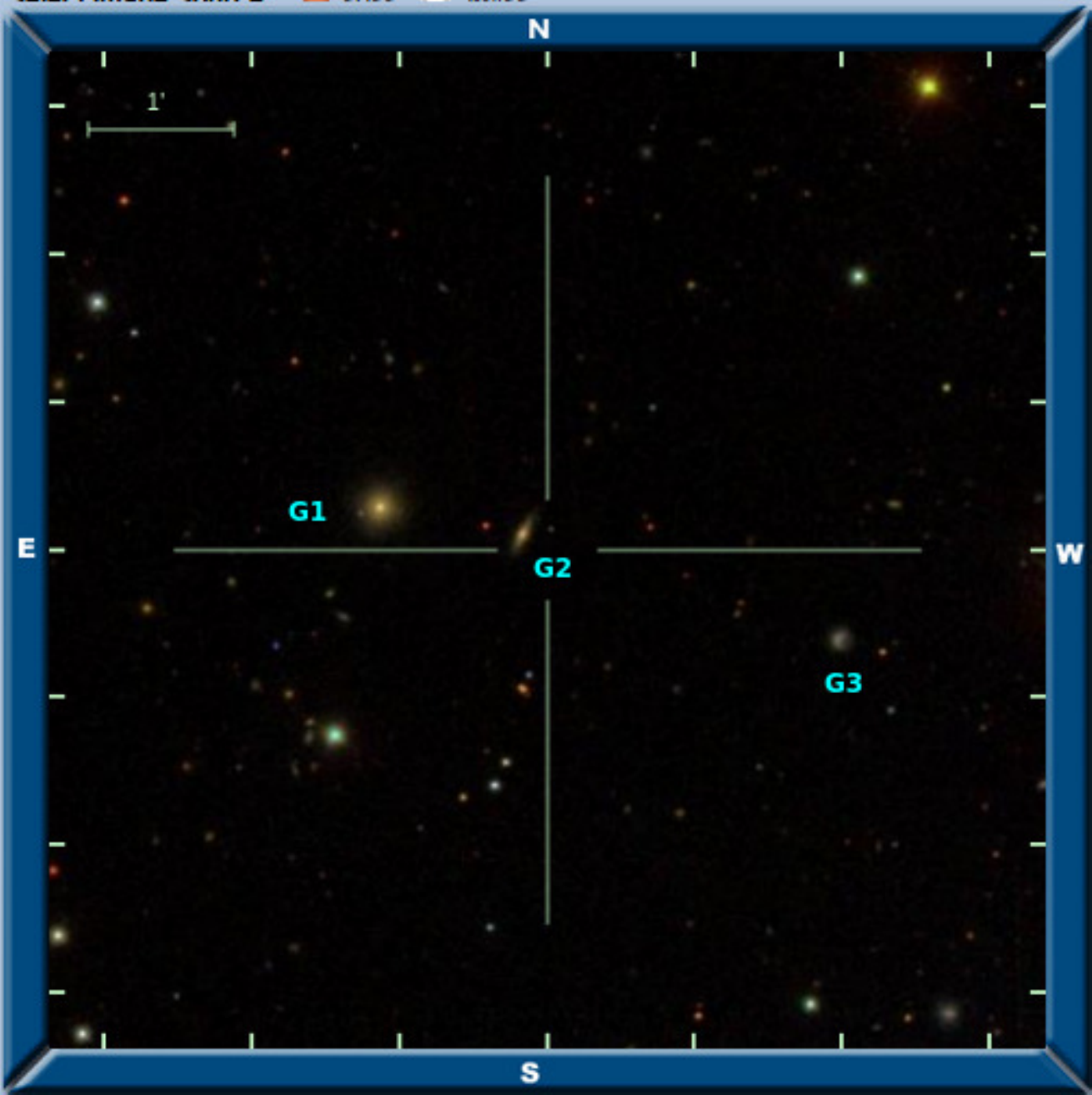}\label{subfig:TW1}}
 \subfloat[\scriptsize SDSS J135442.49+651439.5]{\includegraphics[scale=0.35, 
viewport=0 0 417 417, 
clip]{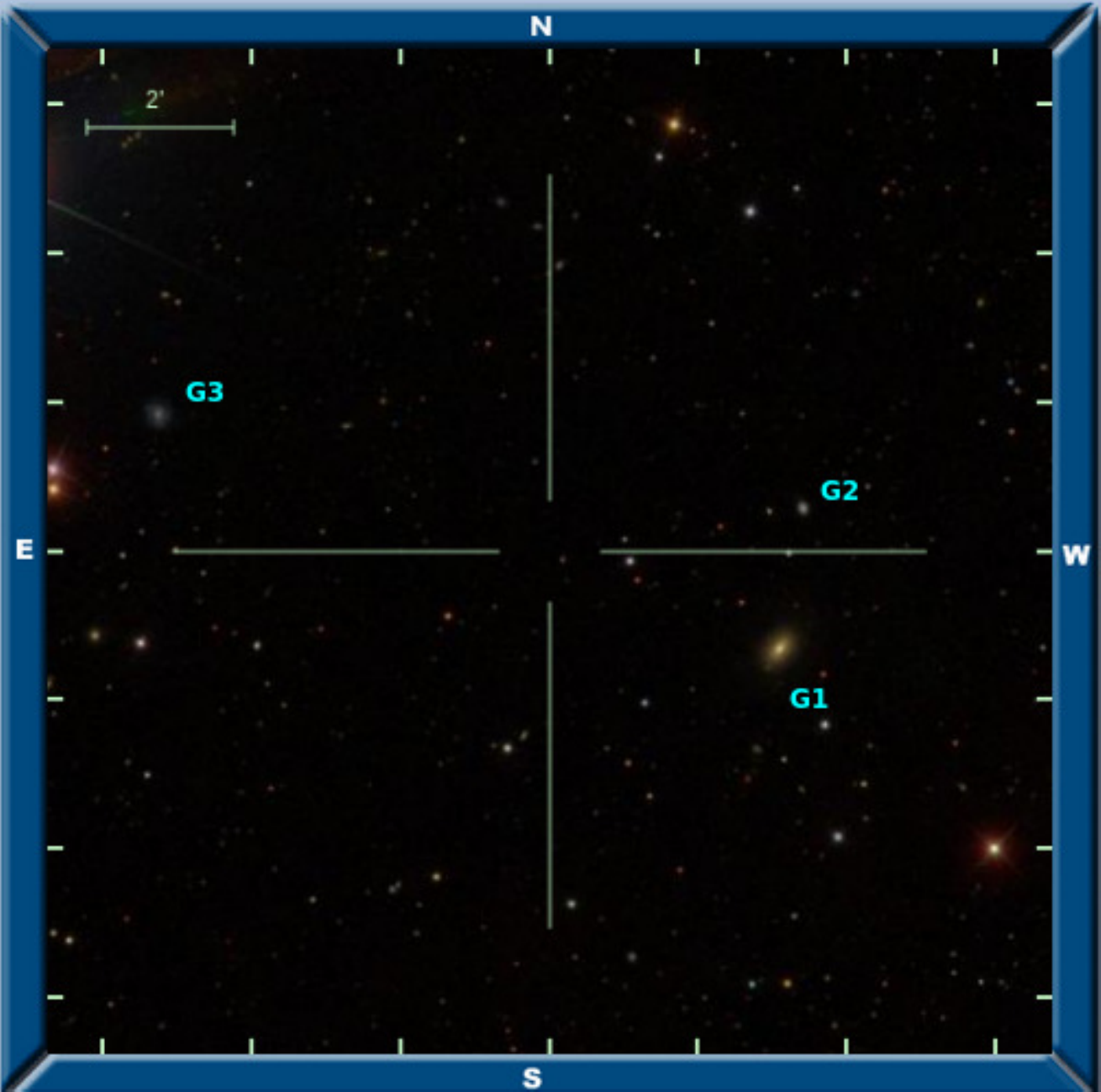}\label{subfig:TW2}}
 \subfloat[\scriptsize SDSS J012853.77+144314.5]{\includegraphics[scale=0.35, 
viewport=0 0 417 417, 
clip]{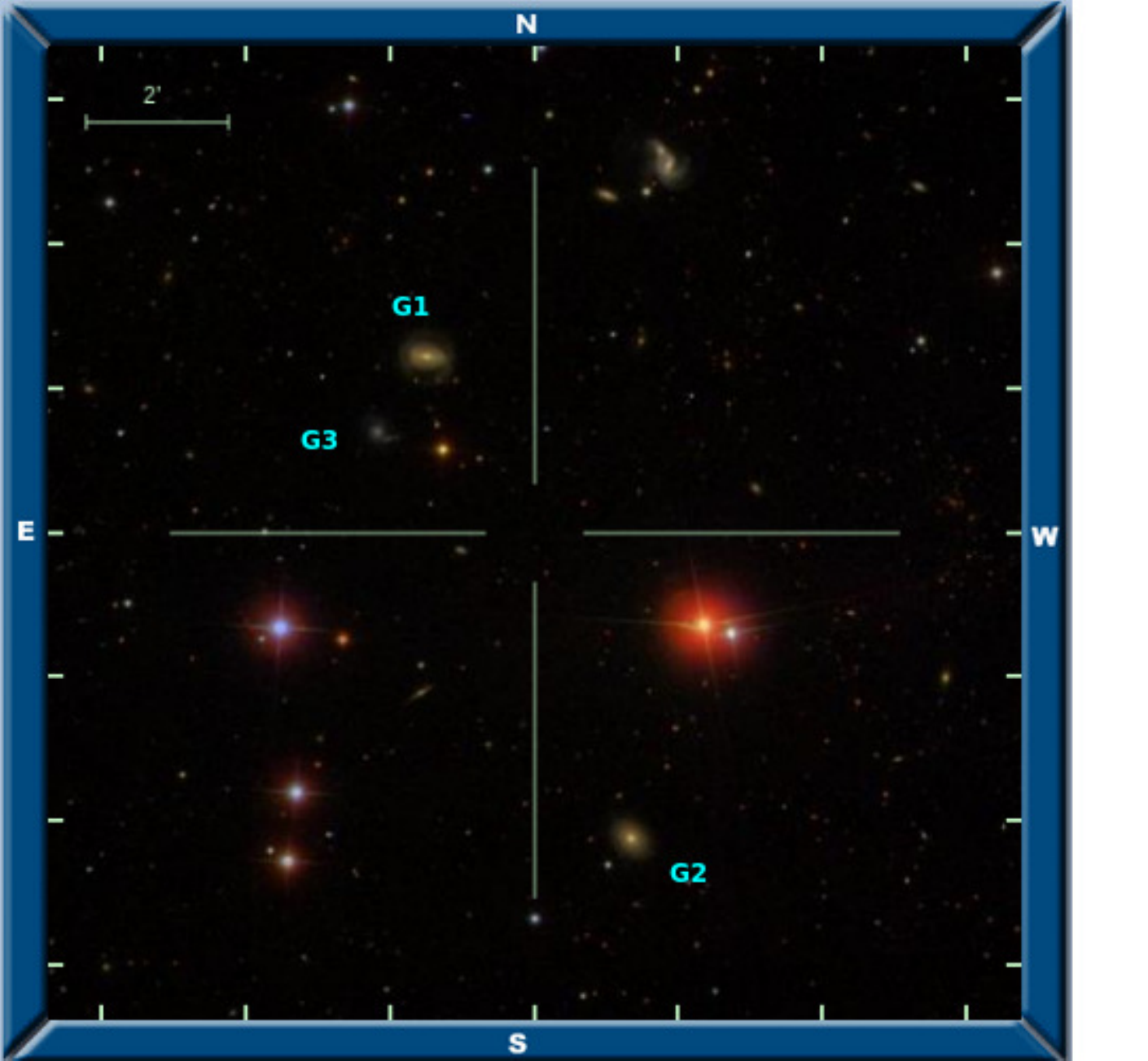}\label{subfig:TW3}}
 \newline
\caption{SDSS color images of three wide triplet systems from the selected catalogue of galaxy triplet sample}
\label{fig:TW}
\end{figure*}

In the following, we show three images of compact triplet systems, in Fig.~\ref{fig:TK}, and three images of wide triplet systems (Fig.~\ref{fig:TW}) from the selected catalogue. In addition, Table ~\ref{tbl:TG} lists the main computed parameters for the 315 studied galaxy triplet systems \footnote{k represents compact systems and w represents wide systems}. 
The columns of the Table are: Index for the systems, position (RA, DEC) of the brightest galaxy in the system, redshift (z), velocity dispersion ($\sigma$), virial mass ($M_{vir}$), mass-to-light ratio ($M_{vir}/L_r$), mean projected separation ($r_p$), mean harmonic separation ($r_h$), and finally a flag indicating the status of compact system (k) and wide (w) triplet.

% \newcounter{Table}
\begin{table*}
 \begin{center}
    \tabcolsep 5.8pt
    \scriptsize
    \caption{Computed basic parameters of the 315 studied triplet systems in the SIT catalogue \label{tbl:TG}. The flag in the last column is (k) for compact galaxy triplets and (w) for wide systems}. 
\begin{tabular}{|r|l|l|l|l|l|l|l|l|l|l|}
\hline
  \multicolumn{1}{|c|}{Index} &
  \multicolumn{1}{c|}{RA} &
  \multicolumn{1}{c|}{DEC} &
  \multicolumn{1}{c|}{z} &
  \multicolumn{1}{c|}{$\sigma$} &
  \multicolumn{1}{c|}{$M_{vir}$} &
  \multicolumn{1}{c|}{$M_{vir}/L_r$} &
  \multicolumn{1}{c|}{$\overline r_p$} & 
  \multicolumn{1}{c|}{$r_h$} &
  \multicolumn{1}{c|}{flag} \\
  \multicolumn{1}{|c|}{} &
  \multicolumn{1}{c|}{} &
  \multicolumn{1}{c|}{} &
  \multicolumn{1}{c|}{} &
  \multicolumn{1}{c|}{$km~s^{-1}$} &
  \multicolumn{1}{c|}{$10^{12}M_{\odot}$} &
  \multicolumn{1}{c|}{$M_{\odot}/L_\odot$} &
  \multicolumn{1}{c|}{Mpc} &
  \multicolumn{1}{c|}{Mpc} &
  \multicolumn{1}{c|}{} \\
\hline
  1 & 171.315 & -2.076 & 0.063 & 16.66 & 0.03 & 6.19 & 0.44 & 0.38 & w\\
  2 & 208.677 & 65.244 & 0.036 & 63.5 & 0.24 & 61.87 & 0.29 & 0.18 & w\\
  3 & 211.203 & 4.981 & 0.046 & 66.47 & 0.17 & 60.9 & 0.21 & 0.12 & w\\
  4 & 251.468 & 44.437 & 0.07 & 45.82 & 0.24 & 30.57 & 0.35 & 0.34 & w\\
  5 & 133.535 & 0.499 & 0.028 & 20.04 & 0.03 & 13.65 & 0.29 & 0.2 & w\\
  6 & 207.824 & 0.385 & 0.03 & 29.0 & 0.02 & 5.86 & 0.14 & 0.05 & w\\
  7 & 218.956 & 62.421 & 0.048 & 61.6 & 0.42 & 43.3 & 0.36 & 0.34 & w\\
  8 & 197.686 & 0.032 & 0.041 & 33.77 & 0.14 & 52.57 & 0.44 & 0.37 & w\\
  9 & 320.876 & -7.746 & 0.028 & 49.5 & 0.09 & 26.36 & 0.13 & 0.11 & w\\
  10 & 22.224 & 14.721 & 0.037 & 83.66 & 0.29 & 42.87 & 0.23 & 0.13 & w\\
  11 & 127.46 & 48.781 & 0.024 & 14.75 & 0.01 & 4.04 & 0.13 & 0.12 & w\\
  12 & 200.92 & 1.362 & 0.057 & 49.22 & 0.1 & 11.19 & 0.14 & 0.13 & w\\
  13 & 232.209 & 0.777 & 0.04 & 9.89 & 0.01 & 1.68 & 0.33 & 0.3 & w\\
  14 & 225.273 & 0.53 & 0.036 & 59.98 & 0.14 & 76.37 & 0.22 & 0.12 & w\\
  15 & 259.779 & 65.732 & 0.036 & 25.94 & 0.04 & 29.49 & 0.23 & 0.2 & w\\
  16 & 192.983 & -2.078 & 0.023 & 18.53 & 0.02 & 8.58 & 0.25 & 0.14 & w\\
  17 & 155.232 & 1.236 & 0.071 & 97.43 & 0.22 & 36.0 & 0.1 & 0.07 & w\\
  18 & 190.146 & 2.468 & 0.046 & 47.74 & 0.09 & 16.36 & 0.16 & 0.12 & w\\
  19 & 211.262 & -0.251 & 0.033 & 56.9 & 0.04 & 29.31 & 0.09 & 0.04 & k\\
  20 & 189.952 & -2.796 & 0.048 & 31.19 & 0.1 & 17.13 & 0.3 & 0.3 & w\\
  21 & 231.978 & 54.27 & 0.04 & 19.79 & 0.02 & 4.67 & 0.26 & 0.16 & w\\
  22 & 134.774 & 53.772 & 0.008 & 22.61 & 0.04 & 47.78 & 0.23 & 0.22 & w\\
  23 & 145.6 & 63.165 & 0.031 & 50.01 & 0.21 & 74.37 & 0.29 & 0.26 & w\\
  24 & 119.168 & 44.87 & 0.05 & 55.08 & 0.13 & 12.81 & 0.2 & 0.13 & w\\
  25 & 134.242 & 52.066 & 0.013 & 56.46 & 0.16 & 76.55 & 0.21 & 0.16 & w\\
  26 & 136.146 & 52.12 & 0.066 & 76.72 & 0.63 & 58.63 & 0.33 & 0.32 & w\\
  27 & 143.99 & 54.993 & 0.059 & 86.07 & 1.07 & 233.0 & 0.47 & 0.44 & w\\
  28 & 122.384 & 39.888 & 0.064 & 57.56 & 0.32 & 42.68 & 0.33 & 0.29 & w\\
  29 & 162.281 & 4.941 & 0.055 & 77.77 & 0.31 & 76.78 & 0.27 & 0.16 & w\\
  30 & 348.122 & 13.942 & 0.034 & 31.82 & 0.02 & 4.39 & 0.08 & 0.07 & w\\
  31 & 143.019 & 51.551 & 0.034 & 36.74 & 0.18 & 48.82 & 0.42 & 0.41 & w\\
  32 & 138.259 & 49.639 & 0.013 & 20.82 & 0.02 & 16.4 & 0.2 & 0.17 & w\\
  33 & 148.124 & 2.154 & 0.017 & 26.96 & 0.02 & 7.43 & 0.11 & 0.08 & w\\
  34 & 310.094 & -6.359 & 0.036 & 29.78 & 0.09 & 26.14 & 0.4 & 0.32 & w\\
  35 & 188.066 & 3.0 & 0.044 & 57.53 & 0.11 & 26.86 & 0.21 & 0.1 & w\\
  36 & 25.882 & 13.514 & 0.054 & 48.89 & 0.12 & 19.69 & 0.17 & 0.15 & w\\
  37 & 136.811 & 2.443 & 0.051 & 61.26 & 0.09 & 20.41 & 0.29 & 0.07 & w\\
  38 & 206.814 & 62.47 & 0.044 & 27.26 & 0.02 & 5.67 & 0.11 & 0.1 & w\\
  39 & 323.223 & -7.197 & 0.061 & 101.35 & 0.55 & 57.14 & 0.22 & 0.16 & w\\
  40 & 186.865 & 3.302 & 0.049 & 24.52 & 0.04 & 3.61 & 0.2 & 0.19 & w\\
  41 & 176.138 & 61.534 & 0.048 & 92.34 & 0.33 & 51.3 & 0.16 & 0.12 & w\\
  42 & 174.362 & 1.535 & 0.046 & 54.53 & 0.15 & 29.93 & 0.18 & 0.15 & w\\
  43 & 157.973 & 62.401 & 0.052 & 63.89 & 0.11 & 18.02 & 0.16 & 0.08 & w\\
  44 & 126.862 & 41.495 & 0.036 & 12.95 & 0.01 & 2.65 & 0.18 & 0.15 & w\\
  45 & 351.251 & -0.002 & 0.034 & 51.94 & 0.03 & 16.8 & 0.04 & 0.03 & k\\
  46 & 208.032 & 5.432 & 0.079 & 26.85 & 0.12 & 10.93 & 0.56 & 0.51 & w\\
  47 & 233.29 & -1.866 & 0.028 & 54.75 & 0.28 & 109.9 & 0.29 & 0.28 & w\\
  48 & 229.519 & 3.49 & 0.068 & 115.36 & 0.7 & 112.37 & 0.21 & 0.16 & w\\
  49 & 202.904 & 61.628 & 0.044 & 31.19 & 0.11 & 27.28 & 0.42 & 0.33 & w\\
  50 & 215.19 & 58.699 & 0.046 & 28.86 & 0.04 & 11.47 & 0.15 & 0.14 & w\\
  51 & 132.899 & 47.558 & 0.029 & 49.46 & 0.1 & 13.5 & 0.13 & 0.12 & w\\
  52 & 4.924 & 0.376 & 0.041 & 29.2 & 0.02 & 9.82 & 0.16 & 0.07 & w\\
  53 & 210.055 & 3.424 & 0.035 & 60.67 & 0.09 & 21.56 & 0.28 & 0.07 & w\\
  54 & 183.633 & 4.435 & 0.077 & 32.6 & 0.04 & 2.41 & 0.23 & 0.12 & w\\
  55 & 120.005 & 28.742 & 0.041 & 65.42 & 0.2 & 80.12 & 0.18 & 0.14 & w\\
  56 & 168.342 & 61.447 & 0.068 & 54.13 & 0.23 & 32.38 & 0.31 & 0.24 & w\\
  57 & 172.764 & 62.305 & 0.033 & 45.33 & 0.06 & 19.2 & 0.09 & 0.08 & w\\
  58 & 184.541 & 62.802 & 0.05 & 58.97 & 0.25 & 29.48 & 0.29 & 0.21 & w\\
  59 & 181.213 & 63.132 & 0.04 & 76.72 & 0.32 & 36.44 & 0.18 & 0.17 & w\\
  60 & 123.241 & 36.255 & 0.008 & 11.22 & 0.0 & 3.36 & 0.17 & 0.08 & w\\
  61 & 123.449 & 37.05 & 0.031 & 38.24 & 0.11 & 31.84 & 0.28 & 0.23 & w\\
  62 & 171.927 & 55.921 & 0.019 & 25.94 & 0.02 & 28.11 & 0.07 & 0.07 & w\\
  63 & 249.554 & 40.173 & 0.037 & 49.04 & 0.2 & 55.19 & 0.33 & 0.25 & w\\
  64 & 180.241 & 59.385 & 0.033 & 113.93 & 0.5 & 251.83 & 0.15 & 0.12 & w\\
  65 & 181.569 & 61.154 & 0.052 & 9.79 & 0.01 & 1.82 & 0.36 & 0.29 & w\\
  66 & 166.865 & 53.988 & 0.06 & 34.99 & 0.03 & 6.26 & 0.21 & 0.07 & w\\
  67 & 208.745 & 46.945 & 0.028 & 36.74 & 0.06 & 25.53 & 0.17 & 0.14 & w\\
  68 & 354.18 & 13.674 & 0.062 & 49.95 & 0.22 & 33.63 & 0.28 & 0.27 & w\\
  69 & 231.268 & 4.83 & 0.022 & 50.01 & 0.09 & 45.21 & 0.13 & 0.11 & w\\
  70 & 347.753 & 13.632 & 0.023 & 42.47 & 0.04 & 18.22 & 0.15 & 0.07 & w\\
  71 & 139.105 & 42.992 & 0.009 & 34.99 & 0.01 & 90.03 & 0.09 & 0.02 & k\\
  72 & 162.697 & 6.099 & 0.042 & 35.3 & 0.16 & 24.4 & 0.42 & 0.38 & w\\
  73 & 165.592 & 6.077 & 0.022 & 53.87 & 0.23 & 148.5 & 0.24 & 0.24 & w\\
  74 & 167.143 & 8.737 & 0.072 & 66.24 & 0.22 & 37.93 & 0.18 & 0.16 & w\\
  75 & 156.429 & 45.348 & 0.044 & 29.61 & 0.0 & 2.09 & 0.09 & 0.02 & k\\
  76 & 166.154 & 6.395 & 0.032 & 21.34 & 0.03 & 19.05 & 0.22 & 0.2 & w\\
  77 & 170.689 & 7.52 & 0.042 & 49.46 & 0.13 & 34.62 & 0.17 & 0.16 & w\\
\hline\end{tabular}
\end{center}
\end{table*}

\addtocounter{table}{-1}
\begin{table*}
  \begin{center}
    \tabcolsep 5.8pt
    \scriptsize
    \caption{- continued}
\begin{tabular}{|r|l|l|l|l|l|l|l|l|l|l|}
\hline
  \multicolumn{1}{|c|}{Index} &
  \multicolumn{1}{c|}{RA} &
  \multicolumn{1}{c|}{DEC} &
  \multicolumn{1}{c|}{z} &
  \multicolumn{1}{c|}{$\sigma$} &
  \multicolumn{1}{c|}{$M_{vir}$} &
  \multicolumn{1}{c|}{$M_{vir}/L_r$} &
  \multicolumn{1}{c|}{$\overline r_p$} &
  \multicolumn{1}{c|}{$r_h$} &
  \multicolumn{1}{c|}{flag} \\
  \multicolumn{1}{|c|}{} &
  \multicolumn{1}{c|}{} &
  \multicolumn{1}{c|}{} &
  \multicolumn{1}{c|}{} &
  \multicolumn{1}{c|}{$km~s^{-1}$} &
  \multicolumn{1}{c|}{$10^{12}M_{\odot}$} &
  \multicolumn{1}{c|}{$M_{\odot}/L_\odot$} &
  \multicolumn{1}{c|}{Mpc} &
  \multicolumn{1}{c|}{Mpc} &
  \multicolumn{1}{c|}{} \\
\hline
  78 & 168.526 & 55.711 & 0.028 & 36.74 & 0.08 & 24.75 & 0.24 & 0.18 & w\\
  79 & 199.903 & -2.912 & 0.023 & 49.46 & 0.23 & 99.59 & 0.29 & 0.28 & w\\
  80 & 199.148 & -2.091 & 0.019 & 64.76 & 0.24 & 55.51 & 0.22 & 0.17 & w\\
  81 & 183.697 & 10.918 & 0.079 & 59.81 & 0.28 & 20.45 & 0.28 & 0.23 & w\\
  82 & 227.65 & 54.387 & 0.038 & 45.33 & 0.17 & 31.73 & 0.28 & 0.25 & w\\
  83 & 132.899 & 39.59 & 0.041 & 21.94 & 0.03 & 12.23 & 0.33 & 0.21 & w\\
  84 & 197.953 & 10.602 & 0.031 & 72.12 & 0.32 & 99.54 & 0.19 & 0.18 & w\\
  85 & 148.019 & 6.209 & 0.062 & 41.64 & 0.21 & 17.88 & 0.4 & 0.36 & w\\
  86 & 151.284 & 44.514 & 0.026 & 21.76 & 0.01 & 9.94 & 0.09 & 0.08 & w\\
  87 & 149.828 & 52.257 & 0.04 & 41.42 & 0.11 & 20.9 & 0.23 & 0.19 & w\\
  88 & 129.277 & 41.456 & 0.029 & 32.87 & 0.08 & 25.36 & 0.26 & 0.23 & w\\
  89 & 211.807 & 45.89 & 0.04 & 37.15 & 0.05 & 14.83 & 0.18 & 0.1 & w\\
  90 & 176.441 & 47.175 & 0.054 & 56.93 & 0.27 & 42.66 & 0.26 & 0.26 & w\\
  91 & 230.778 & 37.67 & 0.065 & 99.56 & 1.12 & 182.24 & 0.37 & 0.35 & w\\
  92 & 155.243 & 6.642 & 0.033 & 72.24 & 0.66 & 241.78 & 0.42 & 0.38 & w\\
  93 & 236.625 & 32.117 & 0.032 & 35.67 & 0.06 & 13.17 & 0.19 & 0.14 & w\\
  94 & 178.758 & 49.9 & 0.059 & 47.23 & 0.2 & 18.99 & 0.3 & 0.27 & w\\
  95 & 233.473 & 45.118 & 0.072 & 57.42 & 0.3 & 46.64 & 0.31 & 0.28 & w\\
  96 & 217.825 & 7.945 & 0.027 & 22.21 & 0.06 & 16.58 & 0.44 & 0.39 & w\\
  97 & 214.386 & 8.856 & 0.037 & 31.19 & 0.13 & 58.02 & 0.42 & 0.4 & w\\
  98 & 135.308 & 6.022 & 0.076 & 69.29 & 0.56 & 60.34 & 0.36 & 0.36 & w\\
  99 & 195.788 & 54.617 & 0.028 & 94.26 & 0.65 & 567.47 & 0.25 & 0.22 & w\\
  100 & 156.661 & 37.775 & 0.051 & 31.25 & 0.04 & 6.42 & 0.14 & 0.13 & w\\
  101 & 206.035 & 45.341 & 0.038 & 33.41 & 0.04 & 13.39 & 0.11 & 0.11 & w\\
  102 & 249.643 & 31.873 & 0.064 & 46.85 & 0.17 & 26.79 & 0.28 & 0.23 & w\\
  103 & 124.187 & 27.592 & 0.04 & 40.3 & 0.07 & 12.53 & 0.18 & 0.14 & w\\
  104 & 128.212 & 30.879 & 0.066 & 55.44 & 0.09 & 14.32 & 0.09 & 0.09 & w\\
  105 & 159.331 & 43.588 & 0.025 & 48.3 & 0.17 & 24.45 & 0.28 & 0.23 & w\\
  106 & 216.903 & 41.258 & 0.009 & 23.35 & 0.01 & 19.28 & 0.23 & 0.07 & w\\
  107 & 212.225 & 10.199 & 0.067 & 122.42 & 1.17 & 192.69 & 0.28 & 0.24 & w\\
  108 & 211.273 & 11.475 & 0.067 & 65.05 & 0.46 & 64.63 & 0.35 & 0.33 & w\\
  109 & 212.137 & 11.816 & 0.019 & 49.52 & 0.13 & 58.11 & 0.18 & 0.16 & w\\
  110 & 134.207 & 35.917 & 0.076 & 41.71 & 0.27 & 30.19 & 0.51 & 0.47 & w\\
  111 & 221.955 & 47.81 & 0.037 & 30.8 & 0.05 & 22.88 & 0.19 & 0.18 & w\\
  112 & 232.647 & 33.632 & 0.065 & 38.73 & 0.1 & 14.78 & 0.24 & 0.21 & w\\
  113 & 191.326 & 42.41 & 0.053 & 43.81 & 0.2 & 19.85 & 0.33 & 0.32 & w\\
  114 & 145.814 & 35.161 & 0.05 & 70.48 & 0.48 & 63.55 & 0.37 & 0.3 & w\\
  115 & 147.83 & 36.178 & 0.052 & 43.81 & 0.1 & 12.75 & 0.18 & 0.15 & w\\
  116 & 172.806 & 40.653 & 0.045 & 42.14 & 0.08 & 17.39 & 0.26 & 0.14 & w\\
  117 & 174.931 & 42.505 & 0.07 & 21.34 & 0.06 & 6.43 & 0.42 & 0.38 & w\\
  118 & 172.828 & 42.813 & 0.045 & 15.74 & 0.01 & 4.16 & 0.21 & 0.14 & w\\
  119 & 215.628 & 11.305 & 0.016 & 73.65 & 0.41 & 308.29 & 0.25 & 0.23 & w\\
  120 & 238.347 & 26.105 & 0.069 & 55.67 & 0.14 & 18.51 & 0.28 & 0.13 & w\\
  121 & 157.067 & 38.935 & 0.055 & 48.22 & 0.24 & 53.78 & 0.33 & 0.31 & w\\
  122 & 186.372 & 47.273 & 0.025 & 43.81 & 0.13 & 31.38 & 0.24 & 0.21 & w\\
  123 & 202.365 & 55.721 & 0.043 & 35.3 & 0.07 & 19.55 & 0.2 & 0.18 & w\\
  124 & 164.403 & 44.071 & 0.034 & 21.94 & 0.02 & 4.68 & 0.17 & 0.11 & w\\
  125 & 209.305 & 12.021 & 0.021 & 47.04 & 0.04 & 6.14 & 0.05 & 0.05 & w\\
  126 & 211.096 & 11.736 & 0.02 & 50.18 & 0.16 & 102.7 & 0.29 & 0.19 & w\\
  127 & 187.741 & 10.803 & 0.049 & 30.57 & 0.04 & 12.19 & 0.16 & 0.13 & w\\
  128 & 197.651 & 39.784 & 0.047 & 31.85 & 0.06 & 9.21 & 0.29 & 0.19 & w\\
  129 & 229.63 & 32.393 & 0.061 & 26.93 & 0.03 & 4.46 & 0.18 & 0.15 & w\\
  130 & 190.309 & 12.989 & 0.047 & 62.45 & 0.11 & 29.38 & 0.22 & 0.08 & w\\
  131 & 157.436 & 40.494 & 0.044 & 24.76 & 0.03 & 4.18 & 0.16 & 0.13 & w\\
  132 & 210.019 & 58.37 & 0.026 & 66.6 & 0.09 & 65.88 & 0.13 & 0.06 & w\\
  133 & 247.486 & 39.944 & 0.059 & 62.61 & 0.06 & 5.69 & 0.1 & 0.04 & w\\
  134 & 150.766 & 12.95 & 0.036 & 58.76 & 0.24 & 62.72 & 0.21 & 0.21 & w\\
  135 & 148.655 & 6.622 & 0.041 & 72.98 & 0.65 & 113.74 & 0.4 & 0.37 & w\\
  136 & 205.26 & 13.163 & 0.056 & 53.01 & 0.15 & 41.25 & 0.18 & 0.17 & w\\
  137 & 202.655 & 14.649 & 0.067 & 40.89 & 0.15 & 21.57 & 0.31 & 0.27 & w\\
  138 & 204.54 & 14.92 & 0.059 & 58.81 & 0.29 & 41.0 & 0.27 & 0.26 & w\\
  139 & 208.419 & 56.134 & 0.035 & 34.3 & 0.07 & 46.78 & 0.24 & 0.19 & w\\
  140 & 154.217 & 39.556 & 0.064 & 26.17 & 0.02 & 2.75 & 0.11 & 0.1 & w\\
  141 & 225.507 & 9.408 & 0.046 & 59.36 & 0.25 & 81.22 & 0.24 & 0.21 & w\\
  142 & 187.588 & 7.488 & 0.067 & 25.94 & 0.01 & 1.26 & 0.06 & 0.05 & w\\
  143 & 163.764 & 42.86 & 0.059 & 18.53 & 0.0 & 0.84 & 0.28 & 0.04 & k\\
  144 & 214.977 & 51.895 & 0.029 & 31.95 & 0.03 & 8.43 & 0.13 & 0.09 & w\\
  145 & 346.954 & 0.941 & 0.042 & 65.73 & 0.22 & 20.31 & 0.17 & 0.16 & w\\
  146 & 170.756 & 47.052 & 0.025 & 56.3 & 0.2 & 62.76 & 0.22 & 0.19 & w\\
  147 & 227.507 & 10.805 & 0.047 & 19.79 & 0.05 & 15.29 & 0.46 & 0.42 & w\\
  148 & 208.949 & 14.422 & 0.052 & 44.53 & 0.22 & 41.1 & 0.4 & 0.34 & w\\
  149 & 136.707 & 26.275 & 0.021 & 47.36 & 0.23 & 106.29 & 0.32 & 0.31 & w\\
  150 & 222.778 & 37.495 & 0.032 & 60.17 & 0.27 & 64.95 & 0.25 & 0.22 & w\\
  151 & 223.662 & 12.245 & 0.077 & 63.78 & 0.16 & 13.63 & 0.3 & 0.12 & w\\
  152 & 216.097 & 13.798 & 0.017 & 43.03 & 0.07 & 161.35 & 0.24 & 0.11 & w\\
  153 & 47.029 & 0.456 & 0.074 & 33.95 & 0.11 & 15.19 & 0.34 & 0.29 & w\\
  154 & 133.705 & 57.67 & 0.014 & 22.21 & 0.04 & 31.16 & 0.24 & 0.23 & w\\
  155 & 163.022 & 41.39 & 0.036 & 21.34 & 0.03 & 6.05 & 0.2 & 0.17 & w\\
  156 & 197.296 & 52.021 & 0.026 & 56.74 & 0.2 & 252.18 & 0.22 & 0.19 & w\\
  157 & 197.053 & 52.466 & 0.055 & 25.71 & 0.05 & 8.02 & 0.26 & 0.23 & w\\
\hline\end{tabular}
\end{center}
\end{table*}

\addtocounter{table}{-1}
\begin{table*}
  \begin{center}
    \tabcolsep 5.8pt
    \scriptsize
    \caption{- continued}
\begin{tabular}{|r|l|l|l|l|l|l|l|l|l|l|}
\hline
  \multicolumn{1}{|c|}{Index} &
  \multicolumn{1}{c|}{RA} &
  \multicolumn{1}{c|}{DEC} &
  \multicolumn{1}{c|}{z} &
  \multicolumn{1}{c|}{$\sigma$} &
  \multicolumn{1}{c|}{$M_{vir}$} &
  \multicolumn{1}{c|}{$M_{vir}/L_r$} &
  \multicolumn{1}{c|}{$\overline r_p$} &
  \multicolumn{1}{c|}{$r_h$} &
  \multicolumn{1}{c|}{flag} \\
  \multicolumn{1}{|c|}{} &
  \multicolumn{1}{c|}{} &
  \multicolumn{1}{c|}{} &
  \multicolumn{1}{c|}{} &
  \multicolumn{1}{c|}{$km~s^{-1}$} &
  \multicolumn{1}{c|}{$10^{12}M_{\odot}$} &
  \multicolumn{1}{c|}{$M_{\odot}/L_\odot$} &
  \multicolumn{1}{c|}{Mpc} &
  \multicolumn{1}{c|}{Mpc} &
  \multicolumn{1}{c|}{} \\
\hline
  158 & 180.132 & 38.611 & 0.037 & 45.49 & 0.13 & 36.69 & 0.28 & 0.19 & w\\
  159 & 236.64 & 41.918 & 0.024 & 7.48 & 0.0 & 4.03 & 0.31 & 0.26 & w\\
  160 & 181.533 & 37.326 & 0.044 & 17.71 & 0.03 & 3.59 & 0.28 & 0.25 & w\\
  161 & 147.358 & 26.638 & 0.052 & 44.53 & 0.08 & 20.71 & 0.15 & 0.12 & w\\
  162 & 230.579 & 21.073 & 0.05 & 54.13 & 0.07 & 22.62 & 0.08 & 0.08 & w\\
  163 & 198.864 & 29.074 & 0.035 & 37.95 & 0.07 & 14.98 & 0.27 & 0.14 & w\\
  164 & 241.584 & 39.244 & 0.039 & 61.59 & 0.27 & 161.46 & 0.22 & 0.22 & w\\
  165 & 225.229 & 8.582 & 0.056 & 27.8 & 0.03 & 2.93 & 0.12 & 0.11 & w\\
  166 & 242.068 & 36.093 & 0.043 & 77.42 & 0.61 & 69.3 & 0.34 & 0.31 & w\\
  167 & 244.578 & 34.111 & 0.047 & 66.0 & 0.35 & 61.09 & 0.35 & 0.24 & w\\
  168 & 231.523 & 9.204 & 0.006 & 35.42 & 0.07 & 810.22 & 0.23 & 0.18 & w\\
  169 & 233.449 & 10.78 & 0.045 & 20.04 & 0.06 & 17.13 & 0.46 & 0.43 & w\\
  170 & 135.198 & 29.99 & 0.054 & 57.42 & 0.33 & 84.66 & 0.35 & 0.3 & w\\
  171 & 200.91 & 31.127 & 0.058 & 36.74 & 0.07 & 17.09 & 0.17 & 0.16 & w\\
  172 & 168.201 & 31.813 & 0.075 & 84.66 & 0.21 & 34.22 & 0.11 & 0.09 & w\\
  173 & 179.952 & 29.641 & 0.029 & 64.9 & 0.26 & 221.97 & 0.32 & 0.19 & w\\
  174 & 143.277 & 33.138 & 0.05 & 60.37 & 0.16 & 44.89 & 0.17 & 0.14 & w\\
  175 & 239.097 & 26.158 & 0.064 & 45.73 & 0.11 & 20.73 & 0.16 & 0.16 & w\\
  176 & 322.365 & 0.033 & 0.052 & 19.43 & 0.01 & 2.64 & 0.1 & 0.1 & w\\
  177 & 148.978 & 27.457 & 0.056 & 64.5 & 0.19 & 27.59 & 0.25 & 0.14 & w\\
  178 & 123.265 & 24.567 & 0.02 & 35.67 & 0.01 & 15.67 & 0.06 & 0.03 & k\\
  179 & 201.143 & 43.752 & 0.047 & 42.84 & 0.08 & 8.15 & 0.23 & 0.13 & w\\
  180 & 225.434 & 27.281 & 0.056 & 54.31 & 0.23 & 50.3 & 0.27 & 0.24 & w\\
  181 & 217.671 & 16.054 & 0.048 & 59.27 & 0.4 & 83.81 & 0.38 & 0.34 & w\\
  182 & 212.874 & 16.755 & 0.027 & 64.87 & 0.57 & 277.89 & 0.47 & 0.41 & w\\
  183 & 218.644 & 14.097 & 0.07 & 56.74 & 0.16 & 19.97 & 0.27 & 0.15 & w\\
  184 & 155.957 & 20.985 & 0.066 & 12.07 & 0.01 & 1.75 & 0.35 & 0.3 & w\\
  185 & 154.049 & 22.26 & 0.065 & 60.47 & 0.17 & 23.48 & 0.15 & 0.14 & w\\
  186 & 165.715 & 36.783 & 0.025 & 49.28 & 0.15 & 43.22 & 0.22 & 0.19 & w\\
  187 & 208.955 & 35.151 & 0.035 & 43.12 & 0.12 & 46.73 & 0.3 & 0.19 & w\\
  188 & 166.148 & 36.654 & 0.021 & 18.53 & 0.01 & 8.73 & 0.14 & 0.06 & w\\
  189 & 169.341 & 25.066 & 0.07 & 21.1 & 0.04 & 4.77 & 0.31 & 0.3 & w\\
  190 & 137.867 & 64.475 & 0.018 & 37.07 & 0.04 & 29.25 & 0.17 & 0.08 & w\\
  191 & 213.171 & 30.855 & 0.014 & 45.29 & 0.05 & 53.08 & 0.07 & 0.07 & w\\
  192 & 237.5 & 24.802 & 0.024 & 70.66 & 0.19 & 56.2 & 0.17 & 0.11 & w\\
  193 & 140.155 & 26.538 & 0.049 & 57.84 & 0.15 & 42.46 & 0.14 & 0.14 & w\\
  194 & 217.968 & 54.722 & 0.043 & 69.08 & 0.43 & 108.72 & 0.29 & 0.27 & w\\
  195 & 219.138 & 20.516 & 0.068 & 45.16 & 0.11 & 14.98 & 0.29 & 0.16 & w\\
  196 & 122.262 & 50.179 & 0.017 & 50.58 & 0.05 & 128.31 & 0.09 & 0.06 & w\\
  197 & 122.56 & 51.873 & 0.061 & 6.16 & 0.0 & 0.14 & 0.08 & 0.08 & w\\
  198 & 162.714 & 17.83 & 0.066 & 19.01 & 0.03 & 1.95 & 0.27 & 0.23 & w\\
  199 & 161.602 & 18.531 & 0.049 & 28.23 & 0.06 & 14.35 & 0.26 & 0.23 & w\\
  200 & 200.035 & 30.45 & 0.048 & 43.26 & 0.02 & 3.28 & 0.13 & 0.04 & k\\
  201 & 151.379 & 15.373 & 0.052 & 90.74 & 0.06 & 9.73 & 0.16 & 0.02 & k\\
  202 & 238.768 & 45.486 & 0.028 & 32.6 & 0.07 & 22.08 & 0.21 & 0.2 & w\\
  203 & 205.289 & 12.772 & 0.047 & 31.95 & 0.08 & 13.63 & 0.31 & 0.23 & w\\
  204 & 195.719 & 7.681 & 0.07 & 58.76 & 0.23 & 26.94 & 0.3 & 0.2 & w\\
  205 & 244.44 & 6.065 & 0.038 & 69.97 & 0.2 & 60.12 & 0.14 & 0.13 & w\\
  206 & 238.247 & 8.016 & 0.041 & 27.84 & 0.05 & 13.04 & 0.22 & 0.21 & w\\
  207 & 193.562 & 9.442 & 0.045 & 58.15 & 0.09 & 13.69 & 0.09 & 0.08 & w\\
  208 & 194.728 & 9.301 & 0.054 & 14.75 & 0.01 & 1.19 & 0.19 & 0.13 & w\\
  209 & 222.111 & 34.998 & 0.029 & 49.52 & 0.34 & 51.11 & 0.46 & 0.42 & w\\
  210 & 222.42 & 35.69 & 0.029 & 30.99 & 0.03 & 9.63 & 0.12 & 0.11 & w\\
  211 & 226.85 & 34.076 & 0.044 & 11.04 & 0.01 & 3.07 & 0.25 & 0.2 & w\\
  212 & 205.825 & 20.314 & 0.05 & 11.22 & 0.0 & 0.84 & 0.25 & 0.09 & w\\
  213 & 157.961 & 20.926 & 0.042 & 35.67 & 0.1 & 29.98 & 0.26 & 0.24 & w\\
  214 & 185.17 & 19.107 & 0.045 & 101.61 & 1.18 & 116.97 & 0.41 & 0.35 & w\\
  215 & 131.428 & 19.726 & 0.055 & 40.89 & 0.09 & 11.91 & 0.3 & 0.16 & w\\
  216 & 133.67 & 20.584 & 0.013 & 61.99 & 0.15 & 159.06 & 0.12 & 0.12 & w\\
  217 & 133.51 & 17.343 & 0.067 & 31.57 & 0.04 & 5.47 & 0.12 & 0.12 & w\\
  218 & 134.144 & 16.809 & 0.052 & 38.1 & 0.09 & 11.64 & 0.21 & 0.18 & w\\
  219 & 210.289 & 21.238 & 0.028 & 30.99 & 0.02 & 9.4 & 0.19 & 0.05 & w\\
  220 & 188.095 & 16.31 & 0.057 & 8.48 & 0.0 & 0.52 & 0.19 & 0.17 & w\\
  221 & 181.669 & 16.477 & 0.062 & 79.58 & 0.12 & 20.61 & 0.07 & 0.06 & w\\
  222 & 166.517 & 14.036 & 0.066 & 10.67 & 0.01 & 0.53 & 0.16 & 0.15 & w\\
  223 & 166.711 & 14.457 & 0.052 & 75.54 & 0.27 & 46.95 & 0.16 & 0.14 & w\\
  224 & 143.69 & 13.816 & 0.026 & 43.03 & 0.12 & 141.76 & 0.26 & 0.2 & w\\
  225 & 142.253 & 12.504 & 0.052 & 29.2 & 0.12 & 19.22 & 0.5 & 0.44 & w\\
  226 & 122.272 & 55.442 & 0.031 & 9.89 & 0.01 & 4.34 & 0.31 & 0.31 & w\\
  227 & 194.629 & 24.974 & 0.076 & 66.63 & 0.37 & 29.15 & 0.34 & 0.25 & w\\
  228 & 138.93 & 15.22 & 0.03 & 31.25 & 0.06 & 19.24 & 0.21 & 0.18 & w\\
  229 & 194.542 & 24.349 & 0.023 & 28.69 & 0.06 & 64.35 & 0.25 & 0.24 & w\\
  230 & 333.072 & 0.268 & 0.045 & 45.49 & 0.16 & 49.91 & 0.25 & 0.23 & w\\
  231 & 120.62 & 20.514 & 0.029 & 40.69 & 0.18 & 113.04 & 0.35 & 0.34 & w\\
  232 & 199.703 & 12.397 & 0.037 & 17.65 & 0.01 & 4.89 & 0.12 & 0.12 & w\\
  233 & 213.152 & -1.036 & 0.054 & 30.61 & 0.05 & 10.65 & 0.2 & 0.16 & w\\
  234 & 125.412 & 8.346 & 0.052 & 48.22 & 0.1 & 20.21 & 0.2 & 0.13 & w\\
  235 & 141.748 & 20.884 & 0.051 & 58.76 & 0.44 & 72.27 & 0.45 & 0.39 & w\\
  236 & 214.13 & 57.81 & 0.01 & 37.39 & 0.03 & 42.66 & 0.13 & 0.07 & w\\
  237 & 147.465 & 10.432 & 0.029 & 34.99 & 0.04 & 18.74 & 0.24 & 0.1 & w\\
\hline\end{tabular}
\end{center}
\end{table*}

\addtocounter{table}{-1}
\begin{table*}
  \begin{center}
    \tabcolsep 5.8pt
    \scriptsize
    \caption{- continued}
\begin{tabular}{|r|l|l|l|l|l|l|l|l|l|l|}
\hline
  \multicolumn{1}{|c|}{Index} &
  \multicolumn{1}{c|}{RA} &
  \multicolumn{1}{c|}{DEC} &
  \multicolumn{1}{c|}{z} &
  \multicolumn{1}{c|}{$\sigma$} &
  \multicolumn{1}{c|}{$M_{vir}$} &
  \multicolumn{1}{c|}{$M_{vir}/L_r$} &
  \multicolumn{1}{c|}{$\overline r_p$} &
  \multicolumn{1}{c|}{$r_h$} &
  \multicolumn{1}{c|}{flag} \\
  \multicolumn{1}{|c|}{} &
  \multicolumn{1}{c|}{} &
  \multicolumn{1}{c|}{} &
  \multicolumn{1}{c|}{} &
  \multicolumn{1}{c|}{$km~s^{-1}$} &
  \multicolumn{1}{c|}{$10^{12}M_{\odot}$} &
  \multicolumn{1}{c|}{$M_{\odot}/L_\odot$} &
  \multicolumn{1}{c|}{Mpc} &
  \multicolumn{1}{c|}{Mpc} &
  \multicolumn{1}{c|}{} \\
\hline
  238 & 191.639 & 43.743 & 0.041 & 19.12 & 0.03 & 10.44 & 0.3 & 0.26 & w\\
  239 & 187.382 & 33.427 & 0.031 & 53.35 & 0.13 & 111.86 & 0.15 & 0.14 & w\\
  240 & 31.683 & 13.372 & 0.061 & 67.67 & 0.36 & 86.8 & 0.26 & 0.24 & w\\
  241 & 235.652 & 21.129 & 0.047 & 19.94 & 0.06 & 15.05 & 0.49 & 0.44 & w\\
  242 & 236.745 & 17.884 & 0.011 & 36.96 & 0.02 & 19.6 & 0.06 & 0.04 & w\\
  243 & 217.107 & 17.924 & 0.019 & 19.79 & 0.01 & 4.44 & 0.28 & 0.12 & w\\
  244 & 158.708 & 17.451 & 0.056 & 91.82 & 0.52 & 72.93 & 0.27 & 0.19 & w\\
  245 & 242.707 & 21.051 & 0.038 & 89.03 & 0.13 & 39.42 & 0.25 & 0.05 & w\\
  246 & 219.967 & 42.742 & 0.008 & 35.67 & 0.04 & 55.35 & 0.19 & 0.09 & w\\
  247 & 197.297 & 46.316 & 0.035 & 65.82 & 0.13 & 25.17 & 0.14 & 0.09 & w\\
  248 & 201.996 & 45.759 & 0.061 & 52.63 & 0.24 & 29.6 & 0.3 & 0.26 & w\\
  249 & 201.44 & 46.162 & 0.036 & 25.79 & 0.04 & 14.48 & 0.18 & 0.17 & w\\
  250 & 215.921 & 47.372 & 0.073 & 39.34 & 0.11 & 14.02 & 0.22 & 0.21 & w\\
  251 & 135.03 & 60.083 & 0.039 & 43.54 & 0.06 & 17.61 & 0.14 & 0.09 & w\\
  252 & 323.752 & -0.511 & 0.03 & 47.21 & 0.08 & 37.76 & 0.32 & 0.1 & w\\
  253 & 245.924 & 24.172 & 0.043 & 60.07 & 0.4 & 73.32 & 0.4 & 0.34 & w\\
  254 & 156.546 & 32.099 & 0.037 & 123.63 & 0.67 & 77.02 & 0.24 & 0.13 & w\\
  255 & 162.117 & 38.294 & 0.033 & 15.55 & 0.01 & 4.05 & 0.16 & 0.15 & w\\
  256 & 171.133 & 39.787 & 0.061 & 46.72 & 0.16 & 26.04 & 0.33 & 0.23 & w\\
  257 & 165.93 & 29.242 & 0.066 & 40.07 & 0.1 & 15.4 & 0.23 & 0.18 & w\\
  258 & 166.433 & 35.884 & 0.068 & 75.41 & 0.49 & 92.75 & 0.34 & 0.26 & w\\
  259 & 145.456 & 25.064 & 0.051 & 17.13 & 0.01 & 1.54 & 0.22 & 0.12 & w\\
  260 & 185.047 & 27.68 & 0.08 & 59.07 & 0.18 & 12.74 & 0.21 & 0.16 & w\\
  261 & 247.377 & 13.083 & 0.033 & 49.99 & 0.08 & 34.67 & 0.13 & 0.1 & w\\
  262 & 245.762 & 26.538 & 0.04 & 51.01 & 0.06 & 27.01 & 0.08 & 0.08 & w\\
  263 & 229.252 & 15.566 & 0.048 & 25.71 & 0.01 & 1.36 & 0.08 & 0.06 & w\\
  264 & 199.061 & 41.494 & 0.021 & 31.82 & 0.02 & 7.29 & 0.05 & 0.05 & w\\
  265 & 223.593 & 27.701 & 0.034 & 25.9 & 0.05 & 17.4 & 0.34 & 0.21 & w\\
  266 & 188.546 & 21.542 & 0.045 & 22.74 & 0.01 & 2.69 & 0.13 & 0.04 & w\\
  267 & 220.738 & 17.637 & 0.056 & 40.69 & 0.11 & 23.26 & 0.23 & 0.21 & w\\
  268 & 143.318 & 23.135 & 0.026 & 33.05 & 0.06 & 33.82 & 0.27 & 0.18 & w\\
  269 & 147.104 & 22.363 & 0.034 & 21.94 & 0.02 & 8.91 & 0.2 & 0.15 & w\\
  270 & 113.634 & 41.215 & 0.043 & 99.2 & 0.11 & 33.67 & 0.04 & 0.03 & k\\
  271 & 162.004 & 28.246 & 0.021 & 40.39 & 0.08 & 21.98 & 0.18 & 0.15 & w\\
  272 & 222.889 & 45.438 & 0.037 & 66.24 & 0.12 & 48.1 & 0.22 & 0.09 & w\\
  273 & 175.759 & 23.944 & 0.023 & 61.99 & 0.2 & 76.86 & 0.17 & 0.16 & w\\
  274 & 143.518 & 17.832 & 0.027 & 41.64 & 0.08 & 27.9 & 0.21 & 0.15 & w\\
  275 & 208.292 & 17.332 & 0.026 & 26.85 & 0.04 & 11.85 & 0.32 & 0.18 & w\\
  276 & 210.442 & 18.808 & 0.028 & 58.05 & 0.12 & 85.61 & 0.12 & 0.11 & w\\
  277 & 178.224 & 21.835 & 0.038 & 83.16 & 0.2 & 56.47 & 0.11 & 0.09 & w\\
  278 & 220.809 & 16.488 & 0.037 & 59.57 & 0.11 & 21.82 & 0.11 & 0.09 & w\\
  279 & 229.829 & 13.126 & 0.034 & 51.89 & 0.31 & 150.8 & 0.37 & 0.35 & w\\
  280 & 128.417 & 9.964 & 0.031 & 26.17 & 0.02 & 5.09 & 0.1 & 0.09 & w\\
  281 & 130.646 & 10.585 & 0.007 & 21.2 & 0.03 & 219.44 & 0.3 & 0.17 & w\\
  282 & 201.334 & 18.453 & 0.022 & 35.56 & 0.08 & 24.98 & 0.36 & 0.19 & w\\
  283 & 228.95 & 18.379 & 0.038 & 46.21 & 0.09 & 18.58 & 0.16 & 0.13 & w\\
  284 & 250.806 & 25.34 & 0.055 & 64.59 & 0.23 & 39.97 & 0.19 & 0.17 & w\\
  285 & 135.797 & 10.152 & 0.03 & 31.19 & 0.04 & 13.6 & 0.21 & 0.12 & w\\
  286 & 194.695 & 32.192 & 0.052 & 96.71 & 0.23 & 31.8 & 0.08 & 0.08 & w\\
  287 & 208.915 & 20.744 & 0.075 & 39.57 & 0.13 & 13.94 & 0.36 & 0.25 & w\\
  288 & 208.53 & 21.975 & 0.063 & 47.21 & 0.16 & 10.72 & 0.22 & 0.22 & w\\
  289 & 234.764 & 13.637 & 0.05 & 65.18 & 0.41 & 60.82 & 0.36 & 0.3 & w\\
  290 & 230.4 & 14.384 & 0.025 & 45.16 & 0.13 & 152.92 & 0.21 & 0.19 & w\\
  291 & 38.702 & -7.684 & 0.022 & 44.06 & 0.04 & 13.5 & 0.27 & 0.07 & w\\
  292 & 128.168 & 19.511 & 0.021 & 24.64 & 0.06 & 85.41 & 0.32 & 0.29 & w\\
  293 & 126.569 & 17.362 & 0.066 & 30.21 & 0.07 & 5.76 & 0.24 & 0.23 & w\\
  294 & 192.894 & 5.864 & 0.049 & 84.45 & 0.33 & 56.25 & 0.17 & 0.14 & w\\
  295 & 162.974 & 10.099 & 0.026 & 50.54 & 0.06 & 34.61 & 0.08 & 0.07 & w\\
  296 & 155.252 & 27.238 & 0.038 & 47.78 & 0.15 & 48.1 & 0.24 & 0.19 & w\\
  297 & 126.523 & 13.009 & 0.044 & 27.07 & 0.02 & 5.32 & 0.26 & 0.09 & w\\
  298 & 177.897 & 17.931 & 0.038 & 36.83 & 0.06 & 17.54 & 0.13 & 0.12 & w\\
  299 & 139.151 & 25.303 & 0.043 & 19.01 & 0.01 & 2.94 & 0.17 & 0.1 & w\\
  300 & 241.399 & 16.526 & 0.016 & 14.13 & 0.03 & 69.25 & 0.54 & 0.5 & w\\
  301 & 209.329 & 15.458 & 0.018 & 96.71 & 0.2 & 50.52 & 0.21 & 0.06 & w\\
  302 & 138.357 & 17.124 & 0.056 & 12.95 & 0.01 & 1.86 & 0.24 & 0.12 & w\\
  303 & 49.846 & 0.823 & 0.034 & 49.14 & 0.03 & 16.08 & 0.3 & 0.04 & k\\
  304 & 129.472 & 12.781 & 0.03 & 33.41 & 0.04 & 13.4 & 0.21 & 0.1 & w\\
  305 & 226.674 & 23.642 & 0.016 & 10.19 & 0.01 & 3.04 & 0.27 & 0.24 & w\\
  306 & 202.446 & 6.465 & 0.054 & 56.74 & 0.37 & 69.09 & 0.39 & 0.35 & w\\
  307 & 165.874 & 16.37 & 0.047 & 36.83 & 0.12 & 13.5 & 0.28 & 0.26 & w\\
  308 & 127.046 & 10.541 & 0.031 & 54.31 & 0.22 & 97.78 & 0.26 & 0.23 & w\\
  309 & 179.361 & 7.461 & 0.032 & 26.4 & 0.1 & 21.91 & 0.48 & 0.43 & w\\
  310 & 130.642 & 11.699 & 0.077 & 51.71 & 0.11 & 6.89 & 0.14 & 0.12 & w\\
  311 & 219.595 & 18.111 & 0.03 & 17.13 & 0.02 & 6.65 & 0.31 & 0.18 & w\\
  312 & 246.783 & 9.096 & 0.046 & 87.03 & 0.8 & 262.79 & 0.34 & 0.32 & w\\
  313 & 223.748 & 18.037 & 0.02 & 20.91 & 0.03 & 9.08 & 0.31 & 0.23 & w\\
  314 & 219.307 & 19.422 & 0.03 & 53.31 & 0.18 & 98.43 & 0.31 & 0.2 & w\\
  315 & 175.664 & 24.823 & 0.021 & 50.3 & 0.18 & 78.2 & 0.3 & 0.22 & w\\

\hline\end{tabular}
\end{center}
\end{table*}

\bsp	% typesetting comment
\label{lastpage}
\end{document}